\newif\iflandscape
\newif\ifportrait
\typein[\lorp]%
{Typein "l" (for landscape, two-column) or
"p" (for portrait, one-column)}
\if l\lorp \landscapetrue \else \portraittrue \fi
\newlength{\extralineskip}
\ifportrait
  \documentstyle[12pt]{article}
\fi
\iflandscape
  \documentstyle[twocolumn]{article}
\fi

\def\bea{\begin{eqnarray}}
\def\eea{\end{eqnarray}}
\def\nn{\nonumber}

\def\beq{\begin{equation}}
\def\eeq{\end{equation}}
\def\ba{\beq\new\begin{array}{c}}
\def\ea{\end{array}\eeq}
\def\be{\ba}
\def\ee{\ea}

\makeatletter
\newdimen\normalarrayskip              
\newdimen\minarrayskip                 
\normalarrayskip\baselineskip
\minarrayskip\jot
\newif\ifold             \oldtrue            \def\new{\oldfalse}
\def\arraymode{\ifold\relax\else\displaystyle\fi} 
\def\eqnumphantom{\phantom{(\theequation)}}     
\def\@arrayskip{\ifold\baselineskip\z@\lineskip\z@
     \else
     \baselineskip\minarrayskip\lineskip2\minarrayskip\fi}
\def\@arrayclassz{\ifcase \@lastchclass \@acolampacol \or
\@ampacol \or \or \or \@addamp \or
   \@acolampacol \or \@firstampfalse \@acol \fi
\edef\@preamble{\@preamble
  \ifcase \@chnum
     \hfil$\relax\arraymode\@sharp$\hfil
     \or $\relax\arraymode\@sharp$\hfil
     \or \hfil$\relax\arraymode\@sharp$\fi}}
\def\@array[#1]#2{\setbox\@arstrutbox=\hbox{\vrule
     height\arraystretch \ht\strutbox
     depth\arraystretch \dp\strutbox
     width\z@}\@mkpream{#2}\edef\@preamble{\halign \noexpand\@halignto
\bgroup \tabskip\z@ \@arstrut \@preamble \tabskip\z@ \cr}%
\let\@startpbox\@@startpbox \let\@endpbox\@@endpbox
  \if #1t\vtop \else \if#1b\vbox \else \vcenter \fi\fi
  \bgroup \let\par\relax
  \let\@sharp##\let\protect\relax
  \@arrayskip\@preamble}
\def\eqnarray{\stepcounter{equation}%
              \let\@currentlabel=\theequation
              \global\@eqnswtrue
              \global\@eqcnt\z@
              \tabskip\@centering
              \let\\=\@eqncr
              $$%
 \halign to \displaywidth\bgroup
    \eqnumphantom\@eqnsel\hskip\@centering
    $\displaystyle \tabskip\z@ {##}$%
    &\global\@eqcnt\@ne \hskip 2\arraycolsep
         $\displaystyle\arraymode{##}$\hfil
    &\global\@eqcnt\tw@ \hskip 2\arraycolsep
         $\displaystyle\tabskip\z@{##}$\hfil
         \tabskip\@centering
    &{##}\tabskip\z@\cr}
\def\theequation{\thesection.\arabic{equation}}
\makeatother
\begin{document}
\begin{titlepage}
\setcounter{footnote}0




\phantom.\hfill ITEP-M3/94 \\
\phantom.\hfill CRM-2202 \\
\phantom.\hfill {\it September} 1994 \\

\bigskip

\centerline{{\Large
  FREE-FIELD REPRESENTATION OF GROUP ELEMENT}}
\centerline{{\Large FOR SIMPLE QUANTUM GROUPS}}

\bigskip

\centerline{Alexei Morozov$^\dagger$  and  Luc Vinet$^\ddagger$}

\centerline{$^\dagger$\it ITEP, Moscow, 117 259, Russia}

\centerline{$^\ddagger$\it Centre de recherche math\'ematiques, Universit\'e de
Montr\'eal}
\centerline{\it C.P. 6128, Succ.\ centre-ville, Montr\'eal (Qu\'ebec) H3C 3J7,
Canada}

\bigskip

\bigskip

\centerline{ABSTRACT}

\bigskip

A representation of the group element (also known as ``universal
${\cal T}$-matrix'') which satisfies $\Delta(g) = g\otimes g$,
is given in the form $$ g = \left(\prod_{s=1}^{d_B}\phantom.^>\
{\cal E}_{1/q_{i(s)}}(\chi^{(s)}T_{-i(s)})\right) q^{2\vec\phi\vec H}
\left(\prod_{s=1}^{d_B}\phantom.^<\ {\cal E}_{q_{i(s)}}(\psi^{(s)}
T_{+i(s)})\right)$$ where $d_B = \frac{1}{2}(d_G - r_G)$, $q_i = q^{||
\vec\alpha_i||^2/2}$ and $H_i = 2\vec H\vec\alpha_i/||\vec\alpha_i||^2$
and $T_{\pm i}$ are the generators of quantum group associated
respectively with Cartan algebra and the {\it simple} roots. The ``free
fields'' $\chi,\ \vec\phi,\ \psi$ form a Heisenberg-like algebra:
$$\begin{array}{llr} \psi^{(s)}\psi^{(s')} = q^{-\vec\alpha_{i(s)}
\vec\alpha_{i(s')}} \psi^{(s')}\psi^{(s)}, & \chi^{(s)}\chi^{(s')} =
q^{-\vec\alpha_{i(s)}\vec\alpha_{i(s')}}\chi^{(s')}\chi^{(s)}& {\rm for}
\ s<s', \\ q^{\vec h\vec\phi}\psi^{(s)} = q^{\vec h\vec\alpha_{i(s)}}
\psi^{(s)}q^{\vec h\vec\phi}, & q^{\vec h\vec\phi}\chi^{(s)} = q^{\vec h
\vec\alpha_{i(s)}}\chi^{(s)}q^{\vec h\vec\phi}, & \\ &\psi^{(s)}
\chi^{(s')} = \chi^{(s')}\psi^{(s)} & {\rm for\ any}\ s,s'.\end{array}$$
We argue that the $d_G$-parametric ``manifold'' which $g$ spans
in the operator-valued universal envelopping algebra, can also be
invariant under the group multiplication $g \rightarrow g'\cdot g''$.
The universal ${\cal R}$-matrix with the property that  ${\cal R}
(g\otimes I)(I\otimes g) = (I\otimes g)(g\otimes I){\cal R}$ is given by
the usual formula $${\cal R} = q^{-\sum_{ij}^{r_G}||\vec\alpha_i||^2||
\vec\alpha_j||^2(\vec\alpha\vec\alpha)^{-1}_{ij}H_i\otimes H_j}\prod_{
\vec\alpha > 0}^{d_B}{\cal E}_{q_{\vec\alpha}}\left(-(q_{\vec\alpha}-
q_{\vec\alpha}^{-1})T_{\vec\alpha}\otimes T_{-\vec\alpha}\right).$$

\end{titlepage}

\tableofcontents

\newpage

\section{Introduction}

The notion of group element is the central one in the theory
of Lie algebras and, once available, it should play the same role in
the theory \cite{FRT,D,J} of quantum groups.\footnote{
For one possible ``physical'' application see \cite{GKLMM}.}
In the quantum case parameters, labeling the ``group manifold''
are no longer commuting $c$-numbers. Moreover, a special care is needed
to find a parametrization where their commutation relations are simple
enough. As usual in modern theoretical physics ``simple enough'' means
the Heisenberg-like (``free-field'') relations
$$
\phantom. [\varphi_a,\varphi_b] = \hbar C_{ab}\ \ {\rm or}\ \
e^{\varphi_a}e^{\varphi_b} = q^{C_{ab}}e^{\varphi_b}e^{\varphi_a},\ \
q = e^\hbar
$$
with $\varphi$-independent matrix $C_{ab}$ (which can be further
diagonalized to bring everything into a form of several independent
Heisenberg algebras).
Besides its general significance for understanding the algebraic
structure, hidden in the notion of quantum group, the Heisenberg-type
realization is the simplest one from the technical point of view
and usualy allows to go much
further with any kind of formal manipulations.

For the first time such representation of group element for $q\neq 1$
was constructed quite recently: by C.Fronsdal and A.Galindo \cite{FG1}
in the case of $G = SL(2)_q$ (see also \cite{S}). Their construction
for $G = SL(N)_q$ with $N>2$ \cite{FG2},
however, is not quite satisfactory,
since non-commuting parameters form a sophisticated algebra,
which itself still needs to be ``bosonized''.

The problem can be also formulated at the level of fundamental
representation of $G = SL(N)_q$. In this case $g$ reduces to the
$N\times N$ matrix with the standard non-commuting entries,
sometime refered to as the elements of ``coordinate ring''
of $G$. The question is then about bosonization (free-field
representation) of this non-commutative ring. This problem
has been recently addressed in \cite{M} for the first non-trivial
case of $SL(3)_q$, but the anzats considered there is partly
degenerate and not represented in the transparent enough form.

The goal of this paper is to get rid of these drawbacks of
the previous approaches and present the simple and transparent
expression for the group element of any simple quantum group.
This opens the possibility of straightforward analysis of all
the properties of the group elements, including the very question
of existence of finite-parametric ``group manifolds''.

At our present level of understanding it looks like only one
of many different parametrizations of group manifold, which
are available at the classical level, is adequate for quantization,
provided one wants the quantum parameters to form Heisenberg-like
algebra. As usual in conformal field/quantum group theory all
the adequate representations involve Gauss decomposition
$$
q = q_L g_D g_U
$$
into a product of lower-triangular, diagonal and upper-triangular
parts, so what will be specific is the representation of $g_U$
(and $g_L$) pieces. In the theory of
Wess-\-Zumino-\-Novikov-\-Witten model (see \cite{GMMOS})
one would take for the fundamental representation of $g_U$
just
$$
g_U = \left( \begin{array}{cccc}
     1 & \psi_1 & \psi_{12} & . \\
     0 & 1      & \psi_2    & . \\
     0 & 0      &  1        & . \\
     . & .      &  .        &
      \end{array} \right)
$$
In the group theory the most naive choice (used also in \cite{FG2}) is
\be
g_U = \prod_{\vec\alpha > 0}^{d_B}
\exp\left(\psi_{\vec\alpha}T_{\vec\alpha}\right)
\label{napa}
\ee
with somehow ordered {\it all} the $d_B = \frac{1}{2}(d_G-r_G)$
positive roots $\vec\alpha$.
Both these representations are in fact not good enough for
quantization: in both cases the algebra of $\psi$-variables
is not of the Heisenderg type.

The adequate representation appears to involve only {\it simple}
roots (which is not a big surprise since the quantum group structure
is most naturaly introduced in exactly this sector). In order
to span the entire Borel subalgebra every simple root should be allowed
to appear several times in the product:
\be
g_U = \prod_{s=1}^{d_B}\phantom.^<
\exp \left(\psi^{(s)} T_{+i(s)}\right),
\ee
where $i = 1,\ldots,r_G$ labels {\it simple} roots $\vec\alpha_i$
and the corresponding generators $T_{\pm i} \equiv
T_{\pm\vec\alpha_i}$. Index $s$ labels in some way all the
positive roots of $G$: it can be considered as a
double index, $s = (p,j)$, where $p$ is the ``height'' of the
root, and for given $p$ the second index $j$ is labeling the
roots of the ``height'' $p$ (there are no more than $r_G$ of
them for any given $p$). The map
\be
i(s) = i (p,j) = j.
\ee
For example, in the case of $SL(r+1)$
the sequence $i(s)$ can be
\be
r;\ r-1,\ r;\ r-2,\ r-1,\ r;\ \ldots\ ; 1,2,\ldots,r-1,\ r
\label{order}
\ee
i.e. $p = r,r-1,\ldots,1$ and for given $p$ index $j$ runs from
$p$ to $r$ so that for $SL(3)$
$$
g_U = e^{\psi^{(1)}T_2}e^{\psi^{(2)}T_1}e^{\psi^{(3)}T_2},
$$
and for $SL(4)$
$$
g_U = e^{\psi^{(1)}T_3}e^{\psi^{(2)}T_2}e^{\psi^{(3)}T_3}
e^{\psi^{(4)}T_1}e^{\psi^{(5)}T_2}e^{\psi^{(6)}T_3}
$$
etc., where in the fundamental representation
$$
T_1 = \left( \begin{array}{cccc}
              0&1&0&.\\ 0&0&0&. \\ .&.&.&
             \end{array} \right), \ \
T_2 = \left( \begin{array}{cccc}
              0&0&0&.\\ 0&0&1&. \\ .&.&.&
             \end{array} \right), \ \ldots
$$
In other words, for adequate quantization the ``canonical basis''
in the universal envelopping algebra should be taken in the form
\be
\{T_A\} = \left\{ \prod_s  T_{i(s)}^{n(s)}\right\}
\ee
rather than the one associated with (\ref{napa}),
\be
\{T_A\} = \left\{ \prod_{\vec\alpha > 0}
      T_{\vec\alpha}^{n(\vec\alpha)} \right\}
\ee
Existence of the ``adequate'' and ``non-adequate'' parametrizations
is somewhat alarming and should imply that the algebras of
$\psi$-variables in other parametrizations, though looking
somewhat sophisticated, still can be straightforwardly
bosonized. This aspect of the theory, as well as its relation to
the theory of quantum double \cite{D}, remain to be further
clarified.

There is always a self-consistent reduction (restriction) of
the free-field representation of the group element:
\be
\psi^{(p,j)} = 0 \ \ {\rm for\ all}\ p>1.
\ee
This reduction can be of special interest for quantization
of KP/Toda integrable hierarchies. It was this reduced
(degenerate) representation that was considered in \cite{M}.

The rest of this paper is organized as follows:

Section 2 is devoted to notations and basic properties of
the $q$-exponents.

The basic formula
$$
\Delta(g) = g\otimes g = (g\otimes I)(I\otimes g)
$$
is then proved in Section 3, with some examples in fundamental
representation discussed in Section 4.

Section \ref{secom} contains explicit evaluation of the group
composition rule (``dual comultiplication'') for the case of
$SL(2)_q$. Preliminary discussion of $SL(3)_q$ is presented in
Appendix B.

Section \ref{uniRsec} and Appendix A are  devoted to evaluation of
$$
(I\otimes g)(g\otimes I) = {\cal R}\Delta(g){\cal R}^{-1}
$$
with the universal ${\cal R}$-matrix from refs.\cite{D,B}. Full
details are presented for all the rank-$2$ simple Lie groups.

\newpage

\section{Quantum groups and $q$-exponentials}
\setcounter{equation}0

In this section we describe the crucial consistency property
between quantum comultiplication and the $q$-exponent.
All the notations to be used throught the paper are also
specified here and in eqs.(\ref{ge})-(\ref{not}).

\subsection{ Comultiplication}

Quantization of the universal envelopping algebra, which provides
one possible description of
the quantum group, is most conveniently described in terms of the
Chevalley generators, associated with the {\it simple} roots
$\vec\alpha_i$, $i = 1,\ldots,r_G$. With every $\vec\alpha_i$
there are associated three generators $H_i$, $\hat T_{\pm i}$
and specific $q$-parameter $q_i \equiv q^{\alpha_{ii}/2}$.
Here $\alpha_{ij} \equiv \vec\alpha_i\vec\alpha_j$ and in the
simply-laced case all $\alpha_{ii} = 2$, $q_i=q$.
We also use
a notation
$\displaystyle{\vec h = \frac{1}{2}\sum_{ij}^{r_G} \vec\alpha_i
\alpha^{-1}_{ij} \alpha_{jj}h_j}$, i.e. $q^{\vec h\vec\alpha_i}
= q_i^{h_i}$ for any vector $\vec h$.\footnote{
Note that such $h_i$ are {\it not} coordinates in an orthonormal
basis! For example, $\vec h \vec\xi = \frac{1}{4}
\sum_{ij}^{r_G}\alpha_{ij}^{-1}\alpha_{ii}\alpha_{jj}h_i\xi_j$,
in particular for $SL(2)$ $2\vec h\vec\xi = h\xi$.}

Chevalley generators satisfy the following algebra:
\be
q_i^{2H_i}\hat T_{\pm j} = q^{\pm\alpha_{ij}}\hat T_{\pm j}q_i^{2H_i},
\ \ \ {\rm or}\ \ \ q^{2\vec H\vec\xi}\hat T_{\pm j} =
q_j^{\pm \xi_j}\hat T_{\pm j}q^{2\vec H\vec\xi};
\label{qT} \\
\hat T_{+i}\hat T_{-j} - \hat T_{-j}\hat T_{+i} =
\frac{q_i^{2H_i} - q_i^{-2H_i}}{q_i - q_i^{-1}}\delta_{ij};
\label{TT} \\
\Delta(H_i) = I\otimes H_i + H_i\otimes I; \\
\Delta(\hat T_{\pm i}) = q_i^{H_i}\otimes \hat T_{\pm i} +
      \hat T_{\pm i}\otimes q_i^{-H_i}.
\label{comuThat}
\ee
They are also subjected to Serre constraints to be explicitly
formulated and used in s.\ref{uniRsec} below.

The adequate Gauss decomposition involves, however, the somewhat
different generators:
\be
T_{+i} \equiv \hat T_{+i}q_i^{-H_i}, \nn \\
T_{-i} \equiv q_i^{H_i}\hat T_{-i}
\label{ThattoT}
\ee
which have asymmetric coproducts:
\be
\Delta(T_{+i}) = I\otimes T_{+i} + T_{+i}\otimes q_i^{-2H_i}, \nn \\
\Delta(T_{-i}) = q_i^{2H_i}\otimes T_{-i} + T_{-i}\otimes I.
\ee

\subsection{$q$-exponential}

We define $\ \ [n] = [n]_q = [n]_{1/q}
\equiv \frac{q^n - q^{-n}}{q - q^{-1}}$,
\be
{\cal E}_{q}(x) \equiv \sum_{n\geq 0}
\frac{x^n}{\left(\frac{1-q^{2n}}{1-q^2}\right)!} =
\sum_{n\geq 0}\frac{x^n}{[n]!}
q^{-\frac{1}{2}n(n-1)}.
\ee
It is a simple combinatorial exercise to show that
\be
{\cal E}_q\left(\frac{x}{q-q^{-1}}\right) =
\prod_{n\geq 0} \frac{1}{1+ q^{2n+1}x},
\label{qexpprod}
\ee
thus\footnote{In general, for any $c$-number $s$ ($sx = xs$)
$$
{\cal E}_{1/q}(-sx){\cal E}_q(x) =
\frac{{\cal E}_q(x)}{{\cal E}_q(sx)} = \sum_{n\geq 0}
\frac{q^{-n(n-1)/2}x^n}{[n]!}(1-s)(1-q^2s)\ldots(1-q^{2(n-1)}s)
$$}
\be
{\cal E}_q\left(q^2\frac{x}{q-q^{-1}}\right) =
(1+qx)\ {\cal E}_q\left(\frac{x}{q-q^{-1}}\right)
\label{qexpprod1}
\ee
The properties of such $q$-exponentials, crucial for our
considerations below, are:
the antipod relation
\be
\left({\cal E}_{q}(x)\right)^{-1} = {\cal E}_{1/q}(-x),
\ee
and the Campbell-Hausdorff formula
in the particular case of $xy = q^2yx$,
when it reduces to especially simple pair of ``addition rule''
\be
{\cal E}_{q}(y){\cal E}_{q}(x) = {\cal E}_{q}(x+y),\ \
{\rm if}\ xy = q^2 yx.
\label{sufo}
\ee
and Faddeev-Volkov cocicle identity \cite{FV}
\be
{\cal E}_q(x){\cal E}_q(y) =
{\cal E}_q(x+y + (1-q^{-2})xy) =
{\cal E}_q(y){\cal E}_q\left((1-q^{-2})xy\right)
{\cal E}_q(x), \ \
{\rm if}\ xy = q^2yx,
\label{FVc}
\ee
which we shall need in Section \ref{secom} in the following form:
\be
{\cal E}_{q}(y){\cal E}_{q}(x) = {\cal E}_q(x+y) = \nn \\ =
{\cal E}_q(x){\cal E}_q\left(
\frac{1}{1+(1-q^{-2})x}\ y\right),\ \
{\rm if}\ xy = q^2yx
\label{FVr}
\ee

\subsection{Comultiplication versus $q$-exponential}

The basic relation for all our reasoning below is:
for any root of the length $||\vec\alpha||^2 = 2$
\be
{\cal E}_{q}\left( \psi \Delta(T_{+})\right) =
{\cal E}_{q}\left(\psi(I\otimes T_{+} + T_{+}\otimes q^{-2H})\right)
= \nn \\ =
{\cal E}_{q}\left(\psi T_+\otimes q^{-2H}\right)
{\cal E}_{q}(\psi I\otimes T_+).
\label{coprEq}
\ee
This is just an application of (\ref{sufo}) for
$x = I\otimes T_+$ and $y = T_+\otimes q^{-2H}$.
Similarly
\be
{\cal E}_{1/q}\left( \chi \Delta(T_{-})\right) =
{\cal E}_{1/q}\left(\chi(q^{2H}\otimes T_{-} + T_{-}\otimes I)\right)
= \nn \\ =
{\cal E}_{1/q}\left(\chi T_-\otimes I\right)
{\cal E}_{1/q}(\chi q^{2H}\otimes T_-).
\label{coprEqi}
\ee

For non-simply-laced algebras when roots of different lenghts are
present, the $q$-parameter of the $q$-exponential should be just
changed for $q_{\vec\alpha} \equiv q^{||\vec\alpha||^2/2}$.
In ss.\ref{secom},\ref{uniRsec} below
we''ll need some other corollaries of (\ref{sufo}).

\newpage

\section{Representation of the group element}
\setcounter{equation}0

Let us now take the natural ansatz
\be
g = \left({\prod_s}^> {\cal E}_{1/q_{i(s)}}(\chi^{(s)}T_{-i(s)})
  \right)\ q^{2\vec\phi\vec H}\ \left(
{\prod_s}^< {\cal E}_{q_{i(s)}}(\psi^{(s)}T_{+i(s)})\right)
\label{ge}
\ee
with the ordering defined by the map $i(s)$ (superscripts ``$<$'' and
``$>$'' stand for the straight and inverse sequences $i(s)$
respectively),
and check the group property
\be
\Delta(g) = g\otimes g = (g\otimes I)(I\otimes g)
\ee
Note that in the limit $q \rightarrow 1$ parameters $\vec\phi$
get large: $\phi \sim 1/\log q$.

In order to simplify the formulas below we shall substitute the
index $i(s)$ by $s$:
\be
q_s \equiv q_{i(s)}, \ \ \ T_{\pm s} \equiv T_{\pm i(s)}, \ \ \
H_s \equiv H_{i(s)},\ \ \ \
\alpha_{ss'} \equiv \vec\alpha_{i(s)}\vec\alpha_{i(s')}.
\label{not}
\ee

By definition
\be
\Delta(g) =
{\prod_s}^> {\cal E}_{1/q_{i(s)}}\left(\chi^{(s)}\Delta(T_{-s})\right)\
  \left(q^{2\vec\phi\vec H}\otimes q^{2\vec\phi\vec H}\right)\
{\prod_s}^< {\cal E}_{q_{s}}\left(\psi^{(s)}\Delta(T_{+s})\right)
\ee
Now we use (\ref{coprEq}), (\ref{coprEqi}):
\be
\Delta(g) =
{\prod_s}^> \left\{ {\cal E}_{1/q_{s}}
\left(\chi^{(s)}T_{-s}\otimes I\right)\
{\cal E}_{1/q_{s}}\left(\chi^{(s)} q_{s}^{2H_{s}}
\otimes T_{-s}\right)\right\}
\cdot\nn\\ \cdot  \left(q^{2\vec\phi\vec H}\otimes I\right)
  \left(I\otimes q^{2\vec\phi\vec H}\right)\
{\prod_s}^< \left\{
    {\cal E}_{q_{s}}\left(\psi^{(s)}T_{+s}
\otimes q_{s}^{-2H_{s}}\right)\
    {\cal E}_{q_{s}}\left(\psi^{(s)}I\otimes T_{+s}\right)\right\}
\label{Delta(g)}
\ee

At the same time
\be
g\otimes g =
\left\{{\prod_s}^> {\cal E}_{1/q_{s}}\left(\chi^{(s)}T_{-s}
\otimes I\right)\
      \left(q^{2\vec\phi\vec H}\otimes I\right)\
      {\prod_s}^< {\cal E}_{q_{s}}\left(\psi^{(s)}T_{+s}
\otimes I\right)\right\}
\cdot\nn \\ \cdot
\left\{{\prod_s}^> {\cal E}_{1/q_{s}}
\left(\chi^{(s)} I\otimes T_{-s}\right)\
      \left(I\otimes q^{2\vec\phi\vec H}\right)\
      {\prod_s}^< {\cal E}_{q_{s}}\left(\psi^{(s)}I
\otimes T_{+s}\right)\right\}
\label{gotimesg}
\ee

In order for (\ref{Delta(g)}) and (\ref{gotimesg}) to coincide it
is enough if:

(a) The $q$-exponents in the square brackets in (\ref{Delta(g)})
can be adequately reordered:
\be
{\prod_s}^< \left\{ {\cal E}_{q_{s}}\left(\psi^{(s)}T_{+s}
\otimes q_{s}^{-2H_{s}}\right)\ {\cal E}_{q_{s}}\left(\psi^{(s)}I
\otimes T_{+s}\right)\right\} = \nn \\ =
{\prod_s}^<  {\cal E}_{q_{s}}\left(\psi^{(s)}T_{+s}
\otimes q_{s}^{-2H_{s}}\right)\
{\prod_s}^<  {\cal E}_{q_{s}}\left(\psi^{(s)}I\otimes T_{+s}\right)
\label{acond}
\ee
and similarly for exponents with $\chi^{(s)}$.

(b) The first factor at the r.h.s. of (\ref{acond}) can be pushed
to the left through $\left( I\otimes q^{2\vec\phi\vec H}\right)$ in
(\ref{Delta(g)}), so that
\be
\left( I\otimes q^{2\vec\phi\vec H}\right)
{\prod_s}^<
    {\cal E}_{q_{s}}\left(\psi^{(s)}T_{+s}
\otimes q_{s}^{-2H_{s}}\right) = \nn \\
=     {\prod_s}^<
    {\cal E}_{q_{s}}\left(\psi^{(s)}T_{+s}\otimes I\right)\
\left( I\otimes q^{2\vec\phi\vec H}\right)
\label{bcond1}
\ee
Similarly
\be
{\prod_s}^> {\cal E}_{1/q_{s}}
\left(\chi^{(s)}q_{s}^{2H_{s}}\otimes T_{-s}\right)
\ \left(q^{2\vec\phi\vec H}\otimes I\right) = \nn \\ =
\left(q^{2\vec\phi\vec H}\otimes I\right)\
{\prod_s}^> {\cal E}_{1/q_{s}}
\left(\chi^{(s)} I\otimes T_{-s}\right)
\label{bcond2}
\ee

(c) The first factor at the r.h.s. of (\ref{bcond1}) commutes with
the second factor at the r.h.s. of (\ref{bcond2}).

Conditions (a), (b), (c)  are defining the commutation properties
of the $\chi,\ \phi,\ \psi$ variables.

Indeed, (a) is fulfiled provided the {\it arguments} of the
relevant $q$-exponents commute:
\be
{\rm for\ any}\ s<s'\ \ \
\left[ \psi^{(s')}T_{+s'}\otimes q_{s}^{-2H_{s'}}, \
    \psi^{(s)}I\otimes T_{+s}\right] = 0
\nn
\ee
Since
$$
\left(T_{+s'}\otimes q_{s}^{-2H_{s'}}\right)
    \left(I\otimes T_{+s}\right) = q^{-\alpha_{ss'}}
 \left(I\otimes T_{+s}\right)
\left(T_{+s'}\otimes q_{s}^{-2H_{s'}}\right)
$$
this implies that
\be
\psi^{(s)}\psi^{(s')} = q^{-\alpha_{ss'}}\psi^{(s')}\psi^{(s)},
\ \ s<s'.
\ee
Similarly
\be
{\rm for\ any}\ s<s'\ \ \
\left[ \chi^{(s)}T_{-s}\otimes I, \
    \chi^{(s')} q_{s}^{2H_{s'}}\otimes T_{-s'}\right] = 0
\nn
\ee
and therefore
\be
\chi^{(s)}\chi^{(s')} = q^{-\alpha_{ss'}}\chi^{(s')}\chi^{(s)},
\ \ s<s'.
\ee

Condition (\ref{bcond1}) is fulfiled if
\be
\left( I\otimes q^{2\vec\phi\vec H}\right) \psi^{(s)} =
\psi^{(s)}\left( I\otimes q_{s}^{2H_{s}}\right)
\left( I\otimes q^{2\vec\phi\vec H}\right),
\nn
\ee
i.e. (since $q^{\vec H\vec\alpha_i}=q_i^{H_i}$)
if for any vector $\vec h$
\be
q^{\vec h\vec\phi}\psi^{(s)} = q^{\vec h\vec\alpha_s}
\psi^{(s)}q^{\vec h\vec\phi} \ \ \ \ \ {\rm or}\ \ \ \ \
q_i^{\phi_i}\psi^{(s)} = q^{\alpha_{is}}\psi^{(s)}q_i^{\phi_i}
\ee
Similarly from (\ref{bcond2})
\be
q^{\vec h\vec\phi}\chi^{(s)} = q^{\vec h\vec\alpha_s}
\chi^{(s)}q^{\vec h\vec\phi} \ \ \ \ \ {\rm or}\ \ \ \ \
q_i^{\phi_i}\chi^{(s)} = q^{\alpha_{is}}\chi^{(s)}q_i^{\phi_i}
\ee

Finally (c) implies that all the $\psi$'s commute with all the
$\chi$'s.

These commutation relations are nicely consistent with
``Hermitean representation'', satisfying $g^* = g$, where
operator $*$ is defined so, that
\be
T_{\pm\vec\alpha}^* = T_{\mp\vec\alpha}, \nn \\
\left(q^{H_i}\right)^* = q^{H_i}, \nn \\
q^* = \frac{1}{q}
\ee
and
\be
\forall A,B \ \ \ (AB)^* = B^*A^*, \nn \\
\forall A\ \ \ (A^*)^* = A.
\ee
Thus $\vec\phi^* = \vec\phi$, $\chi_i = \psi_i^*$ and
explicitly ``Hermitean'' group element looks like
\be
{\prod_s}^> {\cal E}_{1/q_s}(\psi_s^*T_{-s})\
q^{(\vec\phi^* + \vec\phi)\vec H}\
{\prod_s}^< {\cal E}_{q_s}(\psi_sT_{s})
\ee

\newpage

\section{Examples of fundamental representation}
\setcounter{equation}0

In any fundamental representation of $G_q$
\be
\frac{q_{i}^{2H_i}-q_{i}^{-2H_i}}{q_i-q_i^{-1}} = 2H_i
\label{frprop}
\ee
and (\ref{qT}), (\ref{TT}) (but not (\ref{comuThat})!) reduce
to the ordinary commutation relations of $G$ (with $q=1$).
Thus $T_{i}$ (or $\hat T_i$)
can be chosen to be just the ordinary $q$-independent matrices.

\subsection{The case of $SL(2)_q$ $\left(A(1)_q\right)$}

Since $2\vec\phi\vec H = \phi H = \frac{1}{2}\phi\sigma_3$ and
$T_{\pm} = \sigma_\pm$,
eq.(\ref{ge}) gives in this case:
\be
g_{{\rm fund}} = \left(\begin{array}{cc}  a&b\\ c&d \end{array}\right) =
{\cal E}_{1/q}(\chi\sigma_-)q^{\phi\sigma_3/2}{\cal E}_q(\psi\sigma_+)
= \nn \\ =
\left(\begin{array}{cc}  1&0\\ \chi&0 \end{array}\right)
\left(\begin{array}{cc}  q^{\phi/2} & 0\\ 0&q^{-\phi/2}
\end{array}\right)
\left(\begin{array}{cc}  1&\psi\\ 0&1 \end{array}\right)
= \left(\begin{array}{cl}  q^{\phi/2}&q^{\phi/2} \psi\\
    \chi q^{\phi/2} & \chi q^{\phi/2}\psi + q^{-\phi/2}
\end{array}\right),
\ee
and
\be
q^{\phi/2}\psi = q\psi q^{\phi/2}, \nn \\
q^{\phi/2}\chi = q\chi q^{\phi/2}, \nn \\
\psi\chi = \chi\psi
\ee
imply the usual relations:
\be
ab = qba, \ \
ac = qca, \nn \\
bd = q db, \ \
cd = qdc, \nn \\
bc= cb, \nn \\
ad - da = (q-q^{-1})bc, \nn \\
ad-qbc = da - q^{-1}bc = 1.
\ee
Comparison with representation
\be
\left(\begin{array}{cc}  a&b\\ c&d \end{array}\right) =
\left(\begin{array}{cl}  u_1v_1 & u_1v_2 \\
       u_2v_1&u_2v_2 + \frac{1}{u_1v_1} \end{array}\right)
\ee
from \cite{M} with
\be
u_iu_j = qu_ju_i, \ \ v_iv_j = qv_jv_i, \ \ i<j
\ee
is also straightforward.

\subsection{The case of $SL(3)_q$ $\left(A(2)_q\right)$}

For $N=3$ we similarly obtain:
\be
g_{{\rm fund}} =
\left( \begin{array}{ccc}  a&b&e \\ c&d&f \\ g&h&k \end{array}\right)
= \nn \\ \bigskip \bigskip =
\left( \begin{array}{ccc}  1&0&0 \\ 0&1&0\\
         0&\chi^{(3)}&1 \end{array}\right)
\left( \begin{array}{ccc}  1&0&0 \\ \chi^{(2)}&1&0\\
         0&0&1 \end{array}\right)
\left( \begin{array}{ccc}  1&0&0 \\ 0&1&0\\
         0&\chi^{(1)}&1 \end{array}\right) \cdot
\left( \begin{array}{ccc}  q^{\phi_{(1)}}&0&0 \\ 0&q^{\phi_{(2)}}&0 \\
         0&0&q^{\phi_{(3)}} \end{array}\right) \cdot\nn \\ \cdot
\left( \begin{array}{ccc}  1&0&0 \\ 0&1&\psi^{(1)}\\
         0&0&1 \end{array}\right)
\left( \begin{array}{ccc}  1&\psi^{(2)}&0 \\ 0&1&0\\
         0&0&1 \end{array}\right)
\left( \begin{array}{ccc}  1&0&0 \\ 0&1&\psi^{(3)}\\
         0&0&1 \end{array}\right)
= \nn \\ \bigskip \bigskip =
\left( \begin{array}{ccc}  1&0&0 \\ \chi^{(2)}&1&0\\
       \chi^{(2)}\chi^{(3)}&\chi^{(1)}+\chi^{(3)}&1 \end{array}\right)
\left( \begin{array}{ccc}  q^{\phi_{(1)}}&0&0 \\ 0&q^{\phi_{(2)}}&0 \\
         0&0&q^{\phi_{(3)}} \end{array}\right)
\left( \begin{array}{ccc}  1&\psi^{(2)}&\psi^{(2)}\psi^{(3)} \\
          0&1&\psi^{(1)}+\psi^{(3)}\\
         0&0&1 \end{array}\right)
= \nn \\ \bigskip \bigskip =
\left( \begin{array}{lll} q^{\phi_{(1)}}& q^{\phi_{(1)}}\psi^{(2)}&
q^{\phi_{(1)}}\psi^{(2)}\psi^{(3)} \\
 \chi^{(2)}q^{\phi_{(1)}} & \chi^{(2)}q^{\phi_{(1)}}\psi^{(2)} +
q^{\phi_{(2)}} &
\chi^{(2)}q^{\phi_{(1)}}\psi^{(2)}\psi^{(3)} +
q^{\phi_{(2)}}(\psi^{(1)} + \psi^{(3)}) \\
\chi^{(3)}\chi^{(2)}q^{\phi_{(1)}} &
\chi^{(3)}\chi^{(2)}q^{\phi_{(1)}}\psi^{(2)} + &
\chi^{(3)}\chi^{(2)}q^{\phi_{(1)}}\psi^{(2)}\psi^{(3)} + \\
& \ \ \ + (\chi^{(1)} +\chi^{(3)})q^{\phi_{(2)}} & \ \ \ +
(\chi^{(1)} +\chi^{(3)})q^{\phi_{(2)}}(\psi^{(1)} + \psi^{(3)})
+ q^{\phi_{(3)}}
\end{array}\right)
\label{sl3fund}
\ee
Now $\vec\alpha(1) = \vec\alpha(3) = \vec\alpha_2$,
$\vec\alpha(2) = \vec\alpha_1$,
\be
2\vec\phi\vec H =
\phi_{(1)}
\left(\begin{array}{ccc}1&0&0\\0&-1&0\\0&0&0\end{array}\right)
-\phi_{(3)}
\left(\begin{array}{ccc}0&0&0\\0&1&0\\0&0&-1\end{array}\right) =\\ =
2\phi_{(1)}H_1 -2\phi_{(3)}H_2 = 2\left(\frac{2\phi_1+\phi_2}{3}H_1 +
\frac{\phi_1 + 2\phi_2}{3} H_2\right),
\ee
i.e.
\be
\phi_{(1)} = \frac{2\phi_1+\phi_2}{3}, \nn \\
\phi_{(2)} = -\phi_{(1)} - \phi_{(3)} = \frac{-\phi_1+\phi_2}{3}, \nn \\
\phi_{(3)} = -\frac{\phi_1 + 2\phi_2}{3}
\ee

Again, Heisenberg-like commutation relations
\be
\psi^{(1)}\psi^{(2)} = q\psi^{(2)}\psi^{(1)}, \ \
\psi^{(2)}\psi^{(3)} = q\psi^{(3)}\psi^{(2)}, \ \
\psi^{(1)}\psi^{(3)} = q^{-2}\psi^{(3)}\psi^{(1)}, \nn \\
q^{\phi_{(1)}}\psi^{(1)} = \psi^{(1)}q^{\phi_{(1)}}, \ \
q^{\phi_{(1)}}\psi^{(2)} = q\psi^{(2)}q^{\phi_{(1)}}, \ \
q^{\phi_{(1)}}\psi^{(3)} = \psi^{(3)}q^{\phi_{(1)}}, \nn \\
q^{\phi_{(2)}}\psi^{(1)} = q\psi^{(1)}q^{\phi_{(2)}}, \ \
q^{\phi_{(2)}}\psi^{(2)} = q^{-1}\psi^{(2)}q^{\phi_{(2)}}, \ \
q^{\phi_{(2)}}\psi^{(3)} = q\psi^{(3)}q^{\phi_{(2)}}, \nn \\
q^{\phi_{(3)}}\psi^{(1)} = q^{-1}\psi^{(1)}q^{\phi_{(3)}}, \ \
q^{\phi_{(3)}}\psi^{(2)} = \psi^{(2)}q^{\phi_{(3)}}, \ \
q^{\phi_{(3)}}\psi^{(3)} = q^{-1}\psi^{(3)}q^{\phi_{(3)}}
\ee
(plus the same with $\psi \rightarrow \chi$)
reproduce the standard  quadratic relations for
$a,b,c,\ldots,k$.

In the reduced case of $\chi^{(1)} = \psi^{(1)} = 0$
eq.(\ref{sl3fund}) reproduces the result of \cite{M}
(with $w_2 = \frac{v_2}{v_1} = \psi^{(2)}$,
$w_3 = \frac{v_3}{v_2} = \psi^{(3)}$,
$q^{\phi_{(1)}} = u_1v_1$, $q^{\phi_{(3)}} =
\frac{u_4v_4}{u_3v_3}$.



\newpage

\section{Group composition rule \label{secom}}
\setcounter{equation}0

The group composition law (the ``dual comultiplication'') is
defined by the following relation:
\be
g\left(\Delta^*(\chi),\Delta^*(\phi),\Delta^*(\psi)\right)
= g(\chi\otimes I,\phi\otimes I, \psi\otimes I)\
g(I\otimes\chi, I\otimes\phi, I\otimes\psi) = \nn \\ =
g(\chi',\phi',\psi')\ g(\chi'',\phi'',\psi'')
\ee
Given representation (\ref{ge}) for $g(\chi,\phi,\psi)$
it should be just straightforward calculation that leads to explicit
formulas for $\Delta^*(\chi),\Delta^*(\phi),\Delta^*(\psi)$.

\subsection{The case of $SL(2)_q$}

Let us begin, as usual, from the case of a single simple root
(i.e. $SL(2)_q$). In order to avoid possible confusion we remind
that in our notation $2\vec\phi\vec H = \phi H$.
In this case we need to evaluate
\be
g(\chi',\phi',\psi')g(\chi'',\phi'',\psi'') = \nn \\ =
{\cal E}_{1/q}(\chi'T_-)q^{\phi'H}{\cal E}_q(\psi'T_+)
{\cal E}_{1/q}(\chi''T_-)q^{\phi''H}{\cal E}_q(\psi''T_+)
\ee

1) The first thing to do is permutation of
${\cal E}_q(\psi'T_+)$ and ${\cal E}_{1/q}(\chi''T_-)$.

Sandwiching (\ref{TT}) between $T_-^a$ and $T_-^b$, we get:
\be
\left[T_+, T_-^n\right] =
\sum_{a+b = n-1}T_-^a \frac{q^{2H}-q^{-2H}}{q-q^{-1}}T_-^b = \nn \\
= \frac{[n]q^{-(n-1)}}{q-q^{-1}}
\left(T_-^{n-1}q^{2H} - q^{-2H}T_-^{n-1}\right)
\ee
Summation over $n$ with the weight $\frac{\chi^n}{[n]!}
q^{+n(n-1)/2}$ produces:
$$
\left(T_+ + \frac{\chi}{q-q^{-1}}q^{-2H}\right)
{\cal E}_{1/q}(\chi T_-) = {\cal E}_{1/q}(\chi T_-)
\left(T_+ + \frac{\chi}{q-q^{-1}}q^{2H}\right)
$$
This in turn implies that
\be
{\cal E}_q\left(\psi' T_+ +
\frac{\psi'\chi''}{q-q^{-1}}q^{-2H}\right)
{\cal E}_{1/q}(\chi'' T_-) = {\cal E}_{1/q}(\chi'' T_-)
{\cal E}_q
\left(\psi'T_+ + \frac{\psi'\chi''}{q-q^{-1}}q^{2H}\right)
\ee
(all the variables with one prime commute with
those with two primes).

Now it is time to use (\ref{sufo}):
\be
{\cal E}_q
\left(\psi'T_+ + \frac{\psi'\chi''}{q-q^{-1}}q^{2H}\right) =
{\cal E}_q(\psi'T_+) {\cal E}_q
\left(\frac{\psi'\chi''}{q-q^{-1}}q^{2H}\right),
\ee
while
\be
{\cal E}_q
\left(\psi'T_+ + \frac{\psi'\chi''}{q-q^{-1}}q^{-2H}\right) =
{\cal E}_q
\left(\frac{\psi'\chi''}{q-q^{-1}}q^{-2H}\right)
{\cal E}_q(\psi'T_+),
\ee
and we obtain:
\be
{\cal E}_q(\psi'T_+){\cal E}_{1/q}(\chi''T_-) =
{\cal E}_{1/q}\left(-\frac{\psi'\chi''}{q-q^{-1}}q^{-2H}\right)
{\cal E}_{1/q}(\chi''T_-){\cal E}_q(\psi'T_+)
{\cal E}_q
\left(\frac{\psi'\chi''}{q-q^{-1}}q^{2H}\right)
\ee

2) The next thing to do is to carry ${\cal E}_q(\psi'T_+)
{\cal E}_q
\left(\frac{\psi'\chi''}{q-q^{-1}}q^{2H}\right)$ to the right
through $q^{\phi''H}$. This is easy:
\be
{\cal E}_q(\psi'T_+)
{\cal E}_q
\left(\frac{\psi'\chi''}{q-q^{-1}}q^{2H}\right)\
q^{\phi''H} \ = \ q^{\phi''H} \
{\cal E}_q(\psi'q^{-\phi''}T_+)
{\cal E}_q
\left(\frac{\psi'\chi''}{q-q^{-1}}\right)
\label{inte2.1}
\ee
Similarly
\be
q^{\phi'H} \ {\cal E}_{1/q}
\left(-\frac{\psi'\chi''}{q-q^{-1}}q^{-2H}\right)
{\cal E}_{1/q}(\chi'T_-) =
{\cal E}_{1/q}\left(-\frac{\psi'\chi''}{q-q^{-1}}\right)
{\cal E}_{1/q}(\chi'q^{-\phi'}T_-)
q^{\phi'H}
\ee

3) Now we need to permute the last two factors at the r.h.s. of
(\ref{inte2.1}). This is where the Faddeev-Volkov relation
(\ref{FVr}) plays its role:
\be
{\cal E}_q(\psi'q^{-\phi''}T_+)
{\cal E}_q
\left(\frac{\psi'\chi''}{q-q^{-1}}\right) =
{\cal E}_q
\left(\frac{\psi'\chi''}{q-q^{-1}}\right)
{\cal E}_q\left(\frac{1}{1 + q^{-1}\psi'\chi''} q^{-\phi''}
\psi'T_+\right) = \nn \\ =
{\cal E}_q
\left(\frac{\psi'\chi''}{q-q^{-1}}\right)
{\cal E}_q\left(q^{-\phi''/2}\frac{1}{1 + \psi'\chi''} q^{-\phi''/2}
\psi'T_+\right)
\ee
Similarly,
\be
{\cal E}_{1/q}
\left(-\frac{\psi'\chi''}{q-q^{-1}}\right)
{\cal E}_{1/q}(\chi''q^{-\phi'}T_-) =
{\cal E}_{1/q}\left(q^{-\phi'/2}\frac{1}{1 + \psi'\chi''}
q^{-\phi'/2}\chi''T_-\right)
{\cal E}_{1/q}
\left(-\frac{\psi'\chi''}{q-q^{-1}}\right)
\ee

4) With the help of (\ref{sufo}) we now get:
\be
{\cal E}_q\left(q^{-\phi''/2}\frac{1}{1 + \psi'\chi''} q^{-\phi''/2}
\psi'T_+\right){\cal E}_q(\psi''T_+) = \nn \\ =
{\cal E}_q\left(\left(\psi'' \ +\
q^{-\phi''/2}\frac{1}{1 + \psi'\chi''} q^{-\phi''/2}
\psi'\right)T_+\right) = {\cal E}_q\left(\Delta^*(\psi)T_+\right),
\ee
i.e.
\be
\Delta_{SL(2)}^*(\psi) = \psi'' +
q^{-\phi''/2}\frac{1}{1 + \psi'\chi''} q^{-\phi''/2}\psi' = \nn \\ =
I\otimes \psi\  +\  \psi\otimes q^{-\phi/2}\frac{1}{
I\otimes I + \psi\otimes \chi}I\otimes q^{-\phi/2}
\label{delstapsi}
\ee
Similarly
\be
\Delta_{SL(2)}^*(\chi) = \chi' +
q^{-\phi'/2}\frac{1}{1 + \psi'\chi''} q^{-\phi'/2}\chi'' = \nn \\ =
\chi\otimes I\  +\  q^{-\phi/2}\otimes I\frac{1}{
I\otimes I + \psi\otimes \chi}q^{-\phi/2}\otimes \chi
\label{delstachi}
\ee

5) What remains to be found is
\be
q^{H\Delta^*(\phi)} =
{\cal E}_{1/q}\left(-\frac{\psi'\chi''}{q-q^{-1}}\right)
q^{H(\phi'+\phi'')}
{\cal E}_{q}\left(\frac{\psi'\chi''}{q-q^{-1}}\right)
\label{del*phi1}
\ee
In order to evaluate this quantity we can use the trick,
usual in the $q$-bosonization theory \cite{MV}.
Let us note, that for $\Phi = \phi'+\phi''$ and
$z = \psi'\chi''$ we have $\Phi z = z(\Phi +4)$,
thus $\Phi z^n = z^n(\Phi +4n)$ and for any function
$f(z)$
$$
\Phi e^{f(z)} = e^{f(z)}\left(\Phi + 4z\frac{df(z)}{dz}\right),
$$
or
\be
e^{-f(z)}q^{H\Phi}e^{f(z)} = q^{H(\Phi + 4zdf/dz)}
\ee
Thus $\Delta^*(\phi) = \Phi + 4z \frac{df(z)}{dz} $, where,
according to (\ref{qexpprod}),
$f(z) = \log{\cal E}_q\left(\frac{z}{q-q^{-1}}\right) =
- \sum_{k=0}^\infty \log(1 + q^{2k+1}z)$, and
$z\frac{df(z)}{dz} = - \sum_{k=0}^\infty \frac{q^{2k+1}z}{1 + q^{2k+1}z}
= -\sum_{l=1}^{\infty}\frac{(-z)^l}{q^l - q^{-l}}$.
Finaly,
\be
\Delta_{SL(2)}^*(\phi) =
\phi' + \phi'' -4\sum_l \frac{(-\psi'\chi'')^l}{q^l - q^{-l}}
= \nn \\ =
\phi\otimes I + I\otimes \phi - \frac{4}{q-q^{-1}}\sum_l
\frac{(-\psi\otimes \chi)^l}{[l]}
\label{delstaphi}
\ee
(It deserves noting that in the classical limit $q\rightarrow 1$
the field $\phi$ gets large: $\phi \sim 1/\log q$.)

6) Alternative description of $\Delta^*(\phi)$ also deserves
mentioning. It is especially useful in considerations of
particular representations and of the higher-rank groups.

Let us rewrite (\ref{del*phi1}) as
\be
q^{H\Delta^*(\phi)} = q^{H\phi'}
{\cal E}_{1/q}\left(-\frac{\psi'\chi''}{q-q^{-1}}q^{-2H}\right)
{\cal E}_{q}\left(\frac{\psi'\chi''}{q-q^{-1}}q^{+2H}\right)
q^{H\phi''}
\label{del*phi2}
\ee
and make use of the infinite-product representation (\ref{qexpprod})
for the $q$-exponentials. This is most straightforward when
$2H$ has integer eigenvalues (i.e. for the finite-dimensional
representations of $SL(2)_q$). If $H=0$ eq.(\ref{del*phi2})
turns into identity $1=1$. Let further $H = \frac{1}{2}$.
Then the product of two exponents at the r.h.s., evaluated with
the help of (\ref{qexpprod1}) is just $(1+\psi'\chi')$ and
$$
q^{\Delta^*(\phi)/2} = q^{\phi'/2}(1+\psi'\chi'')q^{\phi''/2}
$$
If $H = -\frac{1}{2}$ we get
$$
q^{-\Delta^*(\phi)/2} = q^{-\phi'/2}\frac{1}{1+\psi'\chi''}
q^{-\phi''/2} =  q^{-\phi''/2}\frac{1}{1+\psi'\chi''}q^{-\phi'/2}
= \left(q^{\Delta^*(\phi)/2}\right)^{-1}
$$
If, further, $H=1$,
$$
q^{\Delta^*(\phi)} = q^{\phi'}(1+\frac{1}{q}\psi'\chi'')
(1+q\psi'\chi'')q^{\phi''} =
\left(q^{\phi'/2}(1+\psi'\chi'')q^{\phi''/2}\right)^2
$$
and similarly for any {\it integer} $2H>0$
\be
q^{\pm H\Delta^*(\phi)} = q^{\pm H\phi'}
\left(\prod_{n=1}^{2H}(1+q^{2n-2H-1}\psi'\chi'')\right)
q^{\pm H\phi''} =
\left(q^{\phi'/2}(1+\psi'\chi'')q^{\phi''/2}\right)^{\pm 2H}
\label{del*phi3}
\ee
For non-integer $2H$ this formula can be derived by analytical
continuation (though the infinite-product representation
(\ref{qexpprod}) {\it per se} is not of direct use in the general
case).

Formulas (\ref{delstapsi}), (\ref{delstachi}), (\ref{delstaphi})
are not new: they are already found in ref.\cite{FG1}.
They can be easily derived in the fundamental representation for
$T$-generators, when these are just ordinary $q$-independent
matrices $\left(t_{ab}\right)$ and $\Delta^*(t_{ab})
= \sum_c t_{ac}\otimes t_{cb}$. Above derivation demonstrates
explicitly that $\Delta^*$ in coordinates $\chi, \phi, \psi$
is independent of representation. This actualy proves the existence
of a {\it closed} 3-parametric
subgroup in the (operator-valued!) universal enveloping algebra,
i.e. justifies - at least for the case of $SL(2)_q$ -
introduction of the notion of ``group'' for $q\neq 1$.

\subsection{Comments on the general case}

For higher-rank groups the first three steps of calculation literaly
repeat those for $SL(2)_q$, it is only necessary to make use
of the following commutativity properties:
\be
\left[ T_{i}, T_{-j}\right] = 0\ \ \ {\rm for}\ i\neq j,
\ \ \ {\rm so\ that}\ \
\left[ T_{s}, T_{-s'}\right] \neq 0\ \ \ {\rm only\ for}\
i(s)=i(s'), \nn \\
\left[ \psi_sT_{\pm s}, \psi_{s'}q_{s'}^{\mp 2H_{s'}}\right] = 0
\ \ \ {\rm for}\ s<s',\nn \\
\left[ \chi_sT_{\pm s}, \chi_{s'}q_{s'}^{\mp 2H_{s'}}\right] = 0
\ \ \ {\rm for}\ s<s' \nn
\ee
This is enough to get:
\be
g\{\chi'_s,\phi'_s,\psi'_s\}\ g\{\chi''_s,\phi''_s,\psi''_s\} =
\tilde g_L  q^{2\vec H(\vec\phi'+\vec\phi'')} \tilde g_U
\ee
where $\tilde g_U$ can be represented in one of the four
equivalent forms:
\be
\tilde g(u) =
\left(\prod_{\stackrel{s,s':}{i(s)=i(s')}}^{d_B}\phantom.^<
{\cal E}_{q_s}\left(
\frac{\psi'_s\chi''_{s'}}{q_s-q_s^{-1}}\right)\right) \Delta^*(g_U)
= \\ = \prod_s^{d_B}\phantom.^< {\cal E}_{q_s}
\left(\psi_s'q_s^{-\phi''_s}T_{s}\right)
\prod_{\stackrel{s,s':}{i(s)=i(s')}}^{d_B}\phantom.^<
{\cal E}_{q_s}\left(
\frac{\psi'_s\chi''_{s'}}{q_s-q_s^{-1}}\right)
\prod_s^{d_B}\phantom.^< {\cal E}_{q_s}(\psi_s''T_{s}) = \\ =
\prod_s^{d_B}\phantom.^< \left\{
{\cal E}_{q_s}\left( q_s^{-\phi''_s}\psi'_sT_{s}\right)
\prod_{\stackrel{s':}{i(s')=i(s)}}^{d_B}\phantom.^<
{\cal E}_{q_s}\left(\frac{\psi'_s\chi''_{s'}}{q_s-q_s^{-1}}\right)
\right\}\prod_s^{d_B}\phantom.^< {\cal E}_{q_s}(\psi''_sT_s) =\\ =
\prod_s^{d_B}\phantom.^<
{\cal E}_{q_s}\left( \psi'_s q_s^{-\phi''_s}T_{s}
+ \frac{1}{q_s-q_s^{-1}}\sum_{\stackrel{s':}{i(s')=i(s)}}^{d_B}
\psi'_s\chi''_{s'}\right)
\prod_s^{d_B}\phantom.^< {\cal E}_{q_s}(\psi''_sT_s) = \\ =
\prod_s^{d_B}\phantom.^<  \left\{\left(
\prod_{\stackrel{s':}{i(s')=i(s)}}^{d_B}\phantom.^<
{\cal E}_{q_s}\left(\frac{\psi'_s\chi''_{s'}}{q_s-q_s^{-1}}\right)
\right)   {\cal E}_{q_s}\left( q_s^{-\phi''_s/2}\ (1+
\sum_{\stackrel{s':}{i(s')=i(s)}}^{d_B}\psi'_s\chi''_{s'})^{-1}\
q_s^{-\phi''_s/2}\psi'_sT_{s}\right)\right\}\cdot \\ \cdot
\prod_s^{d_B}\phantom.^< {\cal E}_{q_s}(\psi''_sT_s)
\label{d*gu.0}
\ee
(Exact ordering in the double products is unessential:
what is important is that the item with labels $(s_1,s'_1)$
stands before that with $(s_2,s'_2)$ whenever {\it both}
$s_1\leq s_2, s_1'\leq s_2'$. In the ``mixed'' situation the items
just commute.)
To prove the equivalencies one  makes use of relations like
\be
\left[ \psi_s'\chi_s'',\ \psi_{s'}'q^{-\phi''_{s'}}\right] = 0,
\ \ \ {\rm for}\ s<s'
\nn
\ee
as well as Faddeev-Volkov identity.
Similar expressions can be written down for $\tilde g_L$.

Thus the problem is essentially reduced to the study of Borel
elements and bringing them to our ``standard form'',
\be
\Delta^*(g_L) = \prod_s^{d_B}\phantom.^> {\cal E}_{1/q_s}
\left(\Delta^*(\chi_s)T_{-s}\right), \\
\Delta^*(g_U) =
\prod_s^{d_B}\phantom.^< {\cal E}_{q_s}
\left(\Delta^*(\psi_s)T_s\right)
\ee
Also in the Cartan sector one has:
\be
q^{2\Delta^*(\vec\phi)\vec H} =
\left(\prod_{\stackrel{s,s':}{i(s)=i(s')}}^{d_B}\phantom.^>
{\cal E}_{1/q_s}\left(
-\frac{\psi'_s\chi''_{s'}}{q_s-q_s^{-1}}\right)\right)
\ q^{2(\vec\phi'+\vec\phi'')\vec H}\ \left(
\prod_{\stackrel{s,s':}{i(s)=i(s')}}^{d_B}\phantom.^<
{\cal E}_{q_s}\left(
\frac{\psi'_s\chi''_{s'}}{q_s-q_s^{-1}}\right)\right) = \\ =
q^{2\vec\phi'\vec H} \ \left(
\prod_{\stackrel{s,s':}{i(s)=i(s')}}^{d_B}\phantom.^>
{\cal E}_{1/q_s}\left(
-\frac{\psi'_s\chi''_{s'}}{q_s-q_s^{-1}}q^{-2H_s}\right)
\prod_{\stackrel{s,s':}{i(s)=i(s')}}^{d_B}\phantom.^<
{\cal E}_{q_s}\left(
\frac{\psi'_s\chi''_{s'}}{q_s-q_s^{-1}}q^{2H_s}\right)
\right)\ q^{2\vec\phi''\vec H}
\label{del*car}
\ee
Explicit resolution of these equations makes use of the
Serre identities and related generalizations of the
$q$-exponential identities like (\ref{FVc}).
Detailed discussion remains beyond the scope of the present paper.
As already mentioned, it is of crucial importance
for the proof of existence of the notion of the $d_G$-dimensional
``group manifold'' for $q \neq 1$ and will be addressed elsewhere.
For illustrative purposes we consider the example of $SL(3)_q$
in Appendix B at the end of this paper.

\newpage

\section{Universal ${\cal R}$-matrix \label{uniRsec}}
\setcounter{equation}0

Since we now possess an {\it explicit} representation of the group
element, all other relations from the theory of quantum groups
can be straightforwardly {\it derived}. The (universal)
${\cal R}$-matrix should not be an exclusion. Non-trivial
${\cal R}$-matrix appears because of the non-commutativity of
the $\chi,\phi,\psi$-parameters, implying that
$$
(I\otimes g)(g\otimes I) \neq (g\otimes I)(I\otimes g) =
g\otimes g = \Delta(g).
$$
Instead \cite{FRT}
\be
\Delta(g) = {\cal R}^{-1}(I\otimes g)(g\otimes I){\cal R}.
\label{uniR}
\ee
Thus in order to determine ${\cal R}$ one should evaluate
$(I\otimes g)(g\otimes I)$.

\subsection{From group to algebra}

Given (\ref{ge}), this is a straightforward exercise:
\be
(I\otimes g)(g\otimes I) = \nn \\ =
\left( {\prod_s}^> {\cal E}_{1/q_{s}}
\left(\chi^{(s)}I\otimes T_{-s}\right)\
(I\otimes q^{2\vec\phi\vec H})\
{\prod_s}^< {\cal E}_{q_{s}}
\left(\psi^{(s)}I\otimes T_{+s}\right)\right)
\cdot \nn \\ \cdot
\left( {\prod_s}^> {\cal E}_{1/q_{s}}
\left(\chi^{(s)}T_{-s}\otimes I\right)\
(q^{2\vec\phi\vec H}\otimes I)\
{\prod_s}^< {\cal E}_{q_{s}}
\left(\psi^{(s)}T_{+s}\otimes I\right)\right)
\ee
Of the $6$ different factors on the r.h.s. the $4$-th and the $3$-rd
can be freely permuted. Further,
\be
(I\otimes q^{2\vec\phi\vec H})\
{\cal E}_{1/q_{s}}\left(\chi^{(s)}T_{-s}\otimes I\right) = \nn \\ =
{\cal E}_{1/q_{s}}\left(\chi^{(s)}T_{-s}\otimes q^{2H_{s}}\right)\
(I\otimes q^{2\vec\phi\vec H})
\ee
and
\be
{\cal E}_{q_{s}}\left(\psi^{(s)}I\otimes T_{+s}\right)\
(q^{2\vec\phi\vec H}\otimes I) = \nn \\ =
(q^{2\vec\phi\vec H}\otimes I)\
{\cal E}_{q_{s}}\left(\psi^{(s)} q_{s}^{-2H_{s}}\otimes T_{+s}\right)
\ee
Thus
\be
(I\otimes g)(g\otimes I) = \nn \\ =
{\prod_s}^> {\cal E}_{1/q_{s}}\left(\chi^{(s)}I\otimes T_{-s}\right)\
{\prod_s}^> {
\cal E}_{1/q_{s}}\left(\chi^{(s)}T_{-s}\otimes q^{2H_{s}}\right)\
\cdot \nn \\ \cdot ( q^{2\vec\phi\vec H}\otimes  q^{2\vec\phi\vec H})
{\prod_s}^< {
\cal E}_{q_{s}}\left(\psi^{(s)} q_{s}^{-2H_{s}}\otimes T_{+s}\right)\
{\prod_s}^< {\cal E}_{q_{s}}\left(\psi^{(s)}T_{+s}\otimes I\right)
\ee
The two next steps are already familiar from the Section 3 above:
the two first and the two last products can be reordered, e.g.
\be
{\prod_s}^< {\cal E}_{q_{s}}
\left(\psi^{(s)} q_{s}^{-2H_{s}}\otimes T_{+s}\right)\
{\prod_s}^< {\cal E}_{q_{s}}
\left(\psi^{(s)}T_{+s}\otimes I\right) = \nn\\ = {\prod_s}^<
\left\{ {\cal E}_{q_{s}}\left(\psi^{(s)}T_{+s}\otimes I\right)
{\cal E}_{q_{s}}
\left(\psi^{(s)} q_{s}^{-2H_{s}}\otimes T_{+s}\right)\right\}
\label{int1}
\ee
(this is because $ [\psi^{(s)}T_{+s}\otimes I, \
\psi^{(s')} q_{s'}^{-2H_{s'}}\otimes T_{+s'}] = 0 $
for $ s<s'$), and further (\ref{sufo}) can be used to rewrite every
pair product in the square brackets in (\ref{int1}) as a single
$q$-exponent:
\be
{\cal E}_{q_{s}}\left(\psi^{(s)}T_{+s}\otimes I\right)
{\cal E}_{q_{s}}\left(\psi^{(s)} q_{s}^{-2H_{s}}\otimes T_{+s}\right)
= \nn \\ =
{\cal E}_{q_{s}} \left(\psi^{(s)}\left( T_{+s}\otimes I +
   q_{s}^{-2H_{s}}\otimes T_{+s}\right)\right)
\ee

Finally we get:
\be
\Delta(g) = \nn \\ =
{\prod_s}^> {\cal E}_{1/q_{s}}
\left(\chi^{(s)}\Delta(T_{-s})\right)\
(q^{2\vec\phi\vec H}\otimes q^{2\vec\phi\vec H})\
{\prod_s}^< {\cal E}_{q_{s}}\left(\psi^{(s)}
\Delta(T_{+s})\right) = \nn \\ =
{\prod_s}^> {\cal E}_{1/q_{s}}\left(\chi^{(s)}
\left(q_{s}^{2H_{s}}\otimes T_{-s} + T_{-s}\otimes I\right)\right)
\cdot \nn \\ \cdot
(q^{2\vec\phi\vec H}\otimes q^{2\vec\phi\vec H})\
{\prod_s}^< {\cal E}_{q_{s}}\left(\psi^{(s)}
\left(I\otimes T_{+s}+ T_{+s}\otimes
q_{s}^{-2H_{s}}\right)\right)
= \nn \\ = {\cal R}^{-1}(I\otimes g)(g\otimes I){\cal R} = \nn \\ =
{\cal R}^{-1}
{\prod_s}^> {\cal E}_{1/q_{s}}\left(\chi^{(s)}
\left(I\otimes T_{-s} + T_{-s}\otimes
q_{s}^{2H_{s}}\right)\right)\cdot \nn \\ \cdot
(q^{2\vec\phi\vec H}\otimes q^{2\vec\phi\vec H}\
{\prod_s}^< {\cal E}_{q_{s}}\left(\psi^{(s)}
\left(q_{s}^{-2H_{s}}\otimes T_{+s}+ T_{+s}
\otimes I\right)\right){\cal R}.
\ee
It is clear now that the task of ${\cal R}$-matrix is to perform
the transformation:
\be
\Delta(T_{+i}) = I\otimes T_{+i} + T_{+i}\otimes q_{i}^{-2H_i} =
{\cal R}^{-1}\left(q_{i}^{-2H_i}\otimes T_{+i} + T_{+i}\otimes I\right)
{\cal R}, \nn \\
\Delta(T_{-i}) = q_{i}^{2H_i}\otimes T_{-i} + T_{-i}\otimes I =
{\cal R}^{-1}\left(I\otimes T_{-i} + T_{-i}\otimes q_{i}^{2H_i}\right)
{\cal R}
\label{defR}
\ee
for all the {\it simple} roots $i = 1,\ldots,r_G$,
provided
\be
{\cal R} (q^{2\vec\phi\vec H}\otimes q^{2\vec\phi\vec H}) =
(q^{2\vec\phi\vec H}\otimes q^{2\vec\phi\vec H}) {\cal R}.
\label{RcoCa}
\ee
Thus we see that the universal ${\cal R}$-matrix
for the {\it group} elements, defined in (\ref{uniR}), is  just
the same as the ordinary one, introduced by (\ref{defR}),
(\ref{RcoCa}) for the generators $T$ of the {\it algebra}
(see also \cite{FG1,S,FG2} for a more abstract
reasoning).
For the sake of completeness
in the remainder of this section we''ll briefly describe the
solution \cite{D,B} to the relations (\ref{defR}), (\ref{RcoCa}).

\subsection{
The case of $SL(2)_q$}

Condition (\ref{RcoCa}) is  satisfied, if ${\cal R}$
contains the generators $T_{\vec\alpha}$ only in the combinations
$T_{+\vec\alpha}\otimes T_{-\vec\alpha}$, what will be always the
case below. In accordance with (\ref{defR})
${\cal R}$ is naturally decomposed, ${\cal R = \hat Q\hat R}$,
so that the role of $\hat{\cal Q}$ is to move the factors
$q_i^{\mp 2H_i}$ to the ``right place'':
\be
\hat{\cal Q}^{-1}\left(q_i^{-2H_i}\otimes T_{+i} + T_{+i}\otimes I\right)
\hat{\cal Q} = \left(I\otimes T_{+i} + T_{+i}\otimes q_i^{+2H_i}\right),
\nn \\
\hat{\cal Q}^{-1}\left(I\otimes T_{-i} + T_{-i}\otimes q_i^{2H_i}\right)
{\cal Q} = \left(q_i^{-2H_i}\otimes T_{-i} + T_{-i}\otimes I\right),
\label{Qhat2}
\ee
while the task of $\hat{\cal R}$ is to reverse the sign of $2H_i$:
\be
\hat{\cal R}^{-1}\left(I\otimes T_{+i} + T_{+i}\otimes q_i^{+2H_i}\right)
\hat{\cal R} = \left(I\otimes T_{+i} + T_{+i}\otimes q_i^{-2H_i}\right),
\label{Rhat1}
\ee
\be
\hat{\cal R}^{-1}\left(q_i^{-2H_i}\otimes T_{-i} + T_{-i}\otimes I\right)
\hat{\cal R} = \left(q_i^{+2H_i}\otimes T_{-i} + T_{-i}\otimes I\right).
\label{Rhat2}
\ee
Solution to (\ref{Qhat2}) is obviously given by
\be
\hat{\cal Q} = q^{-\sum_{ij} \alpha_{ii}\alpha_{jj}
(\alpha^{-1})_{ij}H_i\otimes H_j}, \ \ \ \alpha_{ij} =
\vec\alpha_i\vec\alpha_j
\ee

Construction of $\hat{\cal R}$ is somewhat more complicated.
The basic relation is the following. Let us fix $i$ and take
$T_\pm = T_{\pm i}$, $q = q_i$. Then
\be
\phantom. [ (I\otimes T_+), (T_+\otimes T_-)^n] =
T_+^n\otimes\left(\frac{q^{2H}-q^{-2H}}{q-q^{-1}}T_-^{n-1} + T_-
\frac{q^{2H}-q^{-2H}}{q-q^{-1}}T_-^{n-2} + \ldots\right) = \nn \\ =
\frac{[n]q^{n-1}}{q-q^{-1}}\left((T_+\otimes q^{2H})
(T_+\otimes T_-)^{n-1}
- (T_+\otimes T_-)^{n-1}(T_+\otimes q^{-2H})\right)
\nn
\ee
Summation over $n$ with the weight
$\frac{\left(-(q-q^{-1})\right)^n}{[n]!}q^{-\frac{1}{2}n(n-1)}$
gives now:
\be
\left(I\otimes T_+ + T_+\otimes q^{2H}\right)
{\cal E}_{q}(-(q-q^{-1})T_+\otimes T_-) = \nn \\ =
{\cal E}_{q}(-(q-q^{-1})T_+\otimes T_-)
\left(I\otimes T_+ + T_+\otimes q^{-2H}\right)
\label{propRi}
\ee
This looks just as (\ref{Rhat1}) for given $i$ and
\be
\hat{\cal R}_i =
{\cal E}_{q_{i}}\left(-(q_i-q_i^{-1})T_{+i}\otimes T_{-i}\right)
\label{Ri}
\ee
Relation (\ref{Rhat2}) is also satisfied.

This provides the complete answer for universal ${\cal R}$-matrix
in the case of $SL(2)_q$ \cite{D} (when there is only one value
the index $i$ can take):
\be
{\cal R}^{SL(2)} = \hat{\cal Q}\hat{\cal R} = q^{-2H\otimes H}
{\cal E}_{q}\left(-(q-q^{-1})T_{+}\otimes T_{-}\right).
\label{Rsl2}
\ee

For groups of higher rank there is additional complication
because $\hat{\cal R}_j$ acts non-trivially on some other $T_i$'s
with $i\neq j$. Moreover this action produces the generators
$T_{\vec\alpha}$, associated with all the {\it non-simple}
roots $\vec\alpha$, which can be futher eliminated
by conjugation with the corresponding $\hat{\cal R}_{\vec\alpha}$
matrices, so that finally
\be
{\cal R} = \hat{\cal Q}
\prod_{\vec\alpha > 0}^{d_B}\hat{\cal R}_{\vec\alpha}
\label{Rgen}
\ee
Note, that the product here - in varience with (\ref{ge}) -
is over {\it all} the positive roots $\vec\alpha$, and,
as an obvious generalization of (\ref{Ri}),
\be
\hat{\cal R}_{\vec\alpha} =
{\cal E}_{q_{\vec\alpha}}\left(-(q_{\vec\alpha}-q_{\vec\alpha}^{-1})
T_{\vec\alpha}\otimes T_{-\vec\alpha}\right),
\label{Ralpha}
\ee
where $q_{\vec\alpha} = q^{||\vec\alpha||^2/2}$ and generators
$T_{\vec\alpha}$ for non-simple $\vec\alpha$ still need to be
defined.

To make {\it la raison d'etre} of this general formula more clear
let us consider the simplest example of the rank-$2$ groups.
This example is enough to clarify all the subtle points which could
cause confusion in explicit check of (\ref{Rgen}) for any simple
group $G_q$.

\subsection{Serre relations and specification of generators for
$q\neq 1$}

Before we can proceed to actual calculations we need the explicit
form of Serre identities:
\be
\begin{array}{ll}
{\rm if}\ \alpha_{ij} = 0, &
T_{i}T_{j} = T_{j}T_{i}; \\
{\rm if}\ \alpha_{ij} = -\frac{1}{2}\alpha_{ii},&
T_{i}^2T_{j} - (q_i+q_i^{-1})T_iT_jT_i + T_jT_i^2 = 0; \\
{\rm if}\ \alpha_{ij} = -\alpha_{ii}, &
T_i^3 T_j - (q_i^2 + 1 + q_i^{-2})(T_i^2T_jT_i - T_iT_jT_i^2)
- T_iT_j^3 = 0;\\
{\rm if}\ \alpha_{ij} = -\frac{3}{2}\alpha_{ii}, &
T_i^4T_j - \left(\begin{array}{c} 1 \\ 4 \end{array}\right)_i
(T_i^3T_jT_i + T_iT_jT_i^3) +
 \left(\begin{array}{c} 2 \\ 4 \end{array}\right)_iT_i^2T_jT_i^2 = 0,
\end{array} \label{serre+}
\ee
with $\left(\begin{array}{c} 1 \\ 4 \end{array}\right)_i =
[4]_i = q_i^3 + q_i + q_i^{-1} + q_i^{-3}$ and
$\left(\begin{array}{c} 2 \\ 4 \end{array}\right)_i =
\frac{[4]_i!}{[2]_i[2]_i} = (q_i^2 + 1 + q_i^{-2})(q_i^2 + q_i^{-2})$.
The same is true for negative generators:
\be
\begin{array}{ll}
{\rm if}\ \alpha_{ij} = 0, &
T_{-i}T_{-j} = T_{-j}T_{-i}; \\
{\rm if}\ \alpha_{ij} = -1, &
T_{-i}^2T_{-j} - (q_i+q_i^{-1})T_{-i}T_{-j}T_{-i} +
T_{-j}T_{-i}^2 = 0;
\end{array} \label{serre-} \\
{\rm etc} \nn
\ee
(note that the transformation (\ref{ThattoT}) has no effect on these
relations: they look just the same for $\hat T$'s).

Whenever $\alpha_{ij}\neq 0$ for $i \neq j$
the commutator of the two generators $T_{i}$ and
$T_{j}$ in the classical ($q=1$)
case is equal to $T_{ij} = T_{\vec\alpha_i+\vec\alpha_j}$.
Serre identities imply that the multiple commutator,
which would produce $T_{(1-a_{ij})\vec\alpha_i+\vec\alpha_j}$
with $a_{ij} \equiv \frac{2\alpha_{ij}}{\alpha_{ii}}$
($-a_{ij} = 0,1,2,3$ for simple groups), vanishes.
When $q\neq 1$ the definition of $T_{\vec\alpha}$ for non-simple
roots $\alpha$ involves $q$-commutators and needs some care.
Let us make some notational agreements. Namely,
assume that the simple roots of the bigger lenght are ascribed
smaller numbers $i$, $\alpha_{ii} \geq \alpha_{jj}$ for $i<j$.
This guarantees that for $i<j$, $\alpha_{ij}$ is either vanishing
or $\alpha_{ij} = -\frac{1}{2}\alpha_{ii}$, while the ratio
$2\alpha_{ij}/\alpha_{jj}$ can still be equal to $0, -1, -2, -3$.
Everywhere below $[n]_{\vec\alpha} = [n]_{q_{\vec\alpha}}$,
$[n]_j = [n]_{q_j}$.

Let us further {\it define}:
$$
{\rm for}\ i<j,\ \ \ \alpha_{ij} = -\alpha_{ii}/2:
$$
\be
T_{ij} \equiv T_iT_j - q_iT_jT_i,
\label{Tij}
\ee
\be
T_{ijj} \equiv \frac{1}{[2]_j}\left(T_{ij}T_j -
\frac{q_i}{q_j^2}T_jT_{ij}\right),
\label{Tijj}
\ee
\be
T_{ijjj} \equiv \frac{1}{[3]_j}\left(T_{ijj}T_j -
\frac{q_i}{q_j^4}T_jT_{ijj}\right),
\label{Tijjj}
\ee
\be
T_{ijjjj} \equiv \frac{1}{[4]_j}\left(T_{ijjj}T_j -
\frac{q_i}{q_j^6}T_jT_{ijjj}\right)
\label{Tijjjj}
\ee
and also
\be
T_{iij}\equiv T_iT_{ij} - q_i^{-1}T_{ij}T_i,
\label{Tiij}
\ee
\be
T_{iijjj} \equiv \frac{1}{[3]_j}\left(T_{ij}T_{ijj} -
\frac{q_i}{q_j^4}T_{ijj}T_{ij}\right)
\label{Tiijjj}
\ee
Similarly for negative generators
$$
{\rm for}\ i<j,\ \ \ \alpha_{ij} = -\alpha_{ii}/2
$$
\be
T_{-ji} \equiv T_{-j}T_{-i} - \frac{1}{q_i}T_{-i}T_{-j},
\label{T-ji}
\ee
\be
T_{-jji} \equiv \frac{1}{[2]_j}\left( T_{-j}T_{-ji} -
\frac{q_j^2}{q_i}T_{-ji}T_{-j}\right),
\label{T-jji}
\ee
\be
T_{-jjji} \equiv  \frac{1}{[3]_j}\left(T_{-j}T_{-jji} -
\frac{q_j^4}{q_i}T_{-jji}T_{-j}\right),
\label{T-jjji}
\ee
\be
T_{-jjjji} \equiv  \frac{1}{[4]_j}\left( T_{-j}T_{-jjji} -
\frac{q_j^6}{q_i}T_{-jjji}T_{-j}\right)
\label{T-jjjji}
\ee
and
\be
T_{-jii}\equiv T_{-ji}T_{-i} - {q_i}T_{-i}T_{-ji} \label{T-jii}, \nn\\
T_{-jjjii} \equiv \frac{1}{[3]_j}\left(T_{-jji}T_{-ji} -
\frac{q_j^4}{q_i}T_{-ji}T_{-jji}\right) \label{T-jjjii}
\ee
Then Serre identities can be represented in the following form:
$$
{\rm for}\ i<j,\ \ \ \alpha_{ij} = -\alpha_{ii}/2:
$$
\be
{\rm always}\ \ \ \ T_{iij} = 0,
\label{serTiij}
\ee
\be
\begin{array}{llcl}
{\rm if}\ \alpha_{ij} = -\frac{1}{2}\alpha_{jj}, & {\rm thus}\ q_i=q_j,
\ &{\rm then}\ & T_{ijj}=0, \\
{\rm if}\ \alpha_{ij} = -\alpha_{jj},& {\rm thus}\ q_i=q_j^2, \
& {\rm then}\  & T_{ijjj} = 0, \\
{\rm if}\ \alpha_{ij} = -\frac{3}{2}\alpha_{jj},& {\rm thus}\ q_i=q_j^3,
\ &{\rm then}\  & T_{ijjjj} =0
\end{array}
\label{ser+}
\ee

\bigskip

and similarly for negative generators. As direct corollary of Serre
identites we also have:
$$
{\rm if}\ \alpha_{ij} = -\frac{1}{2}\alpha_{jj}\ \ {\rm or}\ \ \
\alpha_{ij} = -\alpha_{jj},
$$
\be
T_{iijjj} = T_{-jjjii} = 0
\label{serTiijjj}
\ee

This set of relations, together with the basic definitions
(\ref{qT}), (\ref{TT})  imply natural definitions   of all the
generators $T_{\vec\alpha}$ and their commutations properties.
It is our next task to list the relevant relations.\footnote{
It can deserve noting that above definitions of the generators
$T_{\vec\alpha}$ for $q\neq 1$ is not the
only possible one, consistent with representation of the
Serre identities in the simple form (\ref{ser+}). This
choice is, however, motivated by the desire to obtain the
ingredients of the universal ${\cal R}$ matrix
in the simple form (\ref{Ralpha}).}

\subsection{Commutation relations between $T_{\vec\alpha}$ for
rank-$2$ groups\label{core}}

To begin with, one more remark about notations.
According to our general agreement in the
non-simply laced case the root ``$1$'' is the bigger one.
In most formulas $q_1$ and $q_2$ are preserved as different
variables, so that the formulas can be used in more general
framework (beyond consideration of the simple rank-$2$
algebras). However, some formulas, involving the generators
$T_{1222}$ and $T_{11222}$, are written in the
form where $q_1= q_2^3$ is already substituted. Otherwise
the formulas would look too ugly.
This restriction is of now harm, at least for considerations
of simple groups, since $G_2$ is the only example when
$q_i = q_j^3$ (in simply-laced case always $q_i=q_j$ and
for $B$,$C$-series sometime also $q_i = q_j^2$). In formulas
where substitution $q_1 = q_2^3$ has been made we write ``$\approx$''
instead of ``$=$''.\footnote{
Examples of what happens when $q_1$ is independent
of $q_2$ are given by the following relations:
$$
T_{11222}T_2 - \frac{q_1^2}{q_2^6}T_2T_{11222} =
T_{12}T_{1222} - \frac{q_1^2}{q_2^6}T_{1222}T_{12},
$$
$$
\left(\frac{q_1^2}{q_2^4}[3]_2 - q_2[2]_2\right)T_{122}T_{1222}
= q_1\left([3]_2 - \frac{q_1^2}{q_2^6}q_2[2]_2\right)
T_{1222}T_{122}
$$
$$
{\rm etc.}
$$
}
\be
T_1T_2 - q_1T_2T_1 - T_{12} \equiv 0,
\label{T1T2}
\ee
\be
T_1T_{12} - \frac{1}{q_1}T_{12}T_1 \equiv 0,
\label{T1T12}
\ee
\be
T_1T_{122} - T_{122}T_1 = \frac{1}{q_1[2]_2}\left(1 -
\frac{q_1^2}{q_2^2}\right)T_{12}^2,
\label{T1T122}
\ee
\be
T_1T_{1222} - q_1T_{1222}T_1 + \frac{q_2^2}{q_1}T_{11222} =\nn \\
= \frac{1}{[3]_2}\left(1 - \frac{q_1^2}{q_2^4}\right)
\left(\frac{q_2}{q_1}[2]_2 T_{12}T_{122} + T_{122}T_{12}\right),
\label{T1T1222}
\ee
\be
T_1T_{11222} - \frac{1}{q_1}T_{11222}T_1 =
\frac{1}{q_1^2[2]_2[3]_2}\left(1 - \frac{q_1^2}{q_2^2}\right)
\left(1 - \frac{q_1^2}{q_2^4}\right)T_{12}^3;
\label{T1T11222} \\
\ee

\be
T_{12}T_2 - \frac{q_1}{q_2^2} T_2T_{12} - [2]_2T_{122} \equiv 0,
\label{T12T2}
\ee
\be
T_{122}T_2 - \frac{q_1}{q_2^4}T_2T_{122} - [3]_2T_{1222}
\equiv 0,
\label{T122T2}
\ee
\be
T_{1222}T_2 - \frac{q_1}{q_2^6}T_2T_{1222} = 0,
\label{T1222T2}
\ee
\be
T_{11222}T_2 - T_2T_{11222} \approx
-\left(1 - \frac{1}{q_2^2}\right) T_{122}^2;
\label{T11222T2}
\ee

\be
T_{12}T_{122} - \frac{q_1}{q_2^4}T_{122}T_{12} - [3]_2T_{11222}
\equiv 0,
\label{T12T122}
\ee
\be
T_{12}T_{1222} - T_{1222}T_{12} \approx -\left(1 - \frac{1}{q_2^2}
\right)T_{122}^2,
\label{T12T1222}
\ee
\be
T_{12}T_{11222} - \frac{1}{q_2^3}T_{11222}T_{12} \approx 0;
\label{T12T11222} \\
\ee

\be
T_{122}T_{1222} - \frac{1}{q_2^3}T_{1222}T_{122} \approx 0,
\label{T122T1222}
\ee
\be
T_{122}T_{11222} - q_2^3 T_{11222}T_{122} \approx 0;
\label{T122T11222} \\
\ee

\be
T_{1222}T_{11222} - q_2^3T_{11222}T_{1222}
\approx \frac{(q_2 - q_2^{-1})^2}{[3]_2}T_{122}^3
\label{T1222T11222}
\ee

\bigskip

This list involves all the relations between {\it positive}-root
generators of $G(2)_q$, when $q_1 = q_2^3$. In order to truncate
the case of $B(2)_q \cong C(2)_q$ it is enough to omit
relations with the generators $T_{1222}$ and $T_{11222}$
and put $q_1 = q_2^2$. Further truncation to $A(2)_q = SL(3)_q$
case implies that relations with $T_{122}$ are also omited
and $q_1 = q_2$.

The list is not the complete set of commutation relations
for rank-$2$ algebras. Commutation relations among
negative generators are easily obtained from those for
positive ones: by ``transponing'' $T_{i\ldots j}T_{k\ldots l}
\longrightarrow T_{-l\ldots k}T_{-j\ldots i}$ and changing
$q \rightarrow q^{-1}$.
Commutation rules with Cartan generators $q^{\vec h\vec H}$ are
trivialy deduced from (\ref{qT}): for $\vec\alpha = k_1\vec\alpha_1
+ k_2\vec\alpha_2$
\be
q^{\pm 2H_1}T_{\vec\alpha} =
q^{\pm(k_1\alpha_{11} + k_2\alpha_{12})}\ T_{\vec\alpha}q^{\pm 2H_1} =
q_1^{\pm(2k_1-k_2)}\ T_{\vec\alpha}q^{\pm 2H_1}, \label{H1Tal} \nn \\
q^{\pm 2H_2}T_{\vec\alpha} =
q^{\pm(k_1\alpha_{12} + k_2\alpha_{22})}\ T_{\vec\alpha}q^{\pm 2H_2} =
q_1^{\mp k_1}q_2^{\pm 2k_2}\ T_{\vec\alpha}q^{\pm 2H_2}  \label{H2Tal}
\ee
Relation between negative and positive
generators are simple corollaries of (\ref{TT}). Of these we''ll
need only the following:
\be
\left[ T_2,T_{-21}\right] = \frac{q_1-q_1^{-1}}{q_2-q_2^{-1}}
T_{-1}q_2^{2H_2}, \label{T2T-21}
\ee
\be
\left[ T_2,T_{-221}\right] = \frac{q_1/q_2-q_2/q_1}{q_2-q_2^{-1}}
T_{-21}q_2^{2H_2}, \label{T2T-221}
\ee
\be
\left[ T_2,T_{-2221}\right] =
\frac{q_1/q_2^2-q_2/q_1^2}{q_2-q_2^{-1}}
T_{-221}q_2^{2H_2}, \label{T2T-2221}
\ee
\be
\left[T_2, T_{-22211}\right] = \frac{q_1^2}{q_2^2[2]_2}
\frac{\left(1-{q_2^2}/{q_1^2}\right)
\left(1-{q_2^4}/{q_1^2}\right)}{1-{1}/{q_2^2}}
T_{-21}^2q_2^{2H_2} \approx \nn \\
\approx q_2(q_2 - q_2^{-1})T_{-21}^2q_2^{2H_2}
\label{T2T-22211}
\ee

Also:\footnote{
Note that the coefficients at the r.h.s. of (\ref{T1T-22211}) are
exactly the same as in (\ref{T1T1222}), if the latter one is
rewriten as:
$$
T_1T_{1222} - q_1T_{1222}T_1 = -\frac{q_1}{q_2^4[3]_2}
\left(q_1(1-\frac{q_2^2}{q_1^2} - \frac{q_2^4}{q_1^2})
T_{122}T_{12} +
(1+q_2^2 - \frac{q_2^4}{q_1^2})T_{12}T_{122}\right)
$$
(In varience with (\ref{T1T1222}) the r.h.s. here is
not vanishing in the limit $q\rightarrow 1$.)
}
\be
\left[T_1, T_{-21}\right] = q_1^{-2H_1}T_{-2},
\ee
\be
\left[T_1, T_{-221}\right] = \frac{q_1}{[2]_2}
\left(1-\frac{q_2^2}{q_1^2}\right)
q_1^{-2H_1}T_{-2}^2,
\ee
\be
\left[T_1, T_{-2221}\right] =
\frac{q_1^2}{[2]_2[3]_2}\left(1-\frac{q_2^2}{q_1^2}\right)
\left(1-\frac{q_2^4}{q_1^2}\right)q_1^{-2H_1}T_{-2}^3,
\ee
\be
\left[T_1, T_{-22211}\right] = \frac{1}{[3]_2}
\left(q_1(1-\frac{q_2^2}{q_1^2} - \frac{q_2^4}{q_1^2})T_{-2}T_{-221} +
(1+q_2^2 - \frac{q_2^4}{q_1^2})T_{-221}T_{-2}\right);
\label{T1T-22211}
\ee

\be
\left[T_{12}, T_{-21}\right] =
\frac{1}{q_2-q_2^{-1}}(
q_1^{2H_1}q_2^{2H_2}-q_1^{-2H_1}q_2^{2H_2}) -
\frac{1}{q_1-q_1^{-1}}(
q_1^{-2H_1}q_2^{2H_2}-q_1^{-2H_1}q_2^{-2H_2})
\ee
(for $SL(3)_q$, when $q_1 = q_2 = q$, the r.h.s. turns just into
$\frac{1}{q-q^{-1}}(
q^{2H_1}q^{2H_2}-q^{-2H_1}q^{-2H_2})$).

\bigskip


After all these specifications we are finaly
in a position to check (\ref{defR}).

\subsection{The case of the rank-$2$ groups}

We are going to prove now that
\be
\hat{\cal R} = \hat{\cal R}_1 \hat{\cal R}_{{\rm ns}}
\hat{\cal R}_2
\label{Rhatr2}
\ee
satisfies eqs.(\ref{Rhat1}), (\ref{Rhat2}) for $i=1,2$, provided
\be
\hat{\cal R}_{{\rm ns}} = \hat{\cal R}_{12}
\ \ \ \ {\rm for}\ A(2)_q, \label{RhatA} \\
\hat{\cal R}_{{\rm ns}} = \hat{\cal R}_{12}\hat{\cal R}_{122}
\ \ \ \ {\rm for}\ B(2)_q\cong C(2)_q, \label{RhatBC} \\
\hat{\cal R}_{{\rm ns}} = \hat{\cal R}_{12}
\hat{\cal R}_{11222}\hat{\cal R}_{122}\hat{\cal R}_{1222}
\ \ \ \ {\rm for}\ G(2)_q  \label{RhatG}
\ee
Here $1\ldots 12\ldots 2$ denote the (non-simple) roots $\vec\alpha_1 +
\ldots + \vec\alpha_1 + \vec\alpha_2 + \ldots + \vec\alpha_2$ and
$T_{\vec\alpha},\
T_{-\vec\alpha}$ in (\ref{Ralpha}) are defined in
(\ref{Tij})-(\ref{T-jii}).

Since we already know the property (\ref{propRi}) of
$\hat{\cal R}_i$, it follows that (\ref{Rhat1}) and
(\ref{Rhat2}) are valid with $\hat{\cal R}$ in (\ref{Rhatr2})
provided
\be
\hat{\cal R}_{{\rm ns}}\hat{\cal R}_{2}
\left(I\otimes T_1 + T_1\otimes q_1^{-2H_1}\right)
\hat{\cal R}_{2}^{-1}\hat{\cal R}_{{\rm ns}}^{-1}
= \left(I\otimes T_1 + T_1\otimes q_1^{-2H_1}\right)
\label{condi1.1},
\ee
\be
\hat{\cal R}_{{\rm ns}}\hat{\cal R}_{2}
\left(q_1^{2H_1}\otimes T_{-1} + T_{-1}\otimes I\right)
\hat{\cal R}_{2}^{-1}\hat{\cal R}_{{\rm ns}}^{-1}
= \left(q_1^{2H_1}\otimes T_{-1} + T_{-1}\otimes I\right)
\label{condi1.1-};
\ee
\be
\hat{\cal R}_{{\rm ns}}^{-1}\hat{\cal R}_1^{-1}
\left(I\otimes T_2 + T_2\otimes q_2^{2H_2}\right)
\hat{\cal R}_1\hat{\cal R}_{{\rm ns}}
= \left(I\otimes T_2 + T_2\otimes q_2^{2H_2}\right),
\label{condi2.1}
\ee
\be
\hat{\cal R}_{{\rm ns}}^{-1}\hat{\cal R}_1^{-1}
\left(q_2^{-2H_2}\otimes T_{-2} + T_{-2}\otimes I\right)
\hat{\cal R}_1\hat{\cal R}_{{\rm ns}} =
\left(q_2^{-2H_2}\otimes T_{-2} + T_{-2}\otimes I\right)
\label{condi2.1-}
\ee

We present calculations only for eq.(\ref{condi2.1}), the three
other cases can be analyzed similarly with the same result.

\subsubsection{Congugation by $\hat{\cal R}_{\vec\alpha}$
\label{alg}}

We shall consider separatly conjugation of
$I\otimes T_2$ and
$T_2\otimes q_2^{2H_2}$ by each of the relevant operators
$\hat{\cal R}$-matrices $\hat{\cal R}_{\vec\alpha} =
{\cal E}_{q_{\vec\alpha}}\left(-(q_{\vec\alpha}-q_{\vec\alpha})
V_{\vec\alpha}\right)$, $V_{\vec\alpha}\equiv T_{\vec\alpha}
\otimes T_{-\vec\alpha}$. Every such conjugation is
evaluated in four steps:

1) Let
\be
U_{\vec\alpha}^{R} = \left[ I\otimes T_2, V_{\vec\alpha}\right]
= (I\otimes T_2)(T_{\vec\alpha}\otimes T_{-\vec\alpha}) -
(T_{\vec\alpha}\otimes T_{-\vec\alpha})(I\otimes T_2), \nn \\
U_{\vec\alpha}^{L} = -\left[ T_2\otimes q_2^{2H_2},
V_{\vec\alpha}\right] =
(T_{\vec\alpha}\otimes T_{-\vec\alpha})(T_2\otimes q_2^{2H_2}) -
(T_2\otimes q_2^{2H_2})(T_{\vec\alpha}\otimes T_{-\vec\alpha})
\label{defU}
\ee
Here commutators are the ordinary ones (not $q$-commutators).

2) With our choice of generators $T_{\vec\alpha}$ we always have
(see Appendix A):
\be
U_{\vec\alpha}^RV_{\vec\alpha} = \sigma_{\vec\alpha}^R
\left(V_{\vec\alpha}U_{\vec\alpha}^R + W^R_{\vec\alpha}\right), \nn \\
W_{\vec\alpha}^RV_{\vec\alpha} = \rho_{\vec\alpha}^R
V_{\vec\alpha}W_{\vec\alpha}^R,
\label{defWR}
\ee
with
\be
\sigma_{\vec\alpha}^R = \frac{1}{q_{\vec\alpha}^2},\ \ \
{\rm and}\ \ \ \rho_{\vec\alpha}^R = (\sigma_{\vec\alpha}^R)^2 =
\frac{1}{q_{\vec\alpha}^4}
\label{defsigmaR}
\ee
while
\be
U_{\vec\alpha}^LV_{\vec\alpha} = \sigma_{\vec\alpha}^L
\left(V_{\vec\alpha}U_{\vec\alpha}^L - W^L_{\vec\alpha}\right), \nn \\
W_{\vec\alpha}^LV_{\vec\alpha} = \rho_{\vec\alpha}^L
V_{\vec\alpha}W_{\vec\alpha}^L,
\label{defWL}
\ee
with
\be
\sigma_{\vec\alpha}^L = q_{\vec\alpha}^{2},\ \ \
{\rm and}\ \ \ \rho_{\vec\alpha}^L = (\sigma_{\vec\alpha}^L)^2 =
q_{\vec\alpha}^{4}
\label{defsigmaL}
\ee

3) It now follows that
\be
V_{\vec\alpha}^aU_{\vec\alpha}^R = q_{\vec\alpha}^{2a}
U_{\vec\alpha}^RV_{\vec\alpha}^a - q_{\vec\alpha}^{3a-3}
[a]_{q_{\vec\alpha}}W_{\vec\alpha}^RV_{\vec\alpha}^{a-1}
\ee
and
\be
U_{\vec\alpha}^LV_{\vec\alpha}^b = q_{\vec\alpha}^{2b}
V_{\vec\alpha}^bU_{\vec\alpha}^L - q_{\vec\alpha}^{3b-1}
[b]_{q_{\vec\alpha}}V_{\vec\alpha}^{b-1}W_{\vec\alpha}^L
\ee
so that finaly:
\be
\left[ I\otimes T_2, V_{\vec\alpha}^n\right] =
\sum_{a+b = n-1}  V_{\vec\alpha}^a U_{\vec\alpha}^R
 V_{\vec\alpha}^b = \nn \\ =
q_{\vec\alpha}^{n-1}[n]_{\vec\alpha}
U_{\vec\alpha}^RV_{\vec\alpha}^{n-1}  -
\frac{q_{\vec\alpha}}{[2]_{\vec\alpha}}
q_{\vec\alpha}^{n-1}[n]_{\vec\alpha}
q_{\vec\alpha}^{n-2}[n-1]_{\vec\alpha}
W_{\vec\alpha}^RV_{\vec\alpha}^{n-2}
\label{ZR.prel}
\ee
while
\be
\left[ T_2\otimes q_2^{2H_2}, V_{\vec\alpha}^n\right] =
-\sum_{a+b = n-1}  V_{\vec\alpha}^a U_{\vec\alpha}^L
 V_{\vec\alpha}^b = \nn \\ =
- q_{\vec\alpha}^{n-1}[n]_{\vec\alpha}
V_{\vec\alpha}^{n-1}U_{\vec\alpha}^L  +
\frac{q_{\vec\alpha}^3}{[2]_{\vec\alpha}}
q_{\vec\alpha}^{n-1}[n]_{\vec\alpha}
q_{\vec\alpha}^{n-2}[n-1]_{\vec\alpha}
V_{\vec\alpha}^{n-2}W_{\vec\alpha}^L
\label{ZL.prel}
\ee

4) Finaly summation over $n$ with the weight
$\frac{(-(q_{\vec\alpha}-q_{\vec\alpha}^{-1}))^n}{[n]_{\vec\alpha}!}
q^{-n(n-1)/2}$ gives:
\be
\hat{\cal R}_{\vec\alpha}^{-1} \left(I\otimes T_2 +
(q_{\vec\alpha}-q_{\vec\alpha}^{-1})U^R_{\vec\alpha} +
\frac{1}{q_{\vec\alpha}[2]_{\vec\alpha}}
(q_{\vec\alpha}-q_{\vec\alpha}^{-1})^2W^R_{\vec\alpha}\right)
\hat{\cal R}_{\vec\alpha}  = I\otimes T_2
\label{ZR}
\ee
and
\be
\hat{\cal R}_{\vec\alpha}^{-1} \left(T_2\otimes q_2^{2H_2}
\right)\hat{\cal R}_{\vec\alpha}  =
T_2\otimes q_2^{2H_2} +
(q_{\vec\alpha}-q_{\vec\alpha}^{-1})U^L_{\vec\alpha} +
\frac{q_{\vec\alpha}}{[2]_{\vec\alpha}}
(q_{\vec\alpha}-q_{\vec\alpha}^{-1})^2W^L_{\vec\alpha}
\label{ZL}
\ee

\subsubsection{The rank-$2$ groups}

We now list operators $U$ and $W$ for all the positive roots
of $A(2)_q$, $B(2)_q\cong C(2)_q$, $G(2)_q$.
They can be found - and properties
(\ref{defWR}), (\ref{defWL}) checked - with the help of the
commutation relations from  subsection \ref{core} (see
Appendix A for more details).

$\vec\alpha = \vec\alpha_1,\ \ \ q_{\vec\alpha} = q_1$:
\be
\begin{array}{ll}
U_1^R = 0,
& W_1^R = 0,  \\
U_1^L = T_{12}\otimes T_{-1}q_2^{2H_2},
& W_1^L = 0
\end{array}
\label{alpha1}
\ee \\

$\vec\alpha = \vec\alpha_1+\vec\alpha_2,\ \ \ q_{12} = q_2$:
\be
\begin{array}{ll}
U_{12}^R = \frac{q_1-q_1^{-1}}{q_2-q_2^{-2}}T_{12}
\otimes T_{-1}q_2^{2H_2},
& W_{12}^R = 0,  \\
U_{12}^L = [2]_2T_{122}\otimes T_{-21}q_2^{2H_2},
& W_{12}^L = [2]_2[3]_2T_{11222}\otimes T_{21}^2q_2^{2H_2}
\end{array}
\label{alpha12}
\ee \\

$\vec\alpha = 2\vec\alpha_1 + 3\vec\alpha_2,\ \ \
q_{11222} = \frac{q_2^9}{q_1^2} \approx q_1$:
\be
\begin{array}{ll}
U_{11222}^R \approx (q_2^2-1) T_{11222}\otimes T_{-21}^2q_2^{2H_2},
& W_{11222}^R = 0,  \\
U_{11222}^L = -(1 - q_2^{-2})T_{122}^2
\otimes T_{-22211}q_2^{2H_2},
& W_{11222}^L = 0
\end{array}
\label{alpha11222}
\ee \\
In this case we shall need one more statement:
\be
\left[ T_{122}\otimes T_{-21}q_2^{2H_2},\  V_{11222}\right] = 0,
\ee
thus
\be
\hat{\cal R}_{11222}^{-1}\left( T_{122}\otimes T_{-21}q_2^{2H_2}\right)
\hat{\cal R}_{11222} = \left( T_{122}\otimes T_{-21}q_2^{2H_2}\right)
\label{dopcr11222}
\ee \\

$\vec\alpha = \vec\alpha_1 + 2\vec\alpha_2,\ \ \
q_{122} = \frac{q_2^4}{q_1}$:
\be
U_{122}^R = \frac{q_1/q_2 - q_2/q_1}{q_2-q_2^{-1}}T_{122}
\otimes T_{-21}q_2^{2H_2}, \nn \\
W_{122}^R = -\frac{q_1/q_2 - q_2/q_1}{q_2-q_2^{-1}}[3]_2T_{122}^2
\otimes T_{-22211}q_2^{2H_2} \approx [2]_2[3]_2T_{122}^2
\otimes T_{-22211}q_2^{2H_2}, \nn \\
U_{122}^L = [3]_2T_{1222}\otimes T_{-221}q_2^{2H_2}, \nn \\
W_{122}^L = 0
\label{alpha122}
\ee \\

$\vec\alpha = \vec\alpha_1 + 3\vec\alpha_2,\ \ \
q_{1222} = \frac{q_2^9}{q_1^2} \approx q_1$:
\be
U_{1222}^R = \frac{q_1/q_2^2 - q_2^2/q_1}{q_2-q_2^{-1}}
T_{1222}\otimes T_{-221}q_2^{2H_2}
\approx T_{1222}\otimes T_{-221}q_2^{2H_2}, \nn \\
W_{1222}^R = U_{1222}^L = W_{1222}^L = 0
\label{alpha1222}
\ee \\

Applying now (\ref{ZR}) and (\ref{ZL}) we get the following
chain of transformations:

\be
\hat{\cal R}_{1222}^{-1}\hat{\cal R}_{122}^{-1}
\hat{\cal R}_{11222}^{-1}\hat{\cal R}_{12}^{-1}\hat{\cal R}_1^{-1}
\left\{I\otimes T_2 + T_2\otimes q^{2H_2}\right\}
\hat{\cal R}_1\hat{\cal R}_{12}\hat{\cal R}_{11222}\hat{\cal R}_{122}
\hat{\cal R}_{1222}
= \label{init} \\
\ee
\be
= \hat{\cal R}_{1222}^{-1}\hat{\cal R}_{122}^{-1}
\hat{\cal R}_{11222}^{-1}\hat{\cal R}_{12}^{-1}
\left\{I\otimes T_2 + T_2\otimes q^{2H_2} +
\right. \nn \\  \left.+(q_1-q_1^{-1})
T_{12}\otimes T_{-1}q_2^{2H_2}\right\}
\hat{\cal R}_{12}\hat{\cal R}_{11222}\hat{\cal R}_{122}
\hat{\cal R}_{1222}
= \label{con1}
\ee
\be
= \hat{\cal R}_{1222}^{-1}\hat{\cal R}_{122}^{-1}
\hat{\cal R}_{11222}^{-1}\hat{\cal R}_{12}^{-1}
\left\{I\otimes T_2 + T_2\otimes q^{2H_2} +
\right. \nn \\  \left.+(q_{12}-q_{12}^{-1})
\left(\frac{q_1-q_1^{-1}}{q_2-q_2^{-1}}
T_{12}\otimes T_{-1}q_2^{2H_2}\right)\right\}
\hat{\cal R}_{12}\hat{\cal R}_{11222}\hat{\cal R}_{122}
\hat{\cal R}_{1222}
= \label{con1mod} \\
\ee
\be
= \hat{\cal R}_{1222}^{-1}\hat{\cal R}_{122}^{-1}
\hat{\cal R}_{11222}^{-1}\left\{ I\otimes T_2 + T_2\otimes q^{2H_2}
+ (q_{12}-q_{12}^{-1})[2]_2T_{122}T_{-21}q_2^{2H_2}+
\right.\nn \\ \left. +
\frac{q_2}{[2]_2}(q_2-q_2^{-1})^2\left([2]_2[3]_2T_{11222}\otimes
T_{-21}^2q_2^{2H_2}\right)\right\}
\hat{\cal R}_{11222}\hat{\cal R}_{122} \hat{\cal R}_{1222}
= \label{con12}
\ee
\be
= \hat{\cal R}_{1222}^{-1}\hat{\cal R}_{122}^{-1}
\hat{\cal R}_{11222}^{-1}\left\{I\otimes T_2 + T_2\otimes q^{2H_2}
+ (q_{12}-q_{12}^{-1})[2]_2T_{122}T_{-21}q_2^{2H_2}
 + \right.\nn \\ \left. +
(q_{11222}-q_{11222}^{-1})\left(q_2(q_2-q_2^{-1})T_{11222}\otimes
T_{-21}^2q_2^{2H_2}\right)\right\}
\hat{\cal R}_{11222}\hat{\cal R}_{122}\hat{\cal R}_{1222}
= \label{con12mod} \\
\ee
\be
= \hat{\cal R}_{1222}^{-1}\hat{\cal R}_{122}^{-1}
\left\{I\otimes T_2 + T_2\otimes q^{2H_2}
+ (q_{12}-q_{12}^{-1})[2]_2T_{122}T_{-21}q_2^{2H_2}
- \right.\nn \\ \left. -
(q_{11222}-q_{11222}^{-1})(1- q_2^{-1})
T_{122}^2\otimes T_{-22211}q_2^{2H_2}\right\}
\hat{\cal R}_{122}\hat{\cal R}_{1222}
= \label{con11222}
\ee
\be
= \hat{\cal R}_{1222}^{-1}\hat{\cal R}_{122}^{-1}
\left\{I\otimes T_2 + T_2\otimes q^{2H_2}
+ (q_{122}-q_{122}^{-1})\left(\frac{q_1/q_2-q_2/q_1}{q_2-q_2^{-1}}
T_{122}T_{-21}q_2^{2H_2}\right) - \right.\nn \\ \left. -
\frac{(q_{122}-q_{122}^{-1})^2}{q_{122}[2]_{122}}
\left([2]_2[3]_2T_{122}^2\otimes T_{-22211}q_2^{2H_2}\right)\right\}
\hat{\cal R}_{122}\hat{\cal R}_{1222}
= \label{con11222mod} \\
\ee
\be
= \hat{\cal R}_{1222}^{-1}\left\{I\otimes T_2 + T_2\otimes q^{2H_2}
+ [3]_2(q_{2}-q_{2}^{-1})T_{1222}\otimes T_{-221}q_2^{2H_2}\right\}
\hat{\cal R}_{1222}
= \label{con122}
\ee
\be
= \hat{\cal R}_{1222}^{-1}
\left\{I\otimes T_2 + T_2\otimes q^{2H_2} + (q_{1222}-q_{1222}^{-1})
T_{1222}\otimes T_{-221}q_2^{2H_2}\right\} \hat{\cal R}_{1222}
= \label{con122mod} \\
\ee
\be
=  \left\{I\otimes T_2 + T_2\otimes q^{2H_2}\right\}
\label{final}
\ee

This chain is written down for the most complicated $G(2)_q$
case, for $A(2)_q$ and $B(2)_q\cong C(2)_q$ cases some
steps are just omited (by puting irrelevant $T_{\vec\alpha} = 0$
and $\hat{\cal R}_{\vec\alpha} = I\otimes I$). In these formulas
expressions (\ref{alpha1})-(\ref{alpha1222}) are used with the
``$\approx$'' equalities (valid only if $q_1 = q_2^3$) - but
only in the terms which are irrelevant for the $A,B,C$ groups.

This completes the proof of (\ref{condi2.1}) for all the
rank-$2$ simple groups. The other three relations
(\ref{condi1.1}), (\ref{condi1.1-}), (\ref{condi2.1-})
can be verified just in the same way.\footnote{
In order to obtain (\ref{condi2.1-}) it is enough to switch
between the positive and negative roots, i.e. change
$T_{i\ldots j}T_{\pm k\ldots l} \longrightarrow
T_{\mp l\dots k}T_{-j \ldots i}$ and $q\rightarrow q^{-1}$.
The proof of (\ref{condi1.1}), (\ref{condi1.1-})
is somewhat more tedious: one should deal not only with
double, but also with 3- and 4-fold $q$-commutators at
intermediate stages of calculation.
}
Generalization to higher-rank groups
is also straightforward. It does not involve anything new, except
for the choice of ordering of positive roots $\vec\alpha$ in
(\ref{Rgen}), which is always obvious (see \cite{B} for the
case of $A$-series).

\newpage

\section{Conclusion}

We described a simple explicit formula for the group element
of any simple quantum group. Non-commuting ``coordinates''
on the ``group manifold'' satisfy a Heisenberg-type algebra
and only the Chevalley generators, associated with {\it simple}
roots are involved. There is no reference to particular
representation of Chevalley generators themselves, which can
be substituted not only as matrices but, for example,
in the form of difference operators\footnote{
E.g. for $SL(2)_q$ \cite{FlV2} one can substitute in (\ref{ge})
$$T_+ = \frac{q}{t}\cdot\frac{1-M^{-2}_t}{q-q^{-1}}, \ \ \
q^{\pm 2H} = q^{\pm 2\lambda}M^{\mp 2}_t, \ \ \
T_- = t\ \frac{q^{2\lambda} - q^{-2\lambda}M_t^{+2}}{q-q^{-1}};$$
see \cite{M}, \cite{ANO} and especially \cite{ASh} for general
description of such representations.
}
etc. The most obscure feature of the formalism is
the need to choose a map $s$ from {\it all} the positive roots
to {\it simple} ones, which is ambiguosly defined, thus giving rise to
somewhat different representations, when $r_G > 1$.
A closely related problem is to find out the adequate
group composition rule for $r_G >1$ and explicit formulas for
$\Delta^*(\chi,\phi,\psi)$. Probably  some representations of
group elements involving non-simple roots can be constructed
as well.

We described also an explicit check of the ${\cal RTT}$ relation,
involving explicit $q$-exponential representation of the universal
${\cal R}$-matrix as a product of ``elementary''
$\hat{\cal R}_{\vec\alpha}$-matrices, associated with all the
positive roots $\vec\alpha$. A detailed presentation was
given of the rank-2 case.

All this provides certain evidence that a well-defined notion of
$d_G$-parametric ``group manifold'' can emerge for $q\neq 1$,
just as it exists in the classical case of $q=1$.
See Appendix B below for more comments on this issue.

The two straightforward directions to further develop this technique
is to work out explicit formulas for quantum affine (Kac-Moody)
algebras and to describe the realization \cite{D} of quantum groups
with the trivial ${\cal R}$ matrix, ${\cal R} = q^{\sum_{a=1}^{d_G}
T^a\otimes T_a}$, but non-trivial $q$-associator.

Another task is to find a Lagrangian/functional integral
description, i.e. an adequate version of the geometrical
quantization and deformation of the WZNW model. Given
the simple algebra of the $\chi,\ \phi,\ \psi$-variables
(fields) this should not be a very complicated problem.
Once solved, it can open the way to complete bosonization of
the quantum group, including also some natural choice of
Heisenberg-like representation of the generators $T_{\vec\alpha}$.

\section{Acknowledgements}

We are indebted to A.Gerasimov, S.Khoroshkin, D.Lebedev and A.Mironov
for stimulating discussions.
 A.M. acknowledges the hospitality and support of the CRM at University
of Montreal where this work was done.
The work of L.V. is partialy supported through funds provided by
NSERC (Canada) and FCAR (Quebec).

\newpage

\section{APPENDIX A. ${\cal R}$-matrix for rank-$2$ algebras}
\setcounter{equation}0

This appendix contains some details of the derivation of formulas
(\ref{alpha1})-(\ref{alpha1222}) and (\ref{init})-(\ref{final}).

\subsection{$I\otimes T_2$ versus $\hat{\cal R}_1$}

Since $[T_2, T_{-1}] = 0$, see (\ref{TT}),
\be
U_1^R = \left[ I\otimes T_2,\ T_1\otimes T_{-1}
\right] = 0,
\ee
then also $W_1^R = 0$,
and the steps 2)-4) from subsection \ref{alg} are trivial:
\be
\hat{\cal R}_1^{-1}\left(I\otimes T_2\right)\hat{\cal R}_1 =
I\otimes T_2
\label{conR1}
\ee

\subsection{$T_2\otimes q_2^{2H_2}$ versus $\hat{\cal R}_1$}

1) This time
\be
{\rm 1a)}:\ \ \
q_2^{2H_2} T_{-1} = q_1 T_{-1}q_2^{2H_2}
\label{L1.1a}
\ee
(since $q^{-\alpha_{21}} = q^{\alpha_{11}/2} = q_1$) and,
according to (\ref{T1T2}),
\be
{\rm 1b)}:\ \ \
T_2T_1 = \frac{1}{q_1}(T_1T_2 - T_{12})
\label{L1.1b}
\ee
When combined these two relations imply:\footnote{
It is important that the coefficients at the r.h.s. in
(\ref{L1.1a}) and (\ref{L1.1b}) cancel each other. It is this
cancelation that (given the universal form (\ref{Ralpha}) of
the ${\cal R}$-matrix)
selects the relevant definition of $T_{12}$,
against the alternative $\bar T_{12} = T_1T_2 - \frac{1}{q}
T_2T_1$. (Things work similarly for all other $T_{\vec\alpha}$
with non-simple roots $\vec\alpha$.) If coefficients do not
cancel, $U^L$ in (\ref{L1.1}) would be some $q$-commutator
instead of an ordinary one. This will produce extra $n$-dependence
in (\ref{ZL.prel}) and destroy the reasoning, which leads to relations
(\ref{ZR}), (\ref{ZL}).
}
\be
U_1^L = -
\left[(T_2\otimes q_2^{2H_2}),\ (T_1\otimes T_{-1})\right]
= T_{12}\otimes T_{-1}q_2^{2H_2}
\label{L1.1}
\ee

2) Since:
\be
{\rm 2a)\ =\ 1a)}:\ \ \
(T_{-1}q_2^{2H_2}) T_{-1} = q_1 T_{-1}(T_{-1}q_2^{2H_2})
\nn
\ee
and, according to (\ref{T1T12}),
\be
{\rm 2b)}:\ \ \ \
T_{12}T_1 = qT_1T_{12}
\ee
we have:
\be
(T_{12}\otimes T_{-1}q_2^{2H_2})(T_1\otimes T_{-1}) = q_1^2
(T_1\otimes T_{-1})(T_{12}\otimes T_{-1}q_2^{2H_2})
\label{L1.2}
\ee
i.e. $W_1^L = 0$ and indeed $\sigma_1^L = q_1^2$, as stated
in (\ref{defsigmaL}).

Steps 3) and 4) deserve no comments: we just apply
(\ref{ZL}) to obtain:
\be
\hat{\cal R}_1^{-1}
\left(T_2\otimes q_2^{2H_2}\right)
\hat{\cal R}_1 =
\left(T_2\otimes q_2^{2H_2} + (q_1 - q_1^{-1})
T_{12}\otimes T_{-1}q_2^{2H_2}\right)
\label{conL1}
\ee
Together with (\ref{conR1}) this formula describes the transition
from (\ref{init}) to (\ref{con1}). The result coincides
also with (\ref{con1mod}) because $q_{12}=q_2$ and
$(q_{12}-q_{12}^{-1})\frac{q_1-q_1^{-1}}{q_2-q_2^{-1}}
= q_1 - q_1^{-1}$.

\subsection{$I\otimes T_2$ versus $\hat{\cal R}_{12}$}

1) From (\ref{T2T-21}) we get:
\be
U_{12}^R =
\left[ I\otimes T_2,\ T_{12}\otimes T_{-21}\right] =
T_{12}\otimes [T_2,\ T_{-21}] = \nn \\ =
\frac{q_1-q_1^{-1}}{q_2-q_2^{-1}} T_{12}\otimes T_{-1}q_2^{2H_2}
\label{R12.1}
\ee

2) Since
\be
{\rm 2a)}:\ \ \
q_2^{2H_2} T_{-21} = \frac{q_1}{q_2^2}T_{-21}q_2^{2H_2}
\label{R12.2a}
\ee
(because $q^{-\alpha_{22} - \alpha_{21}} =
q^{-\alpha_{22}+\alpha_{11}/2} = q_1/q_2^2$)
and (according to the Serre identity $T_{-jii}=0$)
\be
{\rm 2b)}:\ \ \
T_{-1}T_{-21} = q_1^{-1}T_{-21}T_{-1}
\label{R12.2b}
\ee
we have:
\be
( T_{12}\otimes T_{-1}q_2^{2H_2})(T_{12}\otimes T_{-21}) =
\frac{1}{q_2^2}
(T_{12}\otimes T_{-21})( T_{12}\otimes T_{-1}q_2^{2H_2})
\label{R12.2}
\ee
Thus $W_{12}^R = 0$ and
indeed $\sigma_{12}^R = q_2^{-2} = q_{12}^{-2}$ -
as stated in (\ref{defsigmaR}).

It follows from (\ref{ZR}) now, that
\be
\hat{\cal R}_{12}^{-1}\left( I\otimes T_2 + (q_1-q_1^{-1})
T_{12}\otimes T_{-1}q_2^{2H_2}\right)\hat{\cal R}_{12} =
\left(I\otimes T_2\right)
\label{conR12}
\ee

\subsection{$T_2\otimes q_2^{2H_2}$ versus $\hat{\cal R}_{12}$.
The end of the proof for $SL(3)_q$}

1) Now
\be
{\rm 1a)\ =\ eq.(\ref{R12.2a})}:\ \ \ q_2^{2H_2}T_{-21} =
\frac{q_1}{q_2^2}T_{-21}q_2^{2H_2}
\label{L12.1a}
\ee
and, according to (\ref{T12T2}),
\be
{\rm 1b)}:\ \ \ T_2T_{12} = \frac{q_2^2}{q_1}
(T_{12}T_2 - [2]_2T_{122})
\label{L12.1b}
\ee
Thus
\be
U_{12}^L = -
\left[ T_2\otimes q_2^{2H_2},\ T_{12}\otimes T_{-21}\right] =
[2]_2T_{122}\otimes T_{-21}q_2^{2H_2}
\label{L12.1}
\ee

Since for $SL(3)_q = A(2)_q$ the Serre identity implies that
$T_{122} = 0$, this is the end of our calculations in this case:
steps 2)-4) are now trivial,
\be
{\rm for}\ \ SL(3)_q:\ \ \ \
\hat{\cal R}_{12}^{-1}(T_2\otimes q_2^{2H_2})\hat{\cal R}_{12}
= T_2\otimes q_2^{2H_2}
\nn
\ee
which together with (\ref{conR12}) gives the desired
(\ref{condi2.1}), i.e. {\bf the proof of (\ref{condi2.1})
for $A(2)_q$ is} already {\bf completed}.
We, however, proceed further with the non-simply-laced case,
when $T_{122}\neq 0$.

2) From (\ref{R12.2a}) it follows that
\be
{\rm 2a)}:\ \ \ (T_{-21}q_2^{2H_2})T_{-21} =
\frac{q_1}{q_2^2}T_{-21}(T_{-21}q_2^{2H_2})
\nn
\ee
and, due to (\ref{T12T122})
\be
{\rm 2b)}:\ \ \ T_{122}T_{12} = \frac{q_2^4}{q_1}(T_{12}T_{122}-
[3]_2T_{11222})
\nn
\ee
we have:
\be
[2]_2(T_{122}\otimes T_{-21}q_2^{2H_2})(T_{12}\otimes T_{-21}) = \nn \\
= q_2^2[2]_2\left((T_{12}\otimes T_{-21})
(T_{122}\otimes T_{-21}q_2^{2H_2})
- [3]_2T_{11222}\otimes T_{21}^2q_2^{2H_2}\right),
\label{L12.2}
\ee
i.e. $\sigma_{12}^L$ is indeed $q_{122}^2 = q_2^2$, and
\be
W_{12}^L = [2]_2[3]_2T_{11222}\otimes T_{21}^2q_2^{2H_2}
\ee

This quantity is non-vanishing for $G(2)_q$ only.
This allows to use ``$\approx$'' formulas (i.e. valid only
for $q_1 = q_2^3$) in the next
lines. Because of the same (\ref{R12.2a})
\be
{\rm 2c)}:\ \ \ (T_{-21}^2q_2^{2H_2})T_{-21} =
\frac{q_1}{q_2^2}T_{-21}(T_{-21}^2q_2^{2H_2})
\approx q_2T_{-21}(T_{-21}^2q_2^{2H_2})
\nn
\ee
and of (\ref{T12T11222})
\be
{\rm 2d)}:\ \ \ T_{11222}T_{12} \approx q_2^3T_{12}T_{11222}
\nn
\ee
we have:
\be
\left(T_{11222}\otimes T_{21}^2q_2^{2H_2}\right)
\left(T_{12}\otimes T_{-21}\right) \approx \nn \\ \approx
q_2^4 \left(T_{12}\otimes T_{-21}\right)
\left(T_{11222}\otimes T_{21}^2q_2^{2H_2}\right)
\ee
and indeed $\rho_{12}^L = q_2^4 =  q_{12}^4$
as stated in (\ref{defsigmaL}).

Eq.(\ref{ZL}) now says that
\be
\hat{\cal R}_{12}^{-1}\left(T_2\otimes q_2^{2H_2}\right)
\hat{\cal R}_{12} =
T_2\otimes q_2^{2H_2} + \nn \\ + [2]_2(q_2-q_2^{-1})
T_{122}\otimes T_{-21}q_2^{2H_2} +
[3]_2q_2(q_2-q_2^{-1})^2T_{11222}\otimes T_{-21}^2q_2^{2H_2}
\label{conL12}
\ee
Together with (\ref{conR12}) this describes the transition from
(\ref{con1}) to (\ref{con12}). The next step - to (\ref{con12mod})
- is to note that the last term at the r.h.s. of
(\ref{conL12}) is present only in $G(2)_q$ case, when
$q_1 = q_2^3$ and thus one can substitute
$q_{11222}-q_{11222}^{-1} = q_{1} - q_1^{-1}
\approx [3]_2 (q_2-q_2^{-1})$.

As mentioned above calculation for $A(2)_q$ is already finished.
The next conjugation - by $\hat{\cal R}_{11222}$ - is present
only in the $G(2)_q$ case. It will eliminate the last term at
the r.h.s. of (\ref{conL12}) and gives rise instead to another
one - also present only in the case of $G(2)_q$. The {\it second}
term in (\ref{conL12}) is however, present in the
$B(2)_q\cong C(2)_q$ case as well - thus it is important
that it is left intact by the $\hat{\cal R}_{11222}$ conjugation.

\subsection{$T_{122}\otimes T_{21}q_2^{2H_2}$ versus
$\hat{\cal R}_{11222}$}

According to (\ref{T12T11222}),
$$
T_{11222}T_{12} \approx q_2^3T_{12}T_{11222},
$$
thus
$$
T_{-21}T_{-22211} \approx q_2^{-3}T_{-22211}T_{-21}
$$
Together with (\ref{T122T11222}),
$$
T_{122}T_{11222} \approx q_2^3 T_{11222}T_{122},
$$
and orthogonality property $\vec\alpha_{11222}\vec\alpha_{2}
= 2\alpha_{12} + 3\alpha_{22} = 3\alpha_{22} - \alpha_{11} \approx 0$
which guarantees that
\be
q_2^{2H_2}T_{-22211} = T_{-22111}q_2^{2H_2}
\label{D11222.3}
\ee
this implies that
\be
\left[ T_{122}\otimes T_{21}q_2^{2H_2},\ T_{11222}\otimes
T_{-22211}\right] = 0
\label{conD11222}
\ee

\subsection{$I\otimes T_2$ versus $\hat{\cal R}_{11222}$}

1) From (\ref{T2T-22211})
\be
U_{12}^R =
\left[ I\otimes T_2,\ T_{11222}\otimes T_{-22211}\right] =
T_{11222}\otimes [T_2,\ T_{-22211}] \approx \nn \\ \approx
(q_2^2-1) T_{11222}\otimes T_{-21}^2q_2^{2H_2}
\label{R11222.1}
\ee

2) From (\ref{T12T11222})
$$
T_{11222}T_{12} \approx q_2^3 T_{12}T_{11222}
$$
thus
$$
T_{-21}T_{-22211} \approx q_2^{-3}T_{-22211}T_{21}
$$
and, combined with eq.(\ref{D11222.3} this means that
\be
(T_{11222}\otimes T_{-21}^2q_2^{2H_2})(T_{11222}\otimes
T_{-22211}) = \nn \\ = q_2^{-6}(T_{11222}\otimes
T_{-22211})(T_{11222}\otimes T_{-21}^2q_2^{2H_2})
\ee
We get $W_{11222}^R = 0$ and
$\sigma_{11222}^R = q_2^{-6} = q_{11222}^{-2}$,
in accordance with (\ref{defsigmaR}).

Finaly
\be
\hat{\cal R}_{11222}^{-1}\left(I\otimes T_2 +
(q_{11222}-q_{11222}^{-1})(q_2-q_2^{-1})q_2
T_{11222}\otimes T_{-21}^2q_2^{2H_2}\right)
\hat{\cal R}_{11222} = \nn \\ = I\otimes T_2
\label{conR11222}
\ee

\subsection{$T_2\otimes q_2^{2H_2}$ versus $\hat{\cal R}_{11222}$}

1) $U^L_{11222} \neq 0$ only in the $G(2)_q$ case so again
equations with ``$\approx$'' can be used.
Since $q^{-3\alpha_{22}-2\alpha_{12}} =  \frac{q_1^2}{q_2^6}
\approx 1$,
\be
{\rm 1a)}:\ \ \ q_2^{2H_2}T_{-22211} = \frac{q_1^2}{q_2^6}
T_{-22211}q_2^{2H_2},
\label{L11222.1a}
\ee
and due to (\ref{T11222T2})
\be
{\rm 1b)}:\ \ \ T_2T_{11222} \approx T_{11222}T_2 +
(1-\frac{1}{q_2^2})T_{122}^2
\ee
we have
\be
U_{11222}^L = -
\left[T_2\otimes q_2^{2H_2},\ T_{11222}\otimes T_{-22211}
\right] \approx \nn \\ \approx
-(1-\frac{1}{q_2^2})T_{122}^2\otimes T_{-22211}q_2^{2H_2}
\ee

2) Again (\ref{L11222.1a}) implies
\be
{\rm 2a)}:\ \ \ T_{-22211}(q_2^{2H_2}T_{-22211}) =
\frac{q_1^2}{q_2^6}
T_{-22211}(T_{-22211}q_2^{2H_2}),
\ee
and from (\ref{T122T11222}) follows:
\be
{\rm 2b)}: \ \ \
T_{122}^2T_{11222} \approx q_2^6T_{11222}T_{122}^2 =
q_{11222}^2 T_{11222}T_{122}^2,
\ee
so that
\be
(T_{122}^2\otimes T_{-22211}q_2^{2H_2})(T_{11222}\otimes T_{-22211})
= \nn \\ = q_{11222}^2 (T_{11222}\otimes T_{-22211})
(T_{122}^2\otimes T_{-22211}q_2^{2H_2}),
\ee
i.e. indeed $\sigma_{11222}^L = q_{11222}^2$ and
$W_{11222}^L = 0$.

Thus we get
\be
\hat{\cal R}_{11222}^{-1}\left(T_2\otimes q_2^{2H_2}\right)
\hat{\cal R}_{11222} = T_2\otimes q_2^{2H_2} - \nn \\ -
(q_{11222}-q_{11222}^{-1})\frac{q_2-q_2^{-1}}{q_2}
T_{122}^2\otimes T_{-22211}q_2^{2H_2}
\label{conL11222}
\ee
Together with (\ref{conD11222}) and
(\ref{conR11222}) this describes transition
from (\ref{con12mod}) to (\ref{con11222}). In order to
perform the next transformation to (\ref{con11222mod})
we need to make two substitutions.

First,
$[2]_2(q_2-q_2^{-1}) = q_2^2-q_2^{-2}$ is actualy equal to
$\frac{(q_1^4/q_2 - q_2/q_1^4)(q_1/q_2-q_2/q_1)}{q_2-q_2^{-1}}$
for both $q_1 = q_2^2$ (the case of $B(2)_q\cong C(2)_q$)
and $q_1=q_2^3$ (the case of $G(2)_q$).

Second, in the case of $G(2)_q$, when $q_{11222} = q_1 =
q_2^3 = q_{122}^3$ we can also change
$(q_{11222}-q_{11222}^{-1})\frac{q_2-q_2^{-1}}{q_2} \approx
\frac{1}{q_2}[3]_2(q_2-q_2^{-1})^2$
for $[2]_2[3]_2\frac{(q_{122}-q_{122}^{-1})^2}{q_{122}[2]_{122}}$

\subsection{$I\otimes T_2$ versus $\hat{\cal R}_{122}$}

1) Because of (\ref{T2T-221})
\be
U_{122}^R = \left[I\otimes T_2,\ T_{122}\otimes T_{-221}\right]
= T_{122}\otimes [T_2, T_{-221}] = \nn \\ =
\frac{q_1/q_2-q_2/q_1}{q_2-q_2^{-1}}
T_{122}\otimes T_{-21}q_2^{2H_2}
\ee

2) First,
\be
{\rm 2a)}:\ \ \ q_2^{2H_2}T_{-221} = \frac{q_1}{q_2^4}
T_{-221}q_2^{2H_2}
\label{R122.2a}
\ee
Second, since $T_{12}T_{122} = \frac{q_1}{q_2^4}T_{122}T_{12}
+ T_{11222}$ (see (\ref{T12T122}),
\be
{\rm 2b)}:\ \ \ T_{-21}T_{-221} = \frac{q_1}{q_2^4}
\left(T_{-221}T_{-21} - [3]_2T_{-22211}\right)
\ee
and therefore
\be
(T_{122}\otimes T_{-21}q_2^{2H_2})(T_{122}\otimes T_{-221}) =
\nn \\ = q_{122}^{-2} \left(
(T_{122}\otimes T_{-221})(T_{122}\otimes T_{-21}q_2^{2H_2})
- [3]_2T_{122}^2\otimes T_{-22211}q_2^{2H_2}\right)
\ee
so that
$\sigma_{122}^R = q_{122}^{-2}$ and
\be
W_{122}^R = [3]_2\frac{q_1/q_2-q_2/q_1}{q_2-q_2^{-1}}
T_{122}^2\otimes T_{-22211}q_2^{2H_2}
\ee
is as usual, non-vanishing only in the case of $G(2)_q$.

Further, from (\ref{R122.2a})
\be
{\rm 2c)}:\ \ \ q_2^{2H_2}T_{-221} = \frac{q_1}{q_2^4}
T_{-221}q_2^{2H_2} \approx \frac{1}{q_2}T_{-221}q_2^{2H_2}
\ee
and from (\ref{T122T11222})
\be
T_{-22211}T_{-221} \approx \frac{1}{q_2^3}T_{-221}T_{-22211}
\ee
so that
\be
(T_{122}^2\otimes T_{-22211}q_2^{2H_2})
(T_{122}\otimes T_{-221}) = \nn \\ = \frac{1}{q_2^4}
(T_{122}\otimes T_{-221})
(T_{122}^2\otimes T_{-22211}q_2^{2H_2}),
\ee
i.e. $\rho_{122}^R \approx q_{122}^{-4}$, in accordance with
(\ref{defsigmaR}). Note that $q_{122} = q_2$ only for $G(2)_q$,
for $B(2)_q\cong C(2)_q$ instead $q_{122} = q_1 = q_2^2$,
but for this group $W_{122}^R = 0$.

Thus
\be
\hat{\cal R}_{122}^{-1}\left(I\otimes T_2 +
(q_{122}-q_{122}^{-1})\frac{q_1/q_2-q_2/q_1}{q_2-q_2^{-1}}
T_{122}\otimes T_{-21}q_2^{2H_2} - \nn \right.\\ \left. -
\frac{(q_{122}-q_{122}^{-1})^2}{q_{122}[2]_2}[2]_2[3]_2
T_{122}^2\otimes T_{-22211}q_2^{2H_2}\right)
\hat{\cal R}_{122} = I\otimes T_2
\label{conR122}
\ee

\subsection{$T_2\otimes q_2^{2H_2}$ versus $\hat{\cal R}_{122}$.
The end of the proof for $B(2)_q = C(2)_q$ }

1) Using (\ref{R122.2a}),
\be
{\rm 1a)}:\ \ \ q_2^{2H_2}T_{-221} = \frac{q_1}{q_2^4}
T_{-221}q_2^{2H_2}
\ee
and (\ref{T122T2}),
\be
{\rm 1b)}:\ \ \ T_2T_{122} = \frac{q_2^4}{q_1}
(T_{122}T_2 - [3]_2T_{1222}),
\ee
we get:
\be
U_{122}^L = -
\left[T_2\otimes q_2^{2H_2},\ T_{122}\otimes T_{-221}\right]
= [3]_2T_{1222}\otimes T_{-221}q_2^{2H_2}
\ee

2) Since $U_{122}^L$ is non-vanishing only for $G(2)_q$,
we can make use of (\ref{T122T1222}) to obtain:
\be
(T_{1222}\otimes T_{-221}q_2^{2H_2})(T_{122}\otimes T_{-221})
\approx \nn \\ \approx q_2^2
(T_{122}\otimes T_{-221})(T_{1222}\otimes T_{-221}q_2^{2H_2})
\ee
with $\sigma_{122}^L = q_{122}^2 \approx q_2^2$ and
$W_{122}^L = 0$. Thus
\be
\hat{\cal R}_{122}^{-1}\left(T_2\otimes q_2^{2H_2}\right)
\hat{\cal R}_{122} = \nn \\ =
T_2\otimes q_2^{2H_2} + [3]_2(q_{122}-q_{122}^{-1})
T_{1222}\otimes T_{-221}q_2^{2H_2}
\label{conL122}
\ee
and, when added to (\ref{conR122}), this describes the transition
from (\ref{con11222mod}) to (\ref{con122}). The next step
to (\ref{con122mod}) is just the substitution  $[3]_2(q_2-q_2^{-1})
\approx q_{11222}-q_{11222}^{-1}$.

At this stage {\bf the proof} of (\ref{condi2.1}) {\bf is finished
for $B(2)_q\cong C(2)_q$}, since the second term in (\ref{conL122})
- to be eliminated by the last conjugation by
$\hat{\cal R}_{1222}$ - is absent for this group.

\subsection{$I\otimes T_2$ versus $\hat{\cal R}_{1222}$}

1) From (\ref{T2T-2221})
\be
U_{1222}^R = \left[I\otimes T_2,\ T_{1222}\otimes T_{-2221}\right]
= T_{1222}\otimes [T_2, T_{-2221}] = \nn \\ =
\frac{q_1/q_2^2-q_2^2/q_1}{q_2-q_2^{-1}}
T_{1222}\otimes T_{-221}q_2^{2H_2} \approx
T_{1222}\otimes T_{-221}q_2^{2H_2}
\ee

2) Next,
\be
{\rm 2a)}:\ \ \ q_2^{2H_2}T_{-2221} = \frac{q_1}{q_2^6}
T_{-2221}q_2^{2H_2} = \frac{1}{q_{1222}}T_{-2221}q_2^{2H_2}
\label{R1222.2a}
\ee
and eq.(\ref{T122T1222}), $T_{1222}T_{122} = q_2^3T_{122}T_{1222}$,
implies
\be
{\rm 2b)}:\ \ \ T_{-221}T_{-2221} = \frac{1}{q_2^3}
T_{-2221}T_{-221} \approx \frac{1}{q_{1222}}T_{-2221}T_{-221}
\ee
Thus
\be
(T_{1222}\otimes T_{-221}q_2^{2H_2})(T_{1222}\otimes T_{-2221})
\approx \nn \\ \approx
\frac{1}{q_{1222}^2}(T_{1222}\otimes T_{-2221})
(T_{1222}\otimes T_{-221}q_2^{2H_2})
\ee
i.e. $W_{1222}^R = 0$ and indeed $\sigma_{1222}^R = q_{1222}^{-2}$,
so that (\ref{ZR}) is appluicable:
\be
\hat{\cal R}_{1222}^{-1}\left(I\otimes T_2 +
(q_{1222} - q_{1222}^{-1})T_{1222}\otimes T_{-221}q_2^{2H_2}\right)
\hat{\cal R}_{1222} = I\otimes T_2
\label{conR1222}
\ee

\subsection{$T_2\otimes q_2^{2H_2}$ versus $\hat{\cal R}_{1222}$}

1) In this case
\be
{\rm 1a)}:\ \ \ q_2^{2H_2})T_{-2221} = \frac{q_1}{q_2^6}
T_{-2221}q_2^{2H_2}
\ee
and (\ref{T1222T2}),
\be
{\rm 1b)}:\ \ \ T_2T_{1222} = \frac{q_2^6}{q_1}T_{1222}T_2,
\ee
so that
\be
U_{1222}^L = -
\left[T_2\otimes q_2^{2H_2},\ T_{1222}\otimes T_{-2221}\right]
= 0
\ee
Thus it follows that
\be
\hat{\cal R}_{1222}^{-1}\left(T_2\otimes q_2^{2H_2}\right)
\hat{\cal R}_{1222} =
T_2\otimes q_2^{2H_2},
\ee
and together with (\ref{conR1222}) this describes the
last transition from (\ref{con122mod}) to (\ref{final}).

This ends the proof of (\ref{condi2.1}) for {\it all} the
three simple rank-$2$ quantum groups.

\newpage

\section{APPENDIX B.
Group multiplication in the case of $SL(3)_q$}

\setcounter{equation}0

In section \ref{secom} of the main text we did not actualy
prove that the group multiplication closes
after restriction to a $d_G$-parametric ``manifold'' in the
(operator-valued) universal enveloping algebra.
We are not aware of any simple proof in the general situation,
but by tedious calculation one can usually perform an
explicit check.
In order to demonstrate the very idea of the calculation,
we consider briefly the example of $SL(3)_q$.
Even in this case we are not describing exhaustive proof,
but rather some pieces of it, which, however, highlight the
main steps of complete proof. More detailed presentation is
hardly necessary, because some more straightforward kind
of arguments should be found to prove the statement,
which can  be also considered as an adequate formulation of the
Campbell-Hausdorff formula for quantum groups.

Below in this Appendix $G = SL(3)_q$.

Our logic in this Appendix is as follows. Assuming
that the group property is true, one can get the expression
for $\Delta^*(\chi,\phi,\psi)$ from computations in the
fundamental representation, just manipulating  $3\times 3$
matrices (section \ref{fursec}). If these formulas are
substituted into the l.h.s. of (\ref{d*gu.0}) and (\ref{del*car})
we obtain the relations (Campbell-Hausdorff formula for
$SL(3)_q$) which should be proved to hold just in the same
form in {\it any} representation (i.e. for any $H_i$, $T_i$
with the right commutation properties and satisfying Serre
identities).  These relations are formulated and proved for
Cartan part (\ref{del*car}) in s.\ref{carsec} and
just formulated (not proved) for Borel part (\ref{d*gu.0})
in s.\ref{borsec}. Instead of proving relation for Borel part
we consider a strongly simplified particular case (when many
of $\chi$, $\psi$ variables are taken to be zero) in ss.
\ref{siesec.1}-\ref{siesec.2}. This allows to make the role
and interplay of q-exponential properties and Serre identities
transparent. Complete proof can be obtained in the same way
but it will be unacceptably lengthy for such a conceptualy simple
statement (as {\it existence of the group}). Finaly in
s.\ref{CHsec} we comment briefly on the Campbell-Hausdorff
formula for $q$-exponentials (still to be discovered),
of which (\ref{sufo}) and
(\ref{FVc}) should be particular simple examples, and which,
with the $T$-generators as arguments should provide the
Campbell-Hausdorff relations for quantum groups.

\subsection{The answer from fundamental representation
\label{fursec}}

The {\it final} formulas for $\Delta^*(\chi,\phi,\psi)$ are most
easily deduced  from explicit multiplication of two matrices
(\ref{sl3fund}) in the fundamental representation:
$$
\left( \begin{array}{ll} q^{\Delta^*(\phi_{(1)})}&
q^{\Delta^*(\phi_{(1)})}\Delta^*(\psi^{(2)}) \\
 \Delta^*(\chi^{(2)})q^{\Delta^*(\phi_{(1)})} &
 \Delta^*(\chi^{(2)})q^{\Delta^*(\phi_{(1)})}\Delta^*(\psi^{(2)}) +
q^{\Delta^*(\phi_{(2)})}  \\
\Delta^*(\chi^{(3)})\Delta^*(\chi^{(2)})q^{\Delta^*(\phi_{(1)})} &
\Delta^*(\chi^{(3)})\Delta^*(\chi^{(2)})q^{\Delta^*(\phi_{(1)})}
\Delta^*(\psi^{(2)}) +  \\
& \ \ \ + (\Delta^*(\chi^{(1)}) +\Delta^*(\chi^{(3)}))
q^{\Delta^*(\phi_{(2)})}
\end{array}
\right.
$$
$$
\left. \begin{array}{l}
 q^{\Delta^*(\phi_{(1)})}\Delta^*(\psi^{(2)})\Delta^*(\psi^{(3)})\\
  \Delta^*(\chi^{(2)})q^{\Delta^*(\phi_{(1)})}
\Delta^*(\psi^{(2)})\Delta^*(\psi^{(3)}) +
q^{\Delta^*(\phi_{(2)})}(\Delta^*(\psi^{(1)}) + \Delta^*(\psi^{(3)}))\\
  \Delta^*(\chi^{(3)})\Delta^*(\chi^{(2)})q^{\Delta^*(\phi_{(1)})}
\Delta^*(\psi^{(2)})\Delta^*(\psi^{(3)}) + \\ \ \ \ \ \ \ \ +
(\Delta^*(\chi^{(1)}) +\Delta^*(\chi^{(3)}))q^{\Delta^*(\phi_{(2)})}
(\Delta^*(\psi^{(1)}) + \Delta^*(\psi^{(3)}))
+ q^{\Delta^*(\phi_{(3)})} \end{array}\right)
=
$$
$$
\bigskip =
\left( \begin{array}{lll} q^{\phi_{(1)}'}& q^{\phi_{(1)}'}\psi_{(2)}'&
q^{\phi_{(1)}'}\psi_{(2)}'\psi_{(3)}' \\
 \chi_{(2)}'q^{\phi_{(1)}'} & \chi_{(2)}'q^{\phi_{(1)}'}\psi_{(2)}' +
q^{\phi_{(2)}'} &
\chi_{(2)}'q^{\phi_{(1)}'}\psi_{(2)}'\psi_{(3)}' +
q^{\phi_{(2)}'}(\psi_{(1)}' + \psi_{(3)}') \\
\chi_{(3)}'\chi_{(2)}'q^{\phi_{(1)}'} &
\chi_{(3)}'\chi_{(2)}'q^{\phi_{(1)}'}\psi_{(2)}' + &
\chi_{(3)}'\chi_{(2)}'q^{\phi_{(1)}'}\psi_{(2)}'\psi_{(3)}' + \\
& \ \ \ + (\chi_{(1)}' +\chi_{(3)}')q^{\phi_{(2)}'} & \ \ \ +
(\chi_{(1)}' +\chi_{(3)}')q^{\phi_{(2)}'}(\psi_{(1)'} + \psi_{(3)}')
+ q^{\phi_{(3)}'}
\end{array}\right)
\cdot
$$
$$ \cdot
\left( \begin{array}{lll} q^{\phi_{(1)}''}&
q^{\phi_{(1)}''}\psi_{(2)}''&
q^{\phi_{(1)}''}\psi_{(2)}''\psi_{(3)}'' \\
 \chi_{(2)}''q^{\phi_{(1)}''} &
\chi_{(2)}''q^{\phi_{(1)}''}\psi_{(2)}'' +
q^{\phi_{(2)}''} &
\chi_{(2)}''q^{\phi_{(1)}''}\psi_{(2)}''\psi_{(3)}'' +
q^{\phi_{(2)}''}(\psi_{(1)}'' + \psi_{(3)}'') \\
\chi_{(3)}''\chi_{(2)}''q^{\phi_{(1)}''} &
\chi_{(3)}''\chi_{(2)}''q^{\phi_{(1)}''}\psi_{(2)}'' + &
\chi_{(3)}''\chi_{(2)}''q^{\phi_{(1)}''}\psi_{(2)}''\psi_{(3)}'' + \\
& \ \ \ + (\chi_{(1)}'' +\chi_{(3)}'')q^{\phi_{(2)}''} & \ \ \ +
(\chi_{(1)}'' +\chi_{(3)}'')q^{\phi_{(2)}''}(\psi_{(1)''} +
\psi_{(3)}'') + q^{\phi_{(3)}''}
\end{array}\right)
$$
\bigskip

\be
q^{\Delta^*(\phi_{(1)})} = q^{\phi_{(1)}'}(1 + \psi_{(2)}'\chi_{(2)}''
+ \psi_{(2)}'\psi_{(3)}'\chi_{(3)}''\chi_{(2)}'')q^{\phi_{(1)}''} = \\
= q^{\phi_{(1)}'}(1 + z_{(22)} + qz_{(33)}z_{(22)})
q^{\phi_{(1)}''}, \\
\Delta^*(\psi_{(2)}) =  \psi_{(2)}'' +
q^{-\Delta^*(\phi_{(1)})} \left( q^{\phi_{(1)}'}\psi_{(2)}'
\{1 + \psi_{(3)}'(\chi_{(1)}''+\chi_{(3)}'')\}q^{\phi_{(2)}''}\right)
= \\ = \psi_{(2)}''  +
q^{-\phi_{(1)}''}\frac{1}{1 + z_{(22)} + qz_{(33)}z_{(22)}}
(1 + z_{(31)}z_{(33)})q^{\phi_{(2)}''}, \\
\Delta^*(\psi_{(3)}) =  \psi_{(3)}'' + \\ +
\left(q^{\Delta^*(\phi_{(1)})}\Delta^*(\psi_{(2)})\right)^{-1}
q^{\phi_{(1)}'}\left\{\psi_{(2)}'
(1 + \psi_{(3)}'(\chi_{(1)}''+\chi_{(3)}''))
q^{\phi_{(2)}'}\psi_{(1)}''
+ \psi_{(2)}'\psi_{(3)}' q^{\phi_{(3)}'}\right\} = \\ =
\psi_{(3)}'' \ +\
\left\{(1 + z_{(22)} + qz_{(33)}z_{(22)})
q^{\phi_{(1)}''}\psi_{(2)}'' + \psi_{(2)}'
(1 + z_{(31)} + z_{(33)})q^{\phi_{(2)}''}\right\}^{-1}\cdot\\ \cdot
\left\{\psi_{(2)}'
(1 + z_{(31)} + z_{(33)})q^{\phi_{(2)}'}\psi_{(1)}''
+ \psi_{(2)}'\psi_{(3)}' q^{\phi_{(3)}'}\right\},
\\
\ldots
\label{sl3del*}
\ee
Here $z_{(s_1s_2)} \equiv \psi_{s_1}'\chi_{s_2}''$.

This calculation, though straightforward,
is in fact not very interesting: by itself it is not enough to
guarantee that $\Delta^*$ indeed exists, i.e. is
independent on particular representation. Still, if
the group property is valid,
$\Delta^*$ can be evaluated in {\it any} representation,
and it is under this assumption that (\ref{sl3del*})
make sense. Our main goal is to  check the {\it assumption}.
For this we come back to representation-independent
discussion of section \ref{secom}.

\subsection{Cartan part \label{carsec}}

We begin with the relatively simple transformation of the
formula (\ref{del*car}). The task is to generalize the
reasoning of p.6) of section 5.1 to the case of $SL(3)_q$.
In this case each double product in (\ref{del*car})
consists of 5 factors:
\be
q^{2\Delta^*(\vec\phi)\vec H} =
{\cal P}^{-1}q^{2\vec\phi'\vec H}q^{2\vec\phi''\vec H}{\cal P},
\label{d*car.1}
\ee
where
$$
{\cal P} = {\cal P}_{(11)}{\cal P}_{(13)}{\cal P}_{(21)}
{\cal P}_{(31)}{\cal P}_{(33)} =
{\cal E}_q\left(\hat z_{(11)} +\hat z_{(13)}\right)
{\cal E}_q(\hat z_{(22)})
{\cal E}_q\left(\hat z_{(31)} + \hat z_{(33)}\right),
$$
and condensed notation are introduced:
$$
{\cal P}_{(s_1s_2)} \equiv {\cal E}_q(\hat z_{(s_1s_2)}), \ \ \ \
\hat z_{(s_1s_2)} \equiv
\frac{z_{(s_1s_2)}}{q-q^{-1}} = \frac{\psi_{(s_1)}'\chi_{(s_2)}''}
{q-q^{-1}}
$$
and $\psi^{(s)}, \chi^{(s)}$, $s = 1,2,3$ are the same as in the
section 4.2.
(Note that ${\cal P}_{(13)}$, ${\cal P}_{(31)}$,
${\cal P}_{(22)}$ mutualy commute.)

In this subsection it will be convenient to expand $\vec H$ in the
basis of fundamental weights $\vec \lambda_i$ ($i=1\ldots r_G$),
$\vec\lambda_i\vec\alpha_j = \delta_{ij}$:
$\ \vec H = \frac{1}{2}\sum_{i=1}^{r_G}n_i\vec\lambda_i$.
In finite-dimensional representations all $n_i$ are integers.
As in the case of $SL(2)_q$ we perfrom all the calculations
in this subsection for {\it integer} $n_i$'s and, then, if
necessary, the result can be analytically continued to all
non-integer values.
Decomposition in fundamental weights is convenient because it
diagonalises the commutation relations:
$$
q^{\vec\phi'\vec\lambda_j}z_{s_1s_2} =
q^{\delta_{i(s_1)j}}\ z_{s_1s_2}q^{\vec\phi'\vec\lambda_j},\ \ \ \
q^{\vec\phi''\vec\lambda_j}z_{s_1s_2} =
q^{\delta_{i(s_2)j}}\ z_{s_1s_2}q^{\vec\phi''\vec\lambda_j}
$$
Let us remind that in our notation for $SL(3)_q$ (see section 4.2)
$i(1) = i(3) = 2,\ i(2)=1$, and $\phi_{(1)} = \vec\phi\vec\lambda_1$,
$\phi_{(3)} = -\vec\phi\vec\lambda_2$
(in general case $\vec\phi\vec\lambda_j =
\sum_{k=1}^{r_G} \alpha_{jk}^{-1}\frac{\alpha_{kk}}{2}\phi_k$).

In p.6) of section 5.1 we essentially showed that
\be
{\cal E}_{1/q}(-\hat z_{kk})q^{2(\vec\phi'+\vec\phi'')\vec H}
{\cal E}_q(\hat z_{kk}) =
q^{(\vec\phi'+\vec\phi'')\sum_{j\neq k}n_j\vec\lambda_j}
\left(q^{\vec\phi'\vec\lambda_k}(1 + z_{kk})
q^{\vec\phi''\vec\lambda_k}\right)^{n_k}
\ee

Keeping all this in mind we can split evaluation of the r.h.s.
of (\ref{d*car.1}) into several steps.
First,
\be
\left({\cal E}_{q}(\hat z_{(11)}+\hat z_{(13)})\right)^{-1}
q^{2(\vec\phi'+\vec\phi'')\vec H}
{\cal E}_q(\hat z_{(11)}+\hat z_{(13)}) = \\ =
q^{n_1(\vec\phi'+\vec\phi'')\vec\lambda_1}
\left(q^{\vec\phi'\vec\lambda_2}\{1 + z_{(11)}+z_{(13)}\}
q^{\vec\phi''\vec\lambda_2}\right)^{n_2}
\ee
Let us now add conjugation by ${\cal E}_q(\hat z_{(22)}$
(we remind that $n_1$, $n_2$ are assumed to be integers):
\be
\left({\cal E}_{q}(\hat z_{(22)})\right)^{-1}
q^{n_1(\vec\phi'+\vec\phi'')\vec\lambda_1}
\left\{q^{\vec\phi'\vec\lambda_2}(1 + z_{(11)}+z_{(13)})
q^{\vec\phi''\vec\lambda_2}\right\}^{n_2}
{\cal E}_{q}(\hat z_{(22)}) = \\ =
\left\{\left({\cal E}_{q}(\hat z_{(22)})\right)^{-1}
q^{n_1(\vec\phi'+\vec\phi'')\vec\lambda_1}
{\cal E}_{q}(\hat z_{(22)})\right\}\cdot \\ \cdot
\left\{q^{\vec\phi'\vec\lambda_2}
\left({\cal E}_{q}(\hat z_{(22)})\right)^{-1}
(1 + z_{(11)}+z_{(13)}){\cal E}_{q}(\hat z_{(22)})
q^{\vec\phi''\vec\lambda_2}\right\}^{n_2} = \\ =
\left\{q^{\vec\phi'\vec\lambda_1}(1 + z_{(22)})
q^{\vec\phi''\vec\lambda_1}\right\}^{n_1}
\left\{q^{\vec\phi'\vec\lambda_2}
\left(1 + (1+qz_{(22)})z_{(11)}+z_{(13)}\right)
q^{\vec\phi''\vec\lambda_2}\right\}^{n_2}
\ee
At the last stage the following transformation was made in
the second bracket:
\be
\left({\cal E}_{q}(\hat z_{(22)})\right)^{-1}
(1 + z_{(11)}+z_{(13)}){\cal E}_{q}(\hat z_{(22)}) = \\ =
\left({\cal E}_{q}(\hat z_{(22)})\right)^{-1}
\left({\cal E}_{q}(\hat z_{(22)})(1+z_{(13)}) +
{\cal E}_{q}(q^2\hat z_{(22)})z_{(11)}\right)
\ee
and application of (\ref{qexpprod1}) gives
$1 + (1+qz_{(22)})z_{(11)}+z_{(13)}$.

It remains to perform the last conjugation by
${\cal P}_{(31)}{\cal P}_{(33)} = {\cal E}_q(\hat z_{(31)} +
\hat z_{(33)})$:
\be
\left({\cal E}_q(\hat z_{(31)} + \hat z_{(33)})\right)^{-1}
\left(\ \ldots \ \right)^{n_1}\left(\ \ldots \ \right)^{n_2}
{\cal E}_q(\hat z_{(31)} + \hat z_{(33)}) = \\ =
\left\{\left({\cal E}_q(\hat z_{(31)} + \hat z_{(33)})\right)^{-1}
\  \left(\ \ldots\ \right)  \
{\cal E}_q(\hat z_{(31)} + \hat z_{(33)}) \right\}^{n_1}
\cdot\\ \cdot
\left\{\left({\cal E}_q(\hat z_{(31)} + \hat z_{(33)})\right)^{-1}
\ \left(\ \ldots \ \right) \
{\cal E}_q(\hat z_{(31)} + \hat z_{(33)})\right\}^{n_2}
\label{d*car.2}
\ee
The entry of the first bracket actualy is:
\be
q^{\vec\phi'\vec\lambda_1}
\left({\cal E}_q(\hat z_{(33)})\right)^{-1}
\left({\cal E}_q(\hat z_{(31)})\right)^{-1}
(1 + z_{(22)})
{\cal E}_q(\hat z_{(31)}){\cal E}_q(\hat z_{(33)})
q^{\vec\phi''\vec\lambda_1} = \\ =
q^{\vec\phi'\vec\lambda_1}
\left\{1 + (1+qz_{(33)})z_{(22)}\right\}
q^{\vec\phi''\vec\lambda_1},
\label{d*car.2.1}
\ee
Evaluation of that of the second bracket is a little more tedious:
\be
\left({\cal E}_q(\hat z_{(31)} + \hat z_{(33)})\right)^{-1}
q^{\vec\phi'\vec\lambda_2}\left(\ \ \ldots \ \ \right)
q^{\vec\phi''\vec\lambda_2}
{\cal E}_q(\hat z_{(31)} + \hat z_{(33)}) = \\ =
q^{\vec\phi'\vec\lambda_2}
\left({\cal E}_q(q^{-1} \hat z_{(33)})\right)^{-1}
\left({\cal E}_q(q^{-1}\hat z_{(31)})\right)^{-1}\cdot\\ \cdot
\left\{1 + (1+qz_{(22)})z_{(11)} + z_{(13)}\right\}
{\cal E}_q(q\hat z_{(31)})
{\cal E}_q(q\hat z_{(33)})
q^{\vec\phi''\vec\lambda_2} = \\ =
q^{\vec\phi'\vec\lambda_2}
\left({\cal E}_q(q^{-1} \hat z_{(33)})\right)^{-1}
\left\{(1+z_{(31)})(1+z_{(13)}) + (1+qz_{(22)})z_{(11)}\right\}
{\cal E}_q(q\hat z_{(33)})
q^{\vec\phi''\vec\lambda_2} = \\ =
q^{\vec\phi'\vec\lambda_2}
\left\{1+z_{(31)}+z_{(13)} + z_{(33)} + qz_{(22)}z_{(11)} +
(z_{(11)}+ z_{(13)}z_{(31)})\frac{1}{1+z_{(33)}}\right\}
q^{\vec\phi''\vec\lambda_2}
\label{d*car.2.2}
\ee

Combination of (\ref{d*car.2}), (\ref{d*car.2.1}) and (\ref{d*car.2.2})
finaly gives:
\be
q^{2\Delta^*(\vec\phi)\vec H} = q^{\Delta^*(\vec\phi\vec\lambda_1)n_1}
q^{\Delta^*(\vec\phi\vec\lambda_2)n_2} =
\left\{q^{\vec\phi'\vec\lambda_1}
\left(1+z_{(22)}+qz_{(33)}z_{(22)}\right)
q^{\vec\phi''\vec\lambda_1}\right\}^{n_1}\cdot\\ \cdot
\left\{q^{\vec\phi'\vec\lambda_2}
\left(1+z_{(31)}+z_{(13)} + z_{(33)} + qz_{(22)}z_{(11)} +
(z_{(11)}+ z_{(13)}z_{(31)})\frac{1}{1+z_{(33)}}\right)
q^{\vec\phi''\vec\lambda_2}\right\}^{n_2}
\ee
in accordance with (\ref{sl3del*}).
The two expressions in brackets at the r.h.s. are of course
mutualy commuting.

We now proceed to more sophisticated analysis of the Borel part.

\subsection{Elimination of ${\cal P}$-factors \label{borsec}}

The last of the four representations (\ref{d*gu.0}) in the
case of $SL(3)_q$ looks like:
\be
\Delta^*(g_U) = {\cal P}_{(33)}^{-1}{\cal P}_{(31)}^{-1}
{\cal P}_{(22)}^{-1}
{\cal E}_q\left(q^{-\phi''_2/2}\frac{1}{1 + \psi_{(1)}'\chi_{(1)}''
+\psi_{(1)}'\chi_{(3)}''}q^{-\phi''_2/2}\psi_{(1)}'T_2\right)
{\cal P}_{(22)}     \cdot \\ \cdot
{\cal E}_q\left(q^{-\phi''_1/2}\frac{1}{1 + \psi_{(2)}'\chi_{(2)}''}
q^{-\phi''_1/2}\psi_{(2)}'T_1\right){\cal P}_{(31)} {\cal P}_{(33)}
\cdot \\ \cdot
{\cal E}_q\left(q^{-\phi''_2/2}\frac{1}{1 + \psi_{(3)}'\chi_{(1)}''
+\psi_{(3)}'\chi_{(3)}''}q^{-\phi''_2/2}\psi_{(3)}'T_2\right)
{\cal E}_q(\psi_{(1)}''T_2)
{\cal E}_q(\psi_{(2)}''T_1)
{\cal E}_q(\psi_{(3)}''T_2)
\label{d*gu.1}
\ee

Our first goal now is to get rid of the factors ${\cal P}_{(22)}$,
${\cal P}_{(31)}$ and
${\cal P}_{(33)}$, which are still present at the r.h.s. of (\ref{d*gu.1}):
\be
\Delta^*(g_U) = {\cal E}_q\left(T_2\Delta^*(\psi_{(1)})\right)
{\cal E}_q\left(T_1\Delta^*(\psi_{(2)})\right)
{\cal E}_q\left(T_2\Delta^*(\psi_{(3)})\right) = \\
= {\cal E}_q(\xi_{(1)}T_2)
{\cal E}_q(\xi_{(2)}T_1)
{\cal E}_q(\xi_{(3)}T_2)
{\cal E}_q(\psi_{(1)}''T_2)
{\cal E}_q(\psi_{(2)}''T_1)
{\cal E}_q(\psi_{(3)}''T_2)
\label{d*gu.2}
\ee
where
\be
\xi_{(1)} =
{\cal P}_{(33)}^{-1}{\cal P}_{(31)}^{-1}{\cal P}_{(22)}^{-1}
\left(q^{-\phi''_2/2}\frac{1}{1 + z_{(11)} + z_{(13)}}
q^{-\phi''_2/2}\psi_{(1)}'\right)
{\cal P}_{(22)}{\cal P}_{(31)}{\cal P}_{(33)}, \nn \\
\xi_{(2)} =
{\cal P}_{(33)}^{-1}{\cal P}_{(31)}^{-1}
\left(q^{-\phi''_1/2}\frac{1}{1 + z_{(22)}}
q^{-\phi''_1/2}\psi_{(2)}'\right){\cal P}_{(31)}{\cal P}_{(33)}, \nn \\
\xi_{(3)} =
q^{-\phi''_2/2}\frac{1}{1 + z_{(31)} + z_{(33)}}
q^{-\phi''_2/2}\psi_{(3)}' = \\ =
q^{-\phi''_2/2}\frac{1}{1 + \psi_{(3)}'\chi_{(1)}''
+\psi_{(3)}'\chi_{(3)}''}q^{-\phi''_2/2}\psi_{(3)}'
\label{defxi}
\ee
It is easy to evaluate $\xi_{(2)}$ explicitly:
since
$$
(\psi_{(2)}'q^{-\phi_1''})z_{(33)} = q^2z_{(33)}
(\psi_{(2)}'q^{-\phi_1''}),\ \ \ \
(\psi_{(2)}'q^{-\phi_1''})z_{(31)} = q^2z_{(31)}
(\psi_{(2)}'q^{-\phi_1''})
$$
and
$$
z_{(22)}z_{(33)} = q^2z_{(33)}z_{(22)},\ \ \ \
z_{(22)}z_{(31)} = z_{(31)}z_{(22)}
$$
we get:
\be
\left({\cal E}_q(\hat z_{(33)})\right)^{-1}
\left({\cal E}_q(\hat z_{(31)})\right)^{-1}
\frac{1}{1 + q^{-1}z_{(22)}}\
{\cal E}_q(q^2\hat z_{(31)}){\cal E}_q(q^2\hat z_{(33)})
 = \nn \\ =
\frac{1}{(1+ q^{-1}z_{(22)}){\cal E}_q(\hat z_{(33)})}
\ \left({\cal E}_q(\hat z_{(31)})\right)^{-1}
{\cal E}_q(q^2\hat z_{(31)}){\cal E}_q(q^2\hat z_{(33)}) = \nn \\ =
\frac{1}{{\cal E}_q(\hat z_{(33)}) +
\frac{1}{q}{\cal E}_q(q^2\hat z_{(33)}) z_{(22)}}\
\left\{\left({\cal E}_q(\hat z_{(31)})\right)^{-1}
{\cal E}_q(q^2\hat z_{(31)})\right\}
{\cal E}_q(q^2 \hat z_{(33)})
\ee
It now remains to apply (\ref{qexpprod1}) and obtain
(taking into account that
$z_{(31)}z_{(33)} = q^{-2}z_{(33)}z_{(31)}$ and
$\phi_1 = \phi_{(1)}- \phi_{(2)}$ ):
\be
\xi_{(2)} =
\frac{1}{1 + q^{-1}(1+qz_{(33)})z_{(22)}}\
\left({\cal E}_q(\hat z_{(33)})\right)^{-1}(1+qz_{(31)})
{\cal E}_q(q^2 \hat z_{(33)})
\ q^{-\phi_1''}\psi_{(2)}' = \nn \\ =
\frac{1}{1 + q^{-1}(1+qz_{(33)})z_{(22)}}\
(1+qz_{(31)}+qz_{(33)})
\ q^{-\phi_1''}\psi_{(2)}' = \nn \\ =
q^{-\phi_{(1)}''}\frac{1}{1+\psi_{(2)}'\chi_{(2)}'' +
\psi_{(2)}'\psi_{(3)}'\chi_{(3)}''\chi_{(2)}''}\
(1 + \psi_{(3)}'\chi_{(1)}'' + \psi_{(3)}'\chi_{(3)}'')\
q^{\phi_{(2)}''}\psi_{(2)}'
\label{xi2}
\ee

Similarly,
\be
\xi_{(1)} =
\left({\cal E}_q(\hat z_{(33)})\right)^{-1}
\left({\cal E}_q(\hat z_{(31)})\right)^{-1}
\left({\cal E}_q(\hat z_{(22)})\right)^{-1}\cdot \\ \cdot
\frac{1}{1 + q^{-1}z_{(11)} + q^{-1}z_{(13)}}
{\cal E}_q(q^2\hat z_{(22)})
{\cal E}_q(q^{-4}\hat z_{(31)}){\cal E}_q(q^{-4}\hat z_{(33)})
\ee
The same trick can be now used, with puting inverse exponentials
into denominator, which is then transformed  to:
\be
\left\{1 + q^{-1}z_{(11)} + q^{-1}z_{(13)}\right\}
{\cal E}_q(\hat z_{(22)})
{\cal E}_q(\hat z_{(31)}){\cal E}_q(\hat z_{(33)}) = \\ =
{\cal E}_q(\hat z_{(22)})
{\cal E}_q(q^{-2}\hat z_{(31)}){\cal E}_q(q^{-4}\hat z_{(33)})\cdot\\
\cdot\left\{(1+q^{-1}z_{(31)})(1+q^{-3}z_{(33)})
(1 + q^{-1}z_{(13)} + q^{-1}z_{(33)}) + \right.\\ \left.
+ q^{-1}(1 + qz_{(22)} + q^{-2}z_{(33)}z_{(22)})z_{(11)}\right\}
\ee
After that in the numerator we get:
\be
\left({\cal E}_q(q^{-4}\hat z_{(33)})\right)^{-1}
\left({\cal E}_q(q^{-2}\hat z_{(31)})\right)^{-1}
\left({\cal E}_q(\hat z_{(22)})\right)^{-1}
{\cal E}_q(q^2\hat z_{(22)})
{\cal E}_q(q^{-4}\hat z_{(31)}){\cal E}_q(q^{-4}\hat z_{(33)})
= \\ =
\left({\cal E}_q(q^{-4}\hat z_{(33)})\right)^{-1}
\frac{1}{1 + q^{-3}z_{(31)}}(1+qz_{(22)})
{\cal E}_q(q^{-4}\hat z_{(33)}) = \\ =
\frac{1}{1 + q^{-3}z_{(31)}+ q^{-5}z_{(33)} }
\frac{1}{1 + q^{-5}z_{(33)}}
(1 + qz_{(22)} + q^{-2}z_{(33)}z_{(22)})
\ee
and finaly
\be
\xi_{(1)} =
\left\{(1+q^{-1}z_{(31)})(1+q^{-3}z_{(33)})
(1 + q^{-1}z_{(13)} + q^{-1}z_{(33)})
\right. + \\ \ \ \ \ \left.
+ q^{-1}(1 + qz_{(22)} + q^{-2}z_{(33)}z_{(22)})z_{(11)}\right\}
\cdot\\ \cdot
\frac{1}{1 + q^{-3}z_{(31)}+ q^{-5}z_{(33)} }
\frac{1}{1 + q^{-5}z_{(33)}}
(1 + qz_{(22)} + q^{-2}z_{(33)}z_{(22)})
\label{xi1}
\ee

As an intermidiate check let us note that in the fundamental
representation (\ref{d*gu.2}) turns into
$$
\Delta^*(g_U) =
\left(\begin{array}{lll}
1&\Delta^*(\psi_{(2)})&\Delta^*(\psi_{(2)})\Delta^*(\psi_{(3)})\\
0&1&\Delta^*(\psi_{(1)}) + \Delta^*(\psi_{(3)})\\
0&0&1 \end{array}\right) =
$$
$$
 \bigskip =
\left(\begin{array}{lll}
1&\xi_{(2)}&\xi_{(2)}\xi_{(3)}\\
0&1&\xi_{(1)}+\xi_{(3)}\\
0&0&1\end{array}\right)
\left(\begin{array}{lll}
1&\psi_{(2)}''&\psi_{(2)}''\psi_{(3)}''\\
0&1&\psi_{(1)}''+\psi_{(3)}''\\
0&0&1\end{array}\right) =
$$
$$
 \bigskip =
\left(\begin{array}{lll}
1& \psi_{(2)}'' + \xi_{(2)} & \psi_{(2)}''\psi_{(3)}''
+ \xi_{(2)}(\psi_{(1)}''+\psi_{(3)}'') + \xi_{(2)}\xi_{(3)}\\
0&1& \psi_{(1)}''+\psi_{(3)}'' + \xi_{(1)}+\xi_{(3)}\\
0&0&1 \end{array}\right)
$$
and we have:
\be
\Delta^*(\psi_{(2)}) = \psi_{(2)}'' + \xi_{(2)}, \\
\Delta^*(\psi_{(3)}) = \psi_{(3)}'' +
   \frac{1}{\psi_{(2)}'' + \xi_{(2)}}\xi_{(2)}
   (\psi_{(1)}''+\xi_{(3)}), \\
\Delta^*(\psi_{(1)}) = \psi_{(1)}'' + \xi_{(1)} + \xi_{(3)} -
\frac{1}{\psi_{(2)}'' + \xi_{(2)}}\xi_{(2)}
   (\psi_{(1)}''+\xi_{(3)})
\label{d*ans}
\ee
Substituting our expression for $\xi$'s one can recognize
(\ref{sl3del*}).

Our next goal is to demonstrate that
(\ref{d*gu.2}), with (\ref{d*ans}) substituted into the r.h.s.,
holds without reference to any
particular representation. The only properties of the
generators $T_{1,2}$ which can be used
are their commutation relations and Serre identities.

\subsection{Toy-problem \label{siesec.1}}

We shall actually consider now a much simpler problem, which
will help to emphasize the role of the Serre identites and
their consistency property with the q-exponentials. Namely,
let us put all $\chi = 0$ and also $\psi_{(1)}' = \psi_{(3)}' = 0$.
(this is of course consistent with the commutation
relations between $\chi, \phi, \psi$ variables).
In this case all $z_{(s_1s_2)} = 0$, thus
$\xi_{(2)} = q^{-\phi_1''}\psi_{(2)}'$,
$\xi_{(1)} = \xi_{(3)} = 0$. Then (\ref{d*gu.2})
- with (\ref{d*ans}) substituted into it - becomes:
\be
\Delta^*(g_U) = {\cal E}_q\left(T_2\Delta^*(\psi_{(1)})\right)
{\cal E}_q\left(T_1\Delta^*(\psi_{(2)})\right)
{\cal E}_q\left(T_2\Delta^*(\psi_{(3)})\right)
= \\ =
{\cal E}_q\left(\frac{1}{\psi_{(2)}'' + q^{-\phi_1''}\psi_{(2)}'}
\psi_{(2)}''\psi_{(1)}''T_2\right)
{\cal E}_q\left((\psi_{(2)}''+ q^{-\phi_1''}\psi_{(2)}')T_1\right)
\cdot \\ \cdot
{\cal E}_q\left(\psi_{(3)}''T_2 +
\frac{1}{\psi_{(2)}'' + q^{-\phi_1''}\psi_{(2)}'}
q^{-\phi_1''}\psi_{(1)}'\psi_{(1)}''
T_2\right) = \\
= {\cal E}_q(\psi_{(2)}'q^{-\phi_1''}T_1)
{\cal E}_q(\psi_{(1)}''T_2)
{\cal E}_q(\psi_{(2)}''T_1)
{\cal E}_q(\psi_{(3)}''T_2)
\ee
and we need to prove that this relation is indeed true.

In order to analyze this simplified problem
introduce first a more reasonable notation: let
$$
u \equiv \xi_{(2)} = \psi_{(2)}'q^{-\phi_1''}, \ \ \ v \equiv \psi_{(2)}'',
$$
$$
x \equiv \psi_{(3)}'', \ \ \ y \equiv \psi_{(1)}''
$$
Then the identity to be proved is
\be
{\cal E}_q\left(\frac{1}{u+v}vyT_2\right)
{\cal E}_q\left((u+v)T_1\right)
{\cal E}_q\left(\left(x +\frac{1}{u+v}uy \right)T_2\right) =
\\ = {\cal E}_q(uT_1)
{\cal E}_q(yT_2)
{\cal E}_q(vT_1)
{\cal E}_q(xT_2)
\label{sire.1}
\ee
provided
\be
uv = \frac{1}{q^2}vu, \ \ \ \ xy = q^2yx, \\
ux  = qxu, \ \ \ \ uy = qyu, \\
vx = qxv, \ \ \ \  vy = \frac{1}{q}yv
\ee
Addition formula (\ref{sufo}) can be immediately applied to
the last two $q$-exponentials at the l.h.s. of (\ref{sire.1}),
and $x$-dependence is completely eliminated:
\be
{\cal E}_q\left(\frac{1}{u+v}vyT_2\right)
{\cal E}_q(uT_1) {\cal E}_q(vT_1)
{\cal E}_q\left(\frac{1}{u+v}uy T_2\right) =
\\ = {\cal E}_q(uT_1)
{\cal E}_q(yT_2)
{\cal E}_q(vT_1)
\label{ide.1}
\ee

\subsection{On the proof of the simplified identity. Algebraic level
\label{algesec}}

Let us begin from expanding (\ref{ide.1}) in a seria in powers
of $v$. The first non-trivial term is:
\be
{\cal E}_q(uT_1)\left[ vT_1, {\cal E}_q(yT_2)\right] +
\left[\frac{1}{u}vyT_2, {\cal E}_q(uT_1)\right]
{\cal E}_q(yT_2) = 0
\label{ide.2}
\ee
In order to handle this relation we need the simplest version
of the Campbell-Hausdorff formula for $q$-exponentials \cite{FlV},
\be
\forall A,B, \ \ \
{\cal E}_{1/q}(-B)\ A\ {\cal E}_q(B) =
\left({\cal E}_q(B)\right)^{-1}A\ {\cal E}_q(B) = \\ =
\sum_{n=0}^\infty \frac{1}{[n]!}\left[\ldots\left[\left[A,B\right],
B\right]_q\ldots, B \right]_{q^{n-1}} = \\ = A +
\left[A,B\right] +
\frac{1}{[2]}\left[\left[A,B\right],B\right]_q +
\frac{1}{[3]!}\left[\left[\left[A,B\right],B\right]_q,B\right]_{q^2}
+\ldots,
\label{EAB}
\ee
where
\be
\left[A,B\right]_{q^k} \equiv \frac{1}{q^k}AB - q^kBA
\ee
(so that $[A,B] \equiv AB-BA = [A,B]_1$ and
$\left[A,B\right]_q = -\left[B,A\right]_{1/q}$).
Similarly,
\be
{\cal E}_{q}(B)\ A\ {\cal E}_{1/q}(-B) =
\sum_{n=0}^\infty \frac{1}{[n]!}
\left[B,\ldots\left[B\left[B,A\right]
\right]_q\ldots \right]_{q^{n-1}} = \\ = A +
\left[B,A\right] +
\frac{1}{[2]}\left[B,\left[B,A\right]\right]_q +
\frac{1}{[3]!}\left[B\left[B\left[B,A\right]\right]_q\right]_{q^2}
+\ldots,
\label{EA-B}
\ee

Let us now apply these results to (\ref{ide.2}):
\be
\left[ vT_1,\ {\cal E}_q(yT_2)\right]{\cal E}_{1/q}(-yT_2) = \\ =
-\left(\left[yT_2, vT_1\right] + \frac{1}{[2]}
\left[ yT_2,\left[yT_2, vT_1\right]\right]_q + \ldots\right)
\ee
The first commutator at the r.h.s. is
$$
\left[yT_2, vT_1\right] = yv T_2T_1 - vyT_1T_2 = -vyT_{12},
$$
where $T_{12}= T_1T_2 - qT_2T_1$ is our familiar generator (\ref{Tij}),
associated with the non-simple positive root of the $SL(3)_q$.
The next commutator is
\be
\frac{1}{[2]}\left[ yT_2,\left[yT_2, vT_1\right]\right]_q
= -\frac{1}{[2]}\left[yT_2, \ vyT_{12}\right]_q = -qvy^2T_{122}
\ee
But Serre identity (\ref{ser+}) for $SL(3)_q$ states that
$T_{122} = \frac{1}{[2]}\left(T_{12}T_2 - \frac{1}{q}T_2T_{12}\right)$
is vanishing!
Because of the apparent recurrent structure of (\ref{EAB}) this
implies, that instead of the infinite sum we have just\footnote{
Of course, in the non-simply-laced case there will be more terms
at the r.h.s.: with $T_{122}$ and $T_{1222}$.}
\be
\left[ vT_1,\ {\cal E}_q(yT_2)\right] = vyT_{12}{\cal E}_q(yT_2)
\ee
Similarly, because of the second Serre identity (\ref{serTiij})
for $SL(3)_q$, $T_{112} = T_1T_{12}- \frac{1}{q}T_{12}T_1 = 0$,
\be
\left[ \frac{1}{u}vy T_2,\ {\cal E}_q(uT_1)\right] =
-{\cal E}_q(uT_1)vy T_{12}
\ee
and these two equations together imply that (\ref{ide.2})
is indeed true.

One can of course consider higher terms of the expansion, as well
as expansions in powers of $u$ and $y$. This is, however, not the best
way to proceed.

\subsection{On the proof of the simplified identity. Group level
\label{siesec.2}}

In the case of $q\neq 1$ the way to reduce the proof of the
{\it commutant}-type (involving permutation of exponentials, i.e.
``group elements'') relation like (\ref{ide.1}) to that of the
{\it commutator}-type like (\ref{EAB}) or (\ref{EA-B}) (involving
permutation of algebra elements or that of an algebra element
with the group element) is provided by differentiation. For
$q\neq 1$ one needs to use shift operators
$M^{\pm k}_zf(z) = f(q^{\pm k}z)$ instead of derivatives.

\subsubsection{Difference equations}

Coming back to our identity (\ref{ide.1})
let us denote its l.h.s. and r.h.s. through $F(u,v,y)$
and $G(u,v,y)$ respectively. Then, due to (\ref{qexpprod1}),
\be
M^{+2}_uG(u,v,y) = G(q^2u,v,y) = \left(1+(q^2-1)uT_1\right)G(u,v,y),\\
M^{+2}_vG(u,v,y) = G(u,q^2v,y) = G(u,v,y)\left(1+(q^2-1)vT_1\right),\\
M^{+2}_yG(u,v,y) = G(u,v,q^2y) =
{\cal E}_q(uT_1){\cal E}_q(yT_2)\left(1 + (q^2-1)yT_2\right)
{\cal E}_q(vT_1) = \\ =
G(u,v,y)\left\{1+(q^2-1)(yT_2 - vyT_{12})\right\}
\label{eq.ide.1}
\ee
In the last case one has to evaluate
${\cal E}_{1/q}(-vT_1)\ yT_2\ {\cal E}_q(vT_1) = yT_2 - vyT_{12}$
with the help of (\ref{EAB}) and the Serre identity, just as we
did in the previous section \ref{algesec}.

Let us note, that also
$$
M^{+2}_yG(u,v,y) = G(u,v,q^2y) =
\left\{1+(q^2-1)(yT_2 + uy\bar T_{12})\right\}G(u,v,y)
$$
where $\bar T_{12} \equiv T_1T_2 - \frac{1}{q}T_2T_1$ is the alternative
choice of the generator $T_{\vec\alpha_{12}}$.\footnote{In section 6
we dealt with $T_{12}$ only, since it is $T_{12}$ and not
$\bar T_{12}$ that appears in the
simple universal formula  (\ref{Ralpha}) for the ${\cal R}$-matrix.
This does not prevent $\bar T_{12}$ from appearing in other
contexts, as we shall see below in this section.}
If expressed in terms of $\bar T_{12}$
the {\it same} Serre identities for $SL(3)_q$ state that
\be
\bar T_{12}T_2 = q T_2\bar T_{12},\ \ \ \
T_1\bar T_{12} = q\bar T_{12}T_1
\label{serbar}
\ee

\subsubsection{The action of $M^2_y$ on $F(u,v,y)$}

In order to prove (\ref{ide.1}) one should now demonstrate that
its l.h.s., i.e. $F(u,v,y)$ satisfies the same eqs.(\ref{eq.ide.1})
Let us begin from the last equation in (\ref{eq.ide.1}).
\be
M^2_yF(u,v,y) = F(u,v,q^2y) = \\ =
{\cal E}_q\left(\frac{1}{u+v}vyT_2\right)\left\{1 + (q^2-1)
\frac{1}{u+v}vyT_2\right\}{\cal E}_q\left((u+v)T_1\right)
\cdot\\ \cdot
\left\{1 + (q^2-1)\frac{1}{u+v}uyT_2\right\}
{\cal E}_q\left(\frac{1}{u+v}uyT_2\right)
\ee
((\ref{qexpprod1}) was used at this stage). Since
$(vy)(u+v) = q(u+v)(vy)$ is a multiplicative commutation relation
one can effectively use (\ref{EAB}) and Serre identity to show that
\be
\left\{ 1 + (q^2-1)
\frac{1}{u+v}vyT_2\right\}{\cal E}_q\left((u+v)T_1\right) = \\ =
{\cal E}_q\left((u+v)T_1\right)\left\{1 + (q^2-1)
\left(\frac{1}{u+v}vyT_2-vyT_{12}\right)\right\}
\ee
and it remains to prove that
\be
\left(1 + (q^2-1)
\left(\frac{1}{u+v}vyT_2-vyT_{12}\right)\right)
\left(1 + (q^2-1)\frac{1}{u+v}uyT_2\right) = \\ =
{\cal E}_q\left(\frac{1}{u+v}uyT_2\right)
\left(1 + (q^2-1)
\left(yT_2-vyT_{12}\right)\right)
\left({\cal E}_q\left(\frac{1}{u+v}uyT_2\right)\right)^{-1}
\ee
This time one should apply (\ref{EA-B}) at the r.h.s. and show:
\be
(q^2-1)\left\{\frac{1}{u+v}vy\frac{1}{u+v}uy\ T_2^2\ - \
vy\frac{1}{u+v}uy\ T_{12}T_2\right\} = \\ =
\left[\frac{1}{u+v}uy,\  y \right]T_2^2 -
\left[\frac{1}{u+v}uyT_2,\ vyT_{12}\right] + \ \ldots
\label{ide.1.10}
\ee
where ``$\ldots$'' stands for double and higher commutators.
Since $uy$ commutes with $vy$, we have
$\left(\frac{1}{u+v}uy\right)(vy) =
q(vy)\left(\frac{1}{u+v}uy\right)$, and the second commutator at the
\be
q^{\Delta^*(\phi_{(1)})} = q^{\phi_{(1)}'}(1 + \psi_{(2)}'\chi_{(2)}''
+ \psi_{(2)}'\psi_{(3)}'\chi_{(3)}''\chi_{(2)}'')q^{\phi_{(1)}''} = \\
= q^{\phi_{(1)}'}(1 + z_{(22)} + qz_{(33)}z_{(22)})
q^{\phi_{(1)}''}, \\
\Delta^*(\psi_{(2)}) =  \psi_{(2)}'' +
q^{-\Delta^*(\phi_{(1)})} \left( q^{\phi_{(1)}'}\psi_{(2)}'
\{1 + \psi_{(3)}'(\chi_{(1)}''+\chi_{(3)}'')\}q^{\phi_{(2)}''}\right)
= \\ = \psi_{(2)}''  +
q^{-\phi_{(1)}''}\frac{1}{1 + z_{(22)} + qz_{(33)}z_{(22)}}
(1 + z_{(31)}z_{(33)})q^{\phi_{(2)}''}, \\
\Delta^*(\psi_{(3)}) =  \psi_{(3)}'' + \\ +
\left(q^{\Delta^*(\phi_{(1)})}\Delta^*(\psi_{(2)})\right)^{-1}
q^{\phi_{(1)}'}\left\{\psi_{(2)}'
(1 + \psi_{(3)}'(\chi_{(1)}''+\chi_{(3)}''))
q^{\phi_{(2)}'}\psi_{(1)}''
+ \psi_{(2)}'\psi_{(3)}' q^{\phi_{(3)}'}\right\} = \\ =
\psi_{(3)}'' \ +\
\left\{(1 + z_{(22)} + qz_{(33)}z_{(22)})
q^{\phi_{(1)}''}\psi_{(2)}'' + \psi_{(2)}'
(1 + z_{(31)} + z_{(33)})q^{\phi_{(2)}''}\right\}^{-1}\cdot\\ \cdot
\left\{\psi_{(2)}'
(1 + z_{(31)} + z_{(33)})q^{\phi_{(2)}'}\psi_{(1)}''
+ \psi_{(2)}'\psi_{(3)}' q^{\phi_{(3)}'}\right\},
\\
\ldots
\label{sl3del*}
\ee
Here $z_{(s_1s_2)} \equiv \psi_{s_1}'\chi_{s_2}''$.

r.h.s. of (\ref{ide.1.10}) is equal to
$vy\frac{1}{u+v}uy(qT_1T_{12} - T_{12}T_2)$. It remains to use
Serre identity, $T_2T_{12} = qT_{12}T_2$, in order to see that this
coincides with the second term at the l.h.s. of (\ref{ide.1.10}).
The first terms at both sides also coincide, as a result of
identity
\be
\left[\frac{1}{u+v}uy,\ y\right] =
(q^2-1)\frac{1}{u+v}vy\frac{1}{u+v}uy,
\ee
which is most easily derived just by expansion in powers of $v$:
$$
\sum_{n\geq 0}\left\{(-\frac{1}{u}v)^n y - y(-\frac{1}{u}v)^n\right\} =
-(q^2-1)\sum_{k,l\geq 0}(-\frac{1}{u}v)^{k+1}y(-\frac{1}{u}v)^l
$$
Since $y(\frac{1}{u}v) = q^2(\frac{1}{u}v)y$ the both sides here
are indeed equal:
$$
\sum_{n\geq 0}(1-q^{2n})(-\frac{1}{u}v)^n y
= -(q^2-1)\sum_{n\geq 1}\left\{(-\frac{1}{u}v)^n y
\sum_{l=0}^{n-1}q^{2l}\right\}
$$
In the same manner one can easily show that double commutators
and thus all the other terms, substituted by ``$\ldots$'' in
(\ref{ide.1.10}) are vanishing and this completes the check
 of one of the last equation (\ref{eq.ide.1}) for $F(u,v,y)$.

\subsubsection{$F(u,v,y)$ in a factorized form}

In order to check the other two equations
one should first bring $F(u,v,y)$ to the form
which is adequate for application of shift operators.
Let us note that Faddeev-Volkov identity (\ref{FVc}),\footnote{
Of course (\ref{FVcc}) can itself be proved by exactly the same
method of finite-difference equations. Namely, the commutant
$K(x,y) = {\cal E}_{1/q}(-y){\cal E}_q(x){\cal E}_q(y)
{\cal E}_{1/q}(-x)$ satisfies:
$$
M^{+2}_xK(x,y) = K(q^2x,y) =
{\cal E}_{1/q}(-y){\cal E}_q(x)(1+(q^2-1)x){\cal E}_q(y)
{\cal E}_{1/q}(-x)\{1+(q^2-1)x\}^{-1} =
$$
$$
= {\cal E}_{1/q}(-y){\cal E}_q(x){\cal E}_q(y)
\left\{1 + (q^2-1)(x + [x,y] + \ldots)\right\}
{\cal E}_{1/q}(-x)\{1+(q^2-1)x\}^{-1} =
$$
$$
= {\cal E}_{1/q}(-y){\cal E}_q(x){\cal E}_q(y){\cal E}_{1/q}(-x)
\left\{1 + (q^2-1)(x + [x,y] + [x[xy]] + \ldots)\right\}
\{1+(q^2-1)x\}^{-1}
$$
In this calculation (\ref{EAB}) and (\ref{EA-B}) are used and
``$\ldots$'' stand for the terms which contain $[x,y]_q = 0$
(this is why condition $xy=q^2yx$ implies tremendous simplifications).
Also $1 + (q^2-1)(x + [x,y] + [x[xy]]) =
\{1 + (q^2-1)[x,y]\}\{1+(q^2-1)x\}$, and we get:
$$
M^{+2}_xK(x,y) = K(x,y)\{1 + (q^2-1)[x,y]\}
$$
Similarly
$$
M^{+2}_yK(x,y) = \{1 + (q^2-1)[x,y]\}K(x,y)
$$
Thus for $xy = q^2yx$
$$
K(x,y) = {\cal E}_q\left([x,y]\right) =
{\cal E}_q\left((1-q^{-2})xy\right)
$$
Application of addition rule (\ref{sufo}) now gives the
Faddeev-Volkov identity in the form (\ref{FVcc}).
}
\be
{\cal E}_q(x+y + (1-q^{-2})xy) = {\cal E}_q(x){\cal E}_q(y),
\ \ \ {\rm if} \ \ xy = q^2yx
\label{FVcc}
\ee
with the help of (\ref{sufo}) can be transformed into
$$
\left({\cal E}_q(x)\right)^{-1}{\cal E}_q\left(\{1 +
(1-q^{-2})x\}y\right){\cal E}_q(x) = {\cal E}_q(y)
$$
or
$$
{\cal E}_q(y) {\cal E}_q\left(x\{1 +
(1-q^{-2})y\}\right)\left({\cal E}_q(y)\right)^{-1}
= {\cal E}_q(x)
$$
which after the change of variables
$$
\{1 + (1-q^{-2})x\}y \rightarrow y,\ \ \
x \rightarrow \frac{x}{1-q^{-2}} = \frac{qx}{q-q^{-1}}
$$
or
$$
x\{1 + (1-q^{-2})y\} = x\{1 + q^{-2}(q^{2}-1)y\} \rightarrow y,
\ \ \ y \rightarrow \frac{x^{-1}}{q^2-1} =
\frac{q^{-1}x^{-1}}{q-q^{-1}}
$$
respectively implies the two formulas (of which only one was
so far used in this paper):
\be
{\cal E}_q\left(\frac{1}{1+x}y\right) =
{\cal E}_{1/q}\left(-\frac{qx}{q-q^{-1}}\right)
{\cal E}_q(y){\cal E}_q\left(\frac{qx}{q-q^{-1}}\right), \\
{\cal E}_q\left(\frac{1}{1+x^{-1}}y\right) =
{\cal E}_q\left(y\frac{1}{1+q^{-2}x^{-1}}\right) =
{\cal E}_q\left(\frac{q^{-1}x^{-1}}{q-q^{-1}}\right)
{\cal E}_q(y)
{\cal E}_{1/q}\left(-\frac{q^{-1}x^{-1}}{q-q^{-1}}\right)
\ee
Since $(v^{-1}u)y = q^2y(v^{-1}u)$ and $u^{-1}v = (v^{-1}u)^{-1}$
we can use these two identites to get:
\be
{\cal E}_q\left(\frac{1}{u+v}vyT_2\right) =
{\cal E}_q\left(\frac{1}{1+v^{-1}u}yT_2\right) =
{\cal E}_{1/q}\left(-\frac{qv^{-1}u}{q-q^{-1}}\right)
{\cal E}_q(yT_2){\cal E}_q\left(\frac{qv^{-1}u}{q-q^{-1}}\right),\\
{\cal E}_q\left(\frac{1}{u+v}uyT_2\right) =
{\cal E}_q\left(\frac{1}{1+u^{-1}v}yT_2\right) =
{\cal E}_q\left(\frac{q^{-1}u^{-1}v}{q-q^{-1}}\right)
{\cal E}_q(yT_2)
{\cal E}_{1/q}\left(-\frac{q^{-1}u^{-1}v}{q-q^{-1}}\right)
\ee
so that
\be
F(v,u,y) = {\rm l.h.s.\ of}\ (\ref{ide.1}) = \\ =
{\cal E}_{1/q}\left(-\frac{qv^{-1}u}{q-q^{-1}}\right)
{\cal E}_q(yT_2){\cal E}_q\left(\frac{qv^{-1}u}{q-q^{-1}}\right)
{\cal E}_q(uT_1)\cdot\\ \cdot {\cal E}_q(vT_1)
{\cal E}_q\left(\frac{q^{-1}u^{-1}v}{q-q^{-1}}\right)
{\cal E}_q(yT_2)
{\cal E}_{1/q}\left(-\frac{q^{-1}u^{-1}v}{q-q^{-1}}\right)
\label{Fmod}
\ee

\subsubsection{The action of $M^2_u$ on $F(u,v,y)$}

Now, according to (\ref{qexpprod1}), the shift operators $M^2_{u,v}$
act on (\ref{Fmod}) by multiplying or dividing the
relevant items by linear functions of the arguments, and it is
easy to show that, for example,
\be
M_u^2{\cal E}_q\left(\frac{1}{u+v}uyT_2\right) =
{\cal E}_q\left(\frac{1}{q^2u+v}q^2uyT_2\right) = \\ =
\left(1 + (q^2-1)v\frac{1}{(u+v)^2}uy T_2\right)
{\cal E}_q\left(\frac{1}{u+v}uyT_2\right)
\label{Mu.1}
\ee
and
$$
M_u^2{\cal E}_q\left(\frac{1}{u+v}vyT_2\right) =
{\cal E}_q\left(\frac{1}{q^2u+v}vyT_2\right) =
{\cal E}_q\left(\frac{1}{u+v}vyT_2\right)\frac{1}{
1 + (q^2-1)v\frac{1}{(u+v)^2}uyT_2}
$$
or
\be
{\cal E}_q\left(\frac{1}{q^2u+v}vyT_2\right)\left(
1 + (q^2-1)v\frac{1}{(u+v)^2}uyT_2\right) =
{\cal E}_q\left(\frac{1}{u+v}vyT_2\right)
\label{Mu.2}
\ee
Since
\be
M^2_uF(u,v,y) = F(q^2u,v,y) = \\ =
{\cal E}_q\left(\frac{1}{q^2u+v}vyT_2\right)
{\cal E}_q(q^2uT_1){\cal E}_q(vT_1)
{\cal E}_q\left(\frac{1}{q^2u+v}q^2uyT_2\right)
\ee
we substitute (\ref{Mu.1}) and push the new factor to the left
through $T_1$-containing exponentials (with the help of
(\ref{EAB}) and Serre identities):
\be
{\cal E}_q(q^2uT_1){\cal E}_q(vT_1)
\left\{1 + (q^2-1)v\frac{1}{(u+v)^2}uy T_2\right\} = \\ =
\left\{1 + (q^2-1)\left(u T_1 + v\frac{1}{(u+v)^2}uy T_2
+ v\frac{1}{u+v}uy\bar T_{12}\right)\right\}
{\cal E}_q(uT_1){\cal E}_q(vT_1)
\ee
Note that $\bar T_{12} \equiv T_1T_1 - \frac{1}{q}T_2T_1$
appears in this calculation instead
of $T_{12} \equiv T_1T_2 - qT_2T_1$,
and the appropriate Serre identities are (\ref{serbar}).
In order to obtain the desired relation
$M^2_uF(u,v,y) = \left(1+ (q^2-1)uT_1\right)F(u,v,y)$
one still needs to show that
\be
\left(1+ (q^2-1)uT_1\right)
{\cal E}_q\left(\frac{1}{u+v}vyT_2\right) = \\ =
{\cal E}_q\left(\frac{1}{q^2u+v}vyT_2\right)
\left\{1 + (q^2-1)\left(u T_1 + v\frac{1}{(u+v)^2}uy T_2
+ v\frac{1}{u+v}uy\bar T_{12}\right)\right\}
\label{Mu.2.1}
\ee
This is an easy corollary of (\ref{Mu.2}), in fact (\ref{Mu.2.1})
is a sum of two identites: (\ref{Mu.2}) itself and
\be
uT_1{\cal E}_q\left(\frac{1}{u+v}vyT_2\right) =
{\cal E}_q\left(\frac{1}{q^2u+v}vyT_2\right)\left(uT_1 +
v\frac{1}{u+v}uy\bar T_{12}\right)
\ee
This last identity can be proved by expanding $q$-exponential
in a power series and using the fact that due to (\ref{serbar})
$T_1T_2^n = \frac{1}{q^n}T_2^nT_1 + [n]T_2^{n-1}\bar T_{12}$.

This completes the proof of one more equation for $F(u,v,y)$.
The last one - for the action of $M^2_v$ - can be checked just the
same way.

\subsection{On the Campbell-Hausdorff formula \label{CHsec}}

Campbell-Hausdorff formula for the $q$-exponential functions states
that $\forall A,B$
\be
{\cal E}_q(A){\cal E}_q(B) =
{\cal E}_q\left(A + B + \frac{1}{[2]!}[A,B]_{1/q}
+ \right.\\ \left.+
\frac{1}{[3]![2]}\left\{ \left[A,[A,B]_{1/q}\right]_q +
\left[[A,B]_{1/q},B\right]_q\right\} + \right.\\ \left. +
\frac{1}{[4]![2]}\left\{
-\left[A,\left[A,[A,B]_{1/q}\right]_q\right]_{1/q} +
\left[A,\left[A,[A,B]_{1/q}\right]_q\right]_{q} -
\right.\right.\\ \left.\left.
-\left[\left[[A,B]_{1/q},B\right]_q,B\right]_{1/q} +
\left[\left[[A,B]_{1/q},B\right]_q,B\right]_{q} +
\right.\right.\\ \left.\left.
+\left[\left[A,[A,B]_{1/q}\right]_q,B\right]_{1/q^3}
+\left[A,\left[[A,B]_{1/q},B\right]_q\right]_{1/q^3} -
\right.\right.\\ \left.\left.
-\left[\left[A,[A,B]_{q}\right]_{1/q^2},B\right]_{1/q^2}
-\left[A,\left[[A,B]_{1/q},B\right]_q\right]_{1/q} +
\right.\right.\\ \left.\left.
+\left[\left[A,[A,B]_{1/q}\right]_q,B\right]_{q} +
\left[A,\left[[A,B]_{1/q},B\right]_q\right]_{q}
\right\} +\ \ldots\  \right)
\label{CH.1}
\ee
This is just a pure combinatorial relation, no assumption is
made about the properties of ``$A$'' and ``$B$''.
\footnote{
Identities (\ref{sufo})
- for $[A,B]_{1/q}\equiv qAB-\frac{1}{q}BA = 0$ -
and (\ref{FVc}) - for $[A,B]_q \equiv \frac{1}{q}AB - qBA = 0$ -
{\it would be} immediate corollaries of (\ref{CH.1}), if all the
higher commutators in the argument of exponential at the r.h.s.
were proportional to
$\left[A,[A,B]_{1/q}\right]_q =\left[A,[A,B]_{q}\right]_{1/q}$ or
$\left[[A,B]_{1/q},B\right]_q =\left[[A,B]_{q},B\right]_{1/q}$.
However, the story is far less simple, as is clear from occurence
of the item
$\left[\left[A,[A,B]_{q}\right]_{1/q^2},B\right]_{1/q^2}$
in (\ref{CH.1}), which also vanishes in both cases, but does not
contain any vanishing {\it double} commutator.
Representations can be easily found which contain
$q^{\pm 1}$-commutators only, but instead the coefficients in
front of them are no longer $\pm 1$. In particular, the
$A^2B^2$-contribution at the r.h.s. of (\ref{CH.1}) can
be alternatively presented as \cite{KS}
$$
\frac{1}{[4]![2]}\left\{
\frac{1}{4[2]^2}\left(
\left[\left[A,[A,B]_{1/q}\right]_{1/q},B\right]_{1/q}
+\left[A,\left[[A,B]_{1/q},B\right]_{1/q}\right]_{1/q}
-\left[\left[A,[A,B]_{q}\right]_{q},B\right]_{q}
-\left[A,\left[[A,B]_{q},B\right]_{q}\right]_{q}\right) +
\right. $$ $$
+\frac{[3]}{4[2]^2}\left(
\left[\left[A,[A,B]_{q}\right]_{q},B\right]_{1/q}
+\left[A,\left[[A,B]_{q},B\right]_q\right]_{1/q}
-\left[\left[A,[A,B]_{1/q}\right]_{1/q},B\right]_{q}
-\left[A,\left[[A,B]_{1/q},B\right]_{1/q}\right]_{q}\right) +
$$ $$ \left.
+\frac{q^2-1+q^{-2}}{2}\left(
\left[\left[A,[A,B]_{1/q}\right]_q,B\right]_{1/q} +
\left[A,\left[[A,B]_{1/q},B\right]_q\right]_{1/q}
\right)
+\right. $$ $$ \left.
+ \frac{1}{2}\left(
\left[\left[A,[A,B]_{1/q}\right]_q,B\right]_{q} +
\left[A,\left[[A,B]_{1/q},B\right]_q\right]_{q}
\right)   \right\}
$$
Because of occurence of terms like
$\left[\left[[\ ,\ ]_{1/q}\right]_{1/q}\right]_{q^{\pm 1}}$
and
$\left[\left[[\ ,\ ]_{q}\right]_{q}\right]_{q^{\pm 1}}$
this does not make (\ref{sufo}) or (\ref{FVc}) more transparent.
Moreover, now these identities result from cancelation between
different terms. Complicated coefficients (nothing to say about their
dependence on the non-$q$-numbers ``$2$'', ``$4$'' and
``$q^2-1+q^{-2}$'') is another manifestation that such
representations should in fact be irrelevant.

One more identity can be useful for
understanding of the structure of (\ref{CH.1}):
$$
\left[\left[A,[A,B]_{q}\right]_{1/q^2},B\right]_{1/q^2}
+\left[A,\left[[A,B]_{1/q},B\right]_q\right]_{1/q} =
\left[A\left[[A,B]_{q},B\right]_{1/q^2}\right]_{1/q^2}
+\left[\left[A,[A,B]_{1/q}\right]_q,B\right]_{1/q}
$$

It is easy to describe the appearence of the ``highest commutators''
in(\ref{CH.1}).  Let the argument of exponent at the r.h.s. be
$A + B + \frac{1}{[2]}(qAB - q^{-1}BA) +
\sum_{k\geq 2} \gamma_k(A^kB + AB^k ) + \ldots$, where $\ldots$ denote
all the other more complicated terms. Then
coefficients $\gamma_k$ are expressed through Bernulli numbers:
$\gamma(z) = 1 + \frac{1}{[2]}z + \sum_{k\geq 2}\gamma_kz^k =
\frac{z{\cal E}_q(z)}{{\cal E}_q(z)-1} = \frac{z}{1-{\cal E}_{1/q}(-z)} =
1 + \frac{1}{[2]}z + \frac{1}{[2][3]!}z^2 - \frac{q-q^{-1}}{[2][4]!}z^3
+ \frac{q^6 - q^2-2-q^{-2}+q^{-6}}{[2][3]![5]!}z^4 + \ldots$
Note, that while for $q=1$ all the
$\gamma_{2k+1} = 0$ for $k\geq 1$, this is no longer true for $q \neq 1$.
Degrees of the polinomials in $q,q^{-1}$ which appear in the
numerators are easily controlled by consideration of
asymptotics $q^{\pm 1}\rightarrow 0$, which is also useful
for examination of other properties of (\ref{CH.1}), which
drastically simplifies in these limits.
}

In the case of $q=1$ the full expression at the r.h.s. (see,
for example, \cite{Serre}) can be written down in a readable form
with the help of operation $ad$: $\forall A,B\ $
$\ ad_AB \equiv [A,B]$.\footnote{
When $q\neq 1$, there is no such distinguished $ad$:
of equal importance  are $q^n$-commutators with various
$n$'s (also they appear differently in different applications:
say, in (\ref{CH.1}) and in expression for commutants).
If extra gradation with the help of a new variable $t$
is introduced, of real importance is the operation defined by
$t^{n+1}[A,B]_{q^n} = (At\vec M^-_t)(Bt^n) - (Bt^n)(A
\stackrel{\leftarrow}{M^+_t}t)$.
}
To make notations simpler let us also assume that $(ad_A)I \equiv A$.
Then
\be
e^A e^B = \exp\left(\sum_{n=1} \frac{(-)^{n-1}}{n} \left(
\sum_{\stackrel{\{k_i,l_i|i = 1,\ldots, m\}}{k_i+l_i\geq 1}}
\frac{(ad_A)^{k_1}(ad_B)^{l_1}\ldots (ad_A)^{k_m}(ad_B)^{l_m}I}
{k_1!l_1!\ldots k_m!l_m!\sum(k_i+l_i)}\right)\right)
= \\ = \exp\left(
\int_{0}^{1}ds \sum_{n=1} \frac{(-)^{n-1}}{n}
\left(e^{s{\rm ad}_A}e^{s{\rm ad}_B} - 1\right)^{n-1}
\left(A + e^{s{\rm ad}_A}B\right)\right) = \\ =
\exp\left(\int_{0}^{1}ds
\frac{\log\left(e^{s{\rm ad}_A}e^{s{\rm ad}_B}\right)}
{e^{s{\rm ad}_A}e^{s{\rm ad}_B} - 1}
e^{s{\rm ad}_A}(A+B)\right)
\label{CH.cl}
\ee

The crucial feature of (\ref{CH.cl})
is that exponent at the r.h.s. contains only {\it commutators}.
In application to Lie algebras, when $A$ and $B$ are {\it linear}
functions of generators of the {\it algebra} this implies
that exponents map the {\it algebra} into the {\it group}:
group multiplication is again an exponent of a {\it linear}
combination of generators. If one starts not from the whole set
of $d_G$ generators,  but only from those associated with simple
roots, then only finite number ($2(d_G - r_G)$) of new linear
independent combinations ($T_{\vec\alpha}$'s) will be produced -
and this is guaranteed by Serre identities.
Thus, if Lie algebra is taken to be
the original object, it is the Campbell-Hausdorff formula
(supplemented by Serre identities) that ensures the existence of
$d_G$-dimensional group manifold - a small subset of the
universal envelopping algebra, which is invariant under
{\it multiplication}.\footnote{In order to avoid possible
confusion it deserves mentioning that in the fundamental
representation of $GL(N)$ the statement can look trivial,
just because the set of $N^2$ generators form a full basis
in the linear space of $N\times N$ matrices, and it does
not seem surprising that a product of exponents of generators
is again such exponent with some sophisticated mapping of
parameters. However, this argument can not explain, why the
same feature will be preserved in any other representation,
when the entire universal envelopping algebra is no longer
{\it linearly} generated by original generators
of the Lie algebra. Moreover, at the level of fundamental
representation exponential functions do not look distinguished:
the reasoning would work for {\it any} other function. In fact
it is the Campbell-Hausdorff formula that provides real
explanations.}

In order to prove the existence of $d_G$-parametric ``group''
(with operator-valued ``coordinates'')
when $q\neq 1$ one needs a ``quantum'' Campbell-Hausdorff
formula. The new thing is that while for $q = 1$ it was just enough
to obtain  {\it commutators} and nothing else
at the r.h.s. of (\ref{CH.1}),
for $q\neq 1$ it is important that the right $q$-commutators
appear at the right places: they should be adjusted in order to
match with the Serre identites. Moreover, such matching occurs
only after all the $q$-factors coming from permutations of
$\chi, \phi, \psi$ variables are taken into account.
This Appendix, refering to $SL(3)_q$ as the simplest example,
presents some fragments of the future construction, and
can probably convince the reader that the entire construction
can also be built and the notion of ``quantum group'' can indeed
make sense.

\end{document}